\documentclass[longauth]{aa}

\usepackage{graphicx}
\usepackage{txfonts}
\usepackage{gensymb,amssymb}
\usepackage{xcolor}
\usepackage[version=3]{mhchem}
\usepackage{caption,subcaption}
\usepackage{multirow}
\usepackage{ulem}

\usepackage[colorlinks = true,
            linkcolor = blue,
            urlcolor  = blue,
            citecolor = blue,
            anchorcolor = blue]{hyperref}

\begin{document}

\title{MINDS. The DR~Tau disk I: combining JWST-MIRI data with high-resolution \ce{CO} spectra to characterise the hot gas}
\titlerunning{MINDS. The DR~Tau disk I: combining JWST-MIRI data with high-resolution \ce{CO} spectra}
\author{Milou Temmink\orcid{0000-0002-7935-7445}\inst{1} \and
        Ewine F. van Dishoeck\orcid{0000-0001-7591-1907}\inst{1,2} \and
        Sierra L. Grant\orcid{0000-0002-4022-4899}\inst{2} \and 
        Beno\^it Tabone\inst{3} \and
        Danny Gasman\orcid{0000-0002-1257-7742}\inst{4} \and
        Valentin Christiaens\orcid{0000-0002-0101-8814}\inst{4,5} \and
        Matthias Samland\orcid{0000-0001-9992-4067}\inst{6} \and
        Ioannis Argyriou\orcid{0000-0003-2820-1077}\inst{4} \and
        Giulia Perotti\orcid{0000-0002-8545-6175}\inst{6} \and
        Manuel G\"udel\orcid{0000-0001-9818-0588}\inst{7,8} \and
        Thomas Henning\orcid{0000-0002-1493-300X}\inst{6} \and
        Pierre-Olivier Lagage\inst{9} \and 
        Alain Abergel\inst{3} \and
        Olivier Absil\orcid{0000-0002-4006-6237}\inst{5} \and
        David Barrado\orcid{0000-0002-5971-9242}\inst{10} \and
        Alessio Caratti o Garatti\orcid{0000-0001-8876-6614}\inst{11,12} \and
        Adrian M. Glauser\orcid{0000-0001-9250-1547}\inst{8} \and
        Inga Kamp\orcid{0000-0001-7455-5349}\inst{13} \and
        Fred Lahuis \inst{14} \and
        G\"oran Olofsson\orcid{0000-0003-3747-7120}\inst{15} \and
        Tom P. Ray\orcid{0000-0002-2110-1068}\inst{12} \and
        Silvia Scheithauer\orcid{0000-0003-4559-0721}\inst{6} \and
        Bart Vandenbussche\orcid{0000-0002-1368-3109}\inst{4} \and
        L. B. F. M. Waters\orcid{0000-0002-5462-9387}\inst{16,14} \and
        Aditya M. Arabhavi\orcid{0000-0001-8407-4020}\inst{13} \and
        Hyerin Jang\orcid{0000-0002-6592-690X}\inst{16} \and
        Jayatee Kanwar\orcid{0000-0003-0386-2178}\inst{13,17,18} \and
        Maria Morales-Calder\'on\orcid{0000-0001-9526-9499}\inst{10} \and
        Donna Rodgers-Lee\orcid{0000-0002-0100-1297}\inst{12} \and
        J\"urgen Schreiber\inst{6} \and
        Kamber Schwarz\orcid{0000-0002-6429-9457}\inst{6} \and
        Luis Colina\inst{19}}
\institute{Leiden Observatory, Leiden University, 2300 RA Leiden, the Netherlands \\
          \email{temmink@strw.leidenuniv.nl} \and
          Max-Planck-Institut f\"ur Extraterrestrische Physik, Giessenbachstraße 1, D-85748 Garching, Germany \and
          Universit\'e Paris-Saclay, CNRS, Institut d’Astrophysique Spatiale, 91405, Orsay, France \and 
          Institute of Astronomy, KU Leuven, Celestijnenlaan 200D, 3001 Leuven, Belgium \and
          STAR Institute, Universit\'e de Li\`ege, All\'ee du Six Ao\^ut 19c, 4000 Li\`ege, Belgium \and
          Max-Planck-Institut f\"{u}r Astronomie (MPIA), K\"{o}nigstuhl 17, 69117 Heidelberg, Germany \and
          Dept. of Astrophysics, University of Vienna, T\"urkenschanzstr. 17, A-1180 Vienna, Austria \and
          ETH Z\"urich, Institute for Particle Physics and Astrophysics, Wolfgang-Pauli-Str. 27, 8093 Z\"urich, Switzerland \and
          Universit\'e Paris-Saclay, Universit\'e Paris Cit\'e, CEA, CNRS, AIM, F-91191 Gif-sur-Yvette, France \and
          Centro de Astrobiolog\'ia (CAB), CSIC-INTA, ESAC Campus, Camino Bajo del Castillo s/n, 28692 Villanueva de la Ca\~nada, Madrid, Spain \and
          INAF – Osservatorio Astronomico di Capodimonte, Salita Moiariello 16, 80131 Napoli, Italy \and
          Dublin Institute for Advanced Studies, 31 Fitzwilliam Place, D02 XF86 Dublin, Ireland \and
          Kapteyn Astronomical Institute, Rijksuniversiteit Groningen, Postbus 800, 9700AV Groningen, The Netherlands \and
          SRON Netherlands Institute for Space Research, Niels Bohrweg 4, NL-2333 CA Leiden, the Netherlands \and
          Department of Astronomy, Stockholm University, AlbaNova University Center, 10691 Stockholm, Sweden \and
          Department of Astrophysics/IMAPP, Radboud University, PO Box 9010, 6500 GL Nijmegen, The Netherlands \and 
          Space Research Institute, Austrian Academy of Sciences, Schmiedlstr. 6, A-8042, Graz, Austria \and
          TU Graz, Fakultät für Mathematik, Physik und Geodäsie, Petersgasse 16 8010 Graz, Austria \and
          Centro de Astrobiolog\'ia (CAB, CSIC-INTA), Carretera de Ajalvir, E-28850 Torrej\'on de Ardoz, Madrid, Spain}
\date{Received 11/12/2023; accepted 20/03/2024}

%% ---

\abstract
{The MRS mode of the JWST-MIRI instrument has been shown to be a powerful tool to characterise the molecular gas emission of the inner region of planet-forming disks. Investigating their spectra allows to infer the composition of the gas in these regions and, subsequently, the potential atmospheric composition of the forming planets. We present the JWST-MIRI observations of the compact T-Tauri disk, DR~Tau, which are complemented by ground-based, high spectral resolution ($R\sim$60000-90000) \ce{CO} ro-vibrational observations.}
{The aim of this work is to investigate the power of extending the JWST-MIRI \ce{CO} observations with complementary, high-resolution, ground-based observations acquired through the SpExoDisks database, as JWST-MIRI's spectral resolution ($R\sim$1500-3500) is not sufficient to resolve complex \ce{CO} line profiles. In addition, we aim to infer the excitation conditions of other molecular features present in the JWST-MIRI spectrum of DR~Tau and link those with \ce{CO}.}
{The \textnormal{archival }complementary, high-resolution \ce{CO} ro-vibrational observations \textnormal{are analysed with rotational diagrams}. We extend these diagrams to the JWST-MIRI observations by binning and convolution with JWST-MIRI's pseudo-Voigt line profile. In parallel, local thermal equilibrium (LTE) 0D slab models are used to infer the excitation conditions of the detected molecular species.}
%{Various molecular species, including \ce{CO}, \ce{CO_2}, \ce{HCN}, and \ce{C_2H_2}, are detected in the JWST-MIRI spectrum of DR~Tau, with \ce{H_2O} being discussed in a subsequent paper. The high-resolution observations show evidence for two \ce{^{12}CO} components: a broad component (FWHM$\sim$33.5 km s$^{-1}$) tracing the Keplerian disk and a narrow component (FWHM$\sim$11.6 km s$^{-1}$) tracing a slow disk wind. The rotational diagrams yield \ce{CO} excitation temperatures of $T\geq$725 K. Consistently lower excitation temperatures are found for the narrow component, suggesting that the slow disk wind is launched from a larger radial distance. The LTE slab models yield a variety of excitation temperatures ($T\sim$325-900 K) for \ce{^{12}CO_2}, \ce{HCN}, and \ce{C_2H_2}. The inferred excitation temperatures thus suggest that \ce{CO} originates from the highest atmospheric layers close to the host star, followed by \ce{HCN} and \ce{C_2H_2} which emit, together with \ce{^{13}CO}, from slightly deeper layers, whereas the \ce{CO_2} emission originates from even deeper inside or further out the disk. In contrast to the ground-based observations, much higher excitation temperatures are found if only the high-$J$ transitions probed by JWST-MIRI are considered in the rotational diagrams. Additional analysis of the \ce{^{12}CO} line wings suggests a larger emitting area than inferred from the slab models, hinting at a misalignment between inner ($i\sim$20\degree) and outer disk ($i\sim$5\degree). }
{Various molecular species, including \ce{CO}, \ce{CO_2}, \ce{HCN}, and \ce{C_2H_2}, are detected in the JWST-MIRI spectrum of DR~Tau, with \ce{H_2O} being discussed in a subsequent paper. The high-resolution observations show evidence for two \ce{^{12}CO} components: a broad component (FWHM$\sim$33.5 km s$^{-1}$) tracing the Keplerian disk and a narrow component (FWHM$\sim$11.6 km s$^{-1}$) tracing a slow disk wind. The rotational diagrams yield \ce{CO} excitation temperatures of $T\geq$725 K. Consistently lower excitation temperatures are found for the narrow component, suggesting that the slow disk wind is launched from a larger radial distance. \textnormal{In contrast to the ground-based observations, much higher excitation temperatures are found if only the high-$J$ transitions probed by JWST-MIRI are considered in the rotational diagrams.} Additional analysis of the \ce{^{12}CO} line wings suggests a larger emitting area than inferred from the slab models, hinting at a misalignment between inner ($i\sim$20\degree) and outer disk ($i\sim$5\degree). \textnormal{Compared to \ce{CO}, we retrieve lower excitation temperatures of $T\sim$325-900 K for \ce{^{12}CO_2}, \ce{HCN}, and \ce{C_2H_2}.} }
{We show that complementary, high-resolution \ce{CO} ro-vibrational observations are necessary to properly investigate the excitation conditions of the gas in the inner disk and are required to interpret the spectrally unresolved JWST-MIRI \ce{CO} observations. These additional observations, covering the lower-$J$ transitions, are needed to put better constraints on the gas physical conditions and they allow for a proper treatment of the complex line profiles. Comparison with JWST-MIRI requires the use of pseudo-Voigt line profiles in the convolution rather than simple binning. The combined high-resolution \ce{CO} and JWST-MIRI observations can then be used to characterise the emission, and physical and chemical conditions of the other molecules with respect to \ce{CO}. \textnormal{The inferred excitation temperatures suggest that \ce{CO} originates from the highest atmospheric layers close to the host star, followed by \ce{HCN} and \ce{C_2H_2} which emit, together with \ce{^{13}CO}, from slightly deeper layers, whereas the \ce{CO_2} emission originates from even deeper inside or further out the disk.}}

%% ---

\maketitle

\section{Introduction}
The bulk of the exoplanet population are thought to form in the inner regions ($<$10 au) of planet-forming disks \citep{MorbidelliEA12,DJ18}. The chemical composition of these regions and, subsequently, the accreting planetary atmospheres are dictated by their high temperatures and densities, and the location of the \ce{H_2O} and \ce{CO_2} snowlines (e.g. \citealt{PontoppidanEA14,WalshEA15,OB21,BosmanEA22,MolliereEA22}). Snowlines are the radial distances, that dictate what elements are for 50\% in the gas and that are for 50\% locked-up in the icy mantles of dust grains. \\
\indent \textnormal{Over the recent years, the Atacama Large Millimeter/submillimeter Array (ALMA) has allowed us to characterise the chemical composition of the gas, dust, and ice in the outer regions of planet-forming disks with unprecedented sensitivity, as over 30 molecular species and 17 rare isotopologues have been detected (e.g. \citealt{BoothEA21,MarelEA21,BrunkenEA22,FuruyaEA22,McGuire22,ObergEA23,BoothEA24})}. The composition of the inner planet-forming zone is, however, hard to probe with ALMA\textnormal{, as an angular resolution of a few milli-arcseconds is required to resolve structures at 1 au scales in the nearest star forming regions. These resolutions can only be achieved in the most extended configurations (C-10 with baselines of 16.2 km) for the highest frequencies (602-720 GHz in Band 9 and 787-950 GHz in Band 10) and are limited in senstivity the only the dust continuum (see, for example, \citealt{AndrewsEA18}).} \\
\indent Recently, the Medium Resolution Spectroscopy (MRS; \citealt{WellsEA15,ArgyriouEA23}) mode of the Mid-InfraRed Instrument (MIRI; \citealt{RiekeEA15,WrightEA15,WrightEA23}) aboard the \textit{James Webb Space Telescope} (JWST; \citealt{RigbyEA23}) provides a new opportunity to study the inner planet-forming regions. Within its wide wavelength range (4.9 to 28.1 $\mu$m) a large ensemble of molecular species (including \ce{CO}, \ce{CO_2} and \ce{H_2O}) are covered. As JWST-MIRI only covers the highest $J$-transitions of the $P$-branch of \ce{CO} (i.e. $J\geq$25 for the \ce{^{12}CO} v=1-0 $P$-branch) at low spectral resolution ($R\sim$3500) and many T-Tauri disks are too bright for JWST-NIRSpec, complementary observations are required to fully investigate the excitation and kinematics of \ce{CO}. These spectrally-resolved observations can be obtained with ground-based telescopes, such as VLT-CRIRES(+), IRTF-iSHELL and/or Keck-NIRSPEC (e.g. \citealt{SalykEA11,BrownEA13,BanzattiEA22,GrantEA23b}). \\
\indent In this work, we present the JWST-MIRI observations of DR~Tau, an actively accreting ($\Dot{M}\simeq$2$\times$10$^{-7}$ M$_\odot$ yr$^{-1}$; \citealt{ManaraEA22}) T-Tauri star ($M\simeq$0.93 M$_\odot$, $L\simeq$0.63 L$_\odot$; \citealt{LongEA19}) located at a distance of $\sim$195 pc \citep{GC18} in the Taurus star-forming region. In addition, we use complementary, high spectral resolution \ce{CO} ro-vibrational observations to fully investigate the \ce{CO} excitation properties. \\
\indent The chemical composition of its inner disk has been revealed with instruments such as VLT-CRIRES and \textit{Spitzer}. Available spectra show that lines originating from the inner part of the DR~Tau disk have one of the highest line-to-continuum ratio and that a wide variety of molecular species are present (\ce{CO}, \ce{CO_2}, \ce{OH}, \ce{H_2O}, \ce{HCN} and \ce{C_2H_2}; \citealt{BastEA11,MandellEA12,BrownEA13,BanzattiEA20}). As we will show, these are all also detected with JWST-MIRI. Other works on high spectral-resolution, infrared observations have shown that the \ce{^{12}CO} line profile can consist of two components: one narrow component related to a possible disk wind, while the other one is broad and the result of the disk's Keplerian rotation \citep{BastEA11,BrownEA13,BanzattiEA22}. We will revisit this analysis with recent high-resolution observations. As JWST-MIRI also observes part of the \ce{CO} ro-vibrational ladder, it will be of interest to investigate how these observations of a limited number of high-$J$ levels compare with the full range of $J$-levels covered from the ground, especially since the disk of DR~Tau is too bright for JWST-NIRSpec observations. Additionally, we investigate how the emitting radius provided by LTE slab models to fit the \ce{CO} rotational ladder compares with the expected inner and emitting radius of \ce{CO}, that can be derived from optically thin and/or well resolved line profiles using Kepler's law \citep{SalykEA11b,BP15,BanzattiEA20}. \\
The outer disk can, on the other hand, be investigated with millimeter instruments. ALMA observations have shown that the dust disk is very compact, with a 95\% effective radius of $\sim$0.28" or $\sim$54 au \citep{LongEA19}. At the current resolution ($\sim$0.1") of the ALMA data, no dust substructures have been detected, however, visibility- and super-resolution fitting techniques suggest that the disk may be comprised of two rings in the outer disk (respectively at $\sim$22 au and $\sim$42.5 au; \citealt{JenningsEA20,JenningsEA22,ZhangEA23}). Furthermore, \citet{JenningsEA22} note that the visibilities hint at underresolved structures at small spatial scales, which likely point towards a partially resolved inner disk. The structure of the innermost region of the disk can be characterised with instruments on the Very Large Telescope Interferometer (VLTI) and the GRAVITY instrument has, in particular, shown that the inner disk of DR~Tau ($i\sim$18\degree; \citealt{GravityEA23}) may be misaligned with respect to the outer disk ($i$=5.4; \citealt{LongEA19}). \\
\indent The compact size of the millimetre-sized dust disk and potential hints of substructures make DR~Tau a very interesting target to explore the role of the inward drift of icy pebbles in enhancing the emission seen in the inner disk. \ce{H_2O} will be of importance in investigating the significance of radial drift, as shown by \citet{BanzattiEA20}.  In addition, \citet{BraunEA21} derive a gas disk radius of 246 au based on ALMA observations of \ce{^{13}CO} and \ce{C^{18}O}. This yields a gas-to-dust disk radius ratio of $\sim$4.6. As noted by \citet{TrapmanEA19} a ratio of $>$4 is a clear sign of dust evolution and radial drift, suggesting that drift is very effective in the disk of DR~Tau and that the molecular emission in the inner disk may be enhanced. \\
\indent This paper is structured as follows: in \textnormal{Sections \ref{sec:Obs} and \ref{sec:Anal}} we describe\textnormal{, respectively,} the data acquisition, reduction, and analysis methods, while Section \ref{sec:RD} contains our results. The discussions can be found in Section \ref{sec:Disc}. Section \ref{sec:CS} contains the conclusions and a small summary. Furthermore, the \ce{H_2O} emission observed in DR~Tau will be investigated and discussed together with \ce{OH} and first limits on \ce{H_I} and \ce{H_2} in a separate paper (Temmink et al. in prep.). Any \ce{H_2O} slab models displayed will be further explained in that paper.

\section{Observations} \label{sec:Obs}
\textnormal{In the following subsections we describe the used observations (JWST-MIRI in Section \ref{sec:Obs-JM} and IRTF-iSHELL in Section \ref{sec:HighRes-CO}) and data reduction techniques. Additionally, we introduce our continuum subtraction method in Section \ref{sec:ContinuumSubtraction}. }

\subsection{JWST-MIRI observations} \label{sec:Obs-JM}
\begin{figure*}[ht!]
    \centering
    \includegraphics[width=\textwidth]{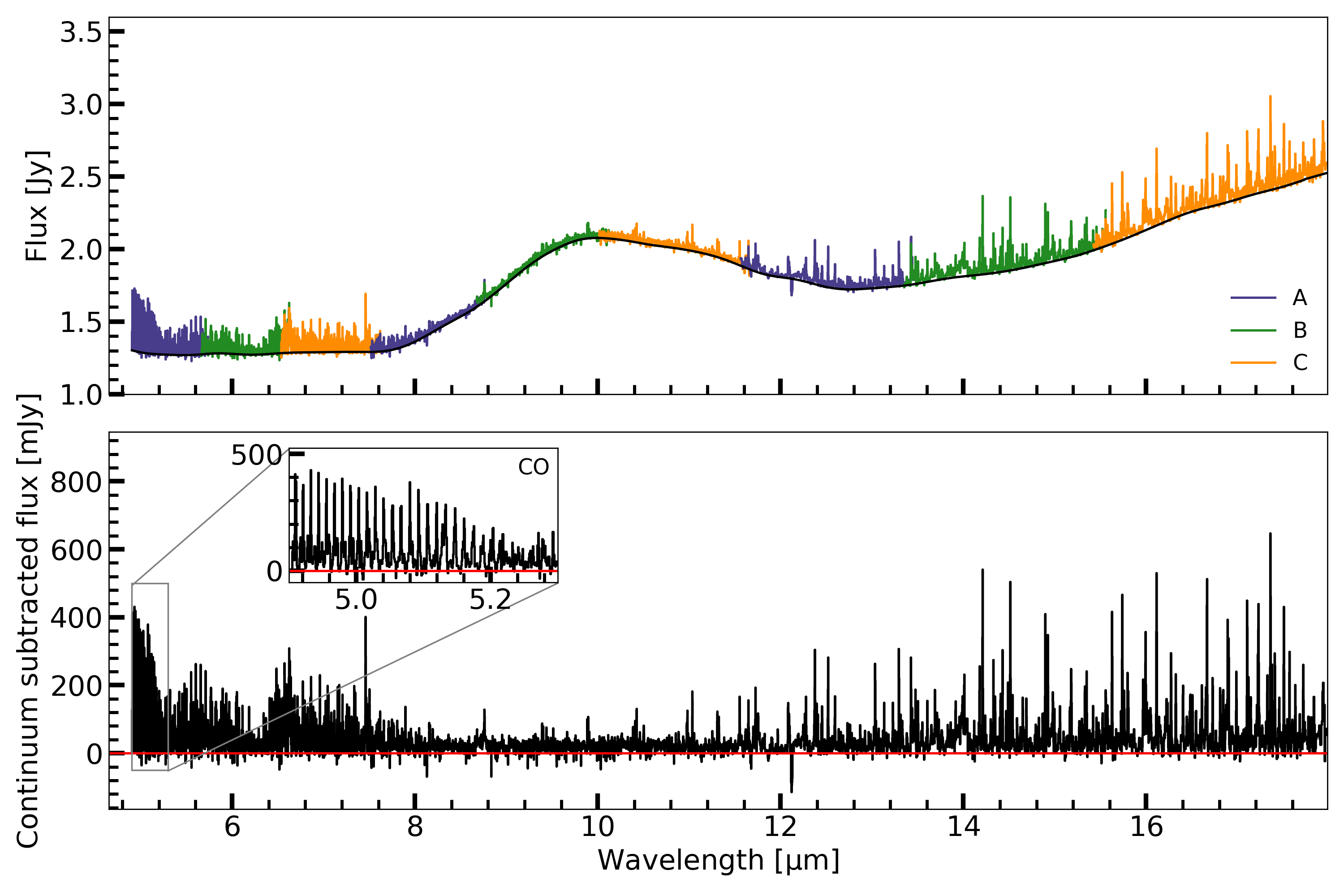}
    \caption{The JWST-MIRI MRS spectrum of DR~Tau over channels 1 through 3 (4.9-17.98 $\mu$m). The different wavelength ranges (subbands) of each MIRI channel are indicated in the top panel in blue (`A'), green (`B'), and orange (`C'), respectively. The black line displays the estimated continuum. The bottom panel shows the continuum subtracted JWST-MIRI spectrum and contains a zoom-in on the \ce{CO} transitions (4.9-5.3 $\mu$m). The red line in the bottom panel highlights the zero flux level.}
    \label{fig:FullSpectrum}
\end{figure*}

The MIRI/MRS observations of DR~Tau have been taken as part of the MIRI mid-INfrared Disk Survey (MINDS) JWST Guaranteed Time Observations Program (PID: 1282, PI: T. Henning). The MRS mode involves 4 Integral Field Units: channel 1 (4.9-7.65 $\mu$m), channel 2 (7.51-11.71 $\mu$m), channel 3 (11.55-18.02 $\mu$m), and channel 4 (17.71-27.9 $\mu$m). As each channel consists of 3 wavelength subbands (respectively, A, B, and C), the MRS observations are comprised of a total of 12 wavelength bands.  The data were taken in \textsc{FASTR1} readout mode on March 4 2023, using a four-point dither pattern and a total on-source integration time of $\sim$27.8 minutes. \\
\indent A combination of the standard JWST pipeline (version 1.11.2, \citealt{BushouseEA23}) and routines from the VIP package \citep{GomezGonzalezEA17,ChristiaensEA23}, to compensate for known issues from the standard pipeline, have been used to reduce the data. Similar to the standard pipeline, the reduction pipeline is structured around the 3 main stages: Detector1, Spec2, and Spec3. We have skipped the background subtraction step in the Spec2 stage, and the outlier detection step in Spec3. The outlier detection step is known to introduce spurious artefacts in unresolved sources due to undersampling of the point-spread function (PSF). To compensate for the outlier detection, we have implemented the bad pixel correction routines from the VIP package. The bad pixels are identified through sigma-filtering and corrected through a Gaussian kernel. VIP-based routines have also been used for the centroiding of the central star. The frames of each spectral cube were aligned through an image cross-correlation algorithm, before a 2D Gaussian fit was used to identify the centroid location in the weighted mean image of the spectral cubes. These centroid locations are, subsequently, used for aperture photometry. We have extracted the spectrum by summing the signal in a 2.0 full-width half maximum (FWHM=1.22 $\lambda/D$, where $D\sim$6.5 m is the diameter of the telescope) aperture centered on the source. To adjust for background emission, we have estimated the background from an annulus directly surrounding the aperture used for the aperture photometry. Aperture correction factors are used to account for the flux of the PSF outside the aperture and inside the annulus \citep{ArgyriouEA23}. \\
\indent The calibrated JWST-MIRI spectrum of DR~Tau over channels 1 through 3 is presented in Figure \ref{fig:FullSpectrum}. The different MIRI subbands are displayed in different colours.  

\subsection{Continuum subtraction} \label{sec:ContinuumSubtraction}
Due to the line-richness of the DR~Tau spectrum, we employed a different continuum subtraction method compared to \citet{GrantEA23} and \citet{GasmanEA23Subm}. The method here relies on a baseline estimation using the \textsc{PyBaselines} package \citep{PyBaselines}. Before estimating the baseline, we masked downwards spikes, such that they cannot influence the estimate. The downward spikes were masked using the following method: the continuum level was roughly estimated using a Savitzky-Golay filter, consisting of window sizes of 100 data points and polynomials of order 3. Through an iterative process, we removed all emission lines which extended above 2$\sigma$ of the standard deviation (STD) of the Savitzky-Golay filter. Once no emission lines remained, we subtracted the final Savitzky-Golay filter and considered all 3$\sigma$ downward outliers from the residuals as downwards spikes. These downwards spikes were subsequently masked from the baseline estimation on the full spectrum. The baseline was estimated using the `Iterative Reweighted Spline Quantiale Regression' (IRSQR) method, which uses penalised splines and an iterative reweighted least squares technique to apply quantile (using a value of 0.05) regression. Within the method, we have used cubic splines with knots every 75 points and differential matrices of order 3. \\
\indent The estimated baseline and continuum-subtracted spectrum are displayed in the top and bottom panel of Figure \ref{fig:FullSpectrum}, respectively. Appendix \ref{app:CS} shows and explains the intermediate steps from the applied continuum subtraction method.

\subsection{High-resolution \ce{CO} observations} \label{sec:HighRes-CO}
\begin{figure*}[ht!]
    \centering
    \includegraphics[width=\textwidth]{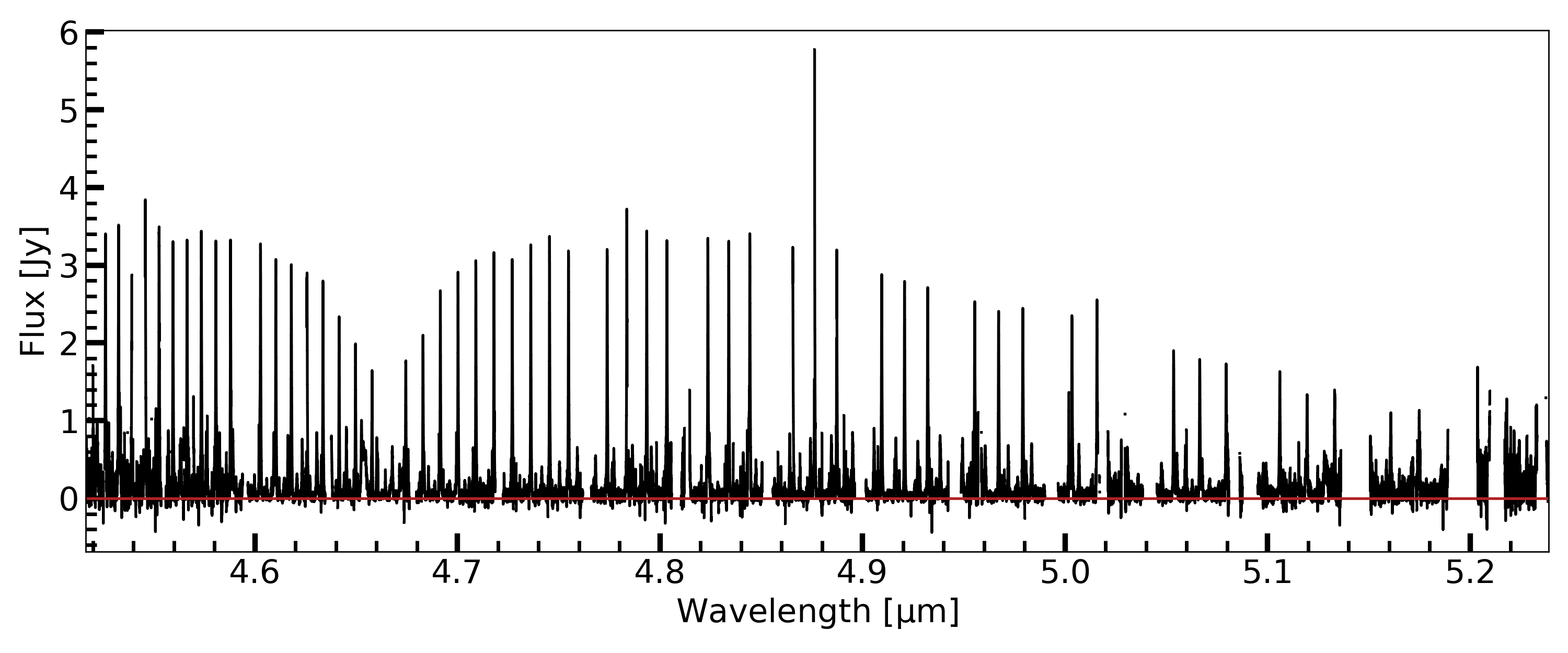}
    \caption{The continuum-subtracted iSHELL spectrum of DR~Tau. The wavelength gaps visible in the spectrum have either not been observed, or were removed due to interference from telluric lines.}
    \label{fig:iSHELL-Spectrum}
\end{figure*}

\begin{figure*}[ht!]
    \centering
    \includegraphics[width=\textwidth]{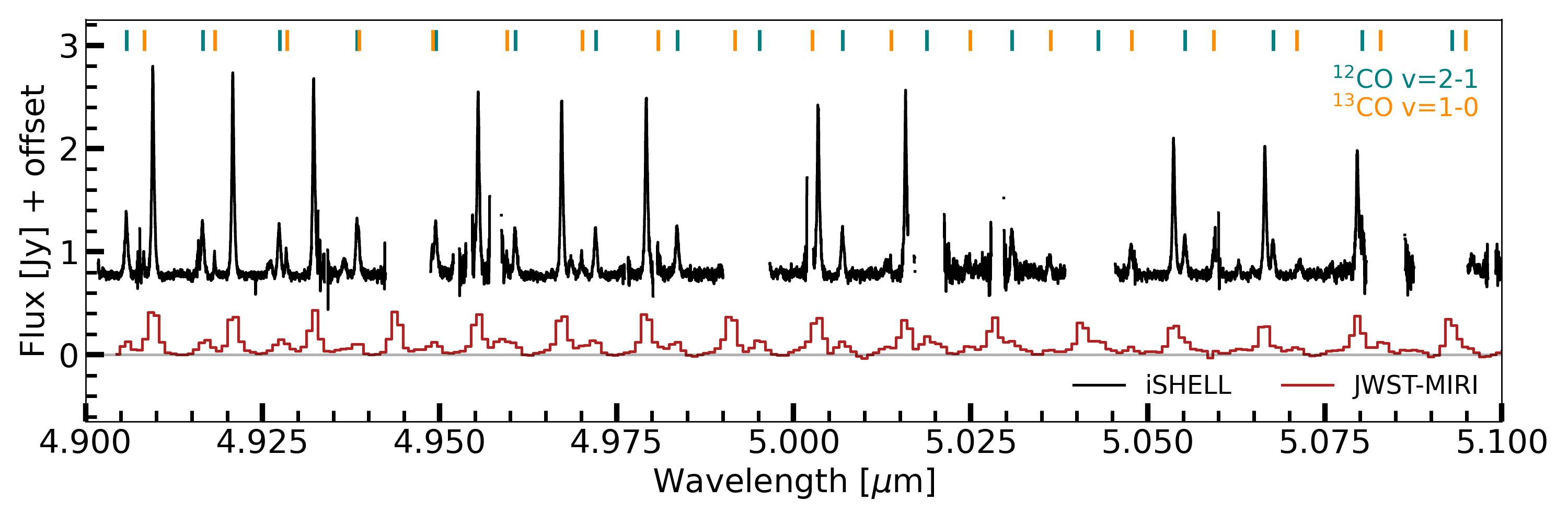}
    \caption{A comparison between the JWST-MIRI (red) and iSHELL (black) data within the wavelength region of 4.9-5.1 $\mu$m, to highlight the differences in flux and resolution. The iSHELL observations are offset by 0.75 Jy. The vertical lines at the top of the plot indicate the positions of the \ce{^{12}CO} v=2-1 (teal) and \ce{^{13}CO} v=1-0 (orange) transitions, to highlight the main contributors of the secondary \ce{CO} emission peaks visible in this wavelength region.}
    \label{fig:iSHELL-MIRI}
\end{figure*}

To complement the high-$J$ levels of the \ce{CO} $P$-branch visible in the JWST spectrum at the shortest wavelengths ($<$5.3 $\mu$m), we have acquired high-resolution ($R\simeq$60000-92000) IRTF-iSHELL data through SpExoDisks\footnote{SpExoDisks: \url{https://www.spexodisks.com/}} (Wheeler et al. in prep.). The iSHELL data were taken on January 23 2022 (see also \citealt{BanzattiEA22}), were flux calibrated using the allWISE catalogue (WISE-W2: $\lambda$=4.6 $\mu$m, $F$=1.91$\pm$0.16 Jy; \citealt{WISE,AllWISE}), and have been corrected for telluric features. As the allWISE flux calibration yields a small offset compared to the JWST-MIRI continuum, we have scaled the iSHELL observations by a factor $\sim$0.71, assuming that the flux at $\sim$5 $\mu$m is consistent across both observations. The required scaling may be the result of variability. However, as the iSHELL data were flux calibrated using WISE-W2 ( $\lambda$=4.6 $\mu$m), JWST-NIRSpec observations are required to properly investigate variability. As DR~Tau is too bright to observe with JWST-NIRSpec, this comparison cannot be made. The data extend over the wavelength range from approximately $\sim$4.516 $\mu$m to $\sim$5.239 $\mu$m and cover ro-vibrational transitions of \ce{^{12}CO} ($v$=1-0, $v$=2-1, and $v$=3-2) and \ce{^{13}CO} ($v$=1-0), several \ce{H_2O} transitions, and the \ce{HI} Pfund $\beta$ ($\sim$4.655 $\mu$m) and Humphreys ($\sim$4.675 $\mu$m) lines. These data complement and extend upon the earlier VLT-CRIRES observations (at $R\simeq$100000), see \citet{BastEA11}, \citet{BrownEA13}, and \citet{BP15}. \\
\indent Similar to the continuum subtraction method described in Section \ref{sec:ContinuumSubtraction}, we subtracted the baseline using the IRSQR-method. However, for the iSHELL data we did not mask any downwards spikes, used a quantile regression with a value of 0.10, and we placed the knots every 100 data points. The continuum-subtracted iSHELL data are presented in Figure \ref{fig:iSHELL-Spectrum}. A comparison between the JWST-MIRI and iSHELL observations in the 4.9-5.1 $\mu$m region is displayed in Figure \ref{fig:iSHELL-MIRI}.

\section{Analysis} \label{sec:Anal}
\textnormal{The following subsections provide an overview of the used techniques for obtaining the excitation properties. In Section \ref{sec:SlabModels} we describe our slab model retrieval process that has been used to characterise the emission from \ce{CO_2}, \ce{HCN}, and \ce{C_2H_2} observed with JWST-MIRI. On the other hand, the techniques invoked for obtaining the \ce{CO} line profiles and the rotational diagrams are described in Section \ref{sec:CO-LPRD}.}

\subsection{LTE slab models} \label{sec:SlabModels}
To extract information about the molecular features seen in the JWST-MIRI observations, we implement the same LTE slab model fitting procedure as described in \citet{GrantEA23, GasmanEA23Subm, PerottiEA23}, and \citet{TaboneEA23}. We have obtained the required spectroscopic data through the HITRAN database \citep{HITRAN}. Our slab models assume Gaussian line profiles with a broadening of $\Delta V$=4.7 km s$^{-1}$ ($\sigma$=2 km s$^{-1}$), following \citet{SalykEA11}. In addition, we have accounted for the mutual shielding of adjacent lines in the Q-branches of \ce{CO_2}, \ce{HCN}, and \ce{C_2H_2} (see \citealt{TaboneEA23} for a description). We have convolved our \ce{CO_2}, \ce{HCN}, and \ce{C_2H_2} models to a resolving power of $\Delta\lambda/\lambda\sim$2500. Finally, the \textsc{SpectRes} package \citep{SpectRes} was used to convolve the slab models to the JWST-MIRI wavelength grid. \\
\indent The models consists of 3 free parameters that need to be varied to obtain the best fit to the molecular features: the column density $N$ (in cm$^{-2}$), the gas temperature $T$ (in K) and the emitting area $\pi R_\textnormal{em}^2$, which is parameterised by the emitting radius $R_\textnormal{em}$ (in au). We note that while the slab models are parameterised by a circular area, the observed emission may have a different emission morphology with an equivalent area. For example, the emission can originate from an annulus with an area equal to $\pi R_\textnormal{em}^2$.  \\
\indent To acquire the best combination of values for $N$, $T$, and $R_\textnormal{em}$, we selected relatively isolated lines or regions of lines and used a reduced $\chi^2$ analysis to fit the slab models to these lines:
\begin{align} \label{eq:Chi2}
    \chi^2 = \frac{1}{N_\textnormal{obs}}\sum^{N_\textnormal{obs}}_{i=1}\frac{\left(F_{\textnormal{obs},i}-F_{\textnormal{mod},i}\right)^2}{\sigma^2}.
\end{align}
Here, $N_\textnormal{obs}$ denotes the number of data points that were used for the fits, while $F_\textnormal{obs}$ and $F_\textnormal{mod}$ are, respectively, the observed and modelled continuum-subtracted fluxes. $\sigma$ is the estimated noise. As opposed to \citet{GrantEA23} and \citet{GasmanEA23Subm}, we have not estimated the noise from line-free regions, as those are not present in the extremely line-rich spectrum of DR~Tau. Instead we base our values for $\sigma$ on the median continuum flux level in each combination of channels and ranges, and divide this by the highly idealised signal-to-noise ratio ($S/N$) acquired through the JWST \textit{Exposure Time Calculator} (ETC)\footnote{ETC: \url{https://jwst.etc.stsci.edu/}}. The median flux, obtained uncertainty and the estimate from the ETC for each subband can be found in Table \ref{tab:Subband-Sigma}. To account for the regions covering multiple subbands, we use a weighted average of the estimated uncertainties for each respective band. The adopted weights represent the number of data points used from each subband in the $\chi^2$-analysis and this yields a different value for $\sigma$ for each molecule. We list the range of $\sigma$ values for the molecules detected in Table \ref{tab:FitParameters}. Our derived values for $\sigma$ are higher than those found for GW~Lup ($\sigma\sim$0.4 mJy in Channel 3; \citealt{GrantEA23}) and Sz~98 ($\sigma\sim$1.8-9.6 mJy; \citealt{GasmanEA23Subm}), due to our adopted method. The confidence intervals for our fits are defined as $\chi^2_\textnormal{min}$+2.3, $\chi^2_\textnormal{min}$+6.2 and $\chi^2_\textnormal{min}$+11.8 for 1$\sigma$, 2$\sigma$, and 3$\sigma$, respectively \citep{Avni76,PressEA92}. \\
\indent Throughout our $\chi^2$ analysis we have used models based on the following grids for $N$, $T$ and $R_\textnormal{em}$: $N$ ranged from 12 to 22 in $\log_{10}$-space, where we used a spacing of $\Delta\left(\log_{10}\left(N\right)\right)$=0.1. For $T$, we have used a spacing of $\Delta T$=25 K and a range of 150$\leq T\leq$2500 K. $R_\textnormal{em}$ was allowed to vary between 0.01 au and 10 au, using steps of $\Delta R$=0.02 au. 

\subsection{\ce{CO} line profiles and rotational diagrams} \label{sec:CO-LPRD}
\indent To obtain information about $T$, $N$, and $R$ from the high-resolution ro-vibrational transitions of \ce{^{12}CO} and \ce{^{13}CO} we carried out a rotational diagram analysis on the integrated fluxes, following the method described in Appendix A of \citet{BanzattiEA12}. We note that rotational diagram analyses and slab model fitting procedures are inherently the same, as long as the assumption of LTE is valid. We have assumed a Doppler broadening of $\delta V$=1 km s$^{-1}$. We note that using the same broadening as for the JWST-MIRI slab models ($\delta V$=4.7 km s$^{-1}$) in the rotational diagram analysis, the same excitation temperatures are inferred. However, the resulting column densities and emitting radii are found to be, respectively, higher and lower, while the total number of molecules is increased by less than 5\%. \\
\indent The required integrated fluxes have been obtained by, first, creating a weighted-averaged, normalised line profile, using the transitions that are unblended, not affected by telluric features, and have a high $S/N$-ratio. For the weighted average, we have used the errors on the flux as the weights. The line profile has been used to acquire one fixed value for the slight velocity offset ($\Delta V$) from the heliocentric velocity ($\sim$27.6 km s$^{-1}$; \citealt{ArdilaEA02}) and for the FWHM. \\
\indent For the ro-vibrational ladders of \ce{^{12}CO} (v=1-0, v=2-1, and v=3-2; see Section \ref{sec:iSHELL-RDAs}), we find that the line profile consists of two components: a narrow one, likely tracing an inner disk wind, and a broad one that traces the Keplerian rotation of the inner disk, as previously observed in \citet{BastEA11} and \citet{BrownEA13}. 
For these double components, we have used a double Gaussian line profile, whereas for the \ce{^{13}CO} v=1-0 transitions we have used a single Gaussian line profile. The line profiles were fitted using the python-package \textsc{lmfit} \citep{lmfit}. These line profiles have been used to obtain the integrated fluxes of a larger set of transitions, including lines that are partly blended with other transitions and/or have a lower $S/N$-ratio. For \ce{^{12}CO} v=3-2, no blended transitions are included, as we were unable to extract reasonable (compared with respect to the unblended lines) fluxes. The integrated fluxes have been acquired using the method described in \citet{BanzattiEA12}, based on the techniques provided in \citet{PascucciEA08,NajitaEA10} and \citet{PontoppidanEA10}. In short, we invoke 1000 iterations of adding normally distributed noise to the observed flux. For each iteration, we determine the integrated flux and, subsequently, acquire a Gaussian distribution of measured line fluxes. We use the median of this distribution as the actual line flux and take the FWHM as the uncertainty. \\
\indent To obtain information about both components, we carried out the rotational diagram analysis for all observed components separately. If the rotational diagrams consist of a straight line, the excitation temperature can be derived from the inverse of the slope of a fitted straight line, whereas the column density is given by the intercept. If the diagrams display curvature, which is an indication of optically thick emission (see Section \ref{sec:OD}), slab models must be used to properly derive the excitation conditions. Similarly as above, we have obtained the \ce{CO} spectroscopic data from the HITRAN database. The best fitting values for $T$, $N$, and $R_\textnormal{em}$ have, similar to the slab-model fitting procedure, been acquired through a $\chi^2$-analysis, where $\sigma$ in equation (\ref{eq:Chi2}) now denotes the acquired error on the integrated flux. For the iSHELL data, we have used slightly different grids for the column density and temperature, but with the same step sizes: 14$\leq\log(N)\leq$22 cm$^{-2}$, 300$\leq T\leq$3000 K. For the emitting radius we adjusted both the grid and the stepsize into, respectively, 0.01$\leq R_\textnormal{em}\leq$1 au and $\Delta R_\textnormal{em}=$0.001 au. We have used the same grid for the different isotopologues and vibrational transitions.

\section{Results} \label{sec:RD}
\textnormal{We present the results of our analysis in the following subsections. In particular, we highlight the observed molecular species with JWST-MIRI (\ce{CO_2}, \ce{HCN}, and \ce{C_2H_2}) in Section \ref{sec:RD-JWST} and we provide an extensive analysis of the \ce{CO} emission in Section \ref{sec:iSHELL_Results}, including our results regarding the excitation conditions, the line profiles, the optical depth, and the emitting radius.}

\begin{figure*}[ht!]
    \centering
    \includegraphics[width=0.95\textwidth]{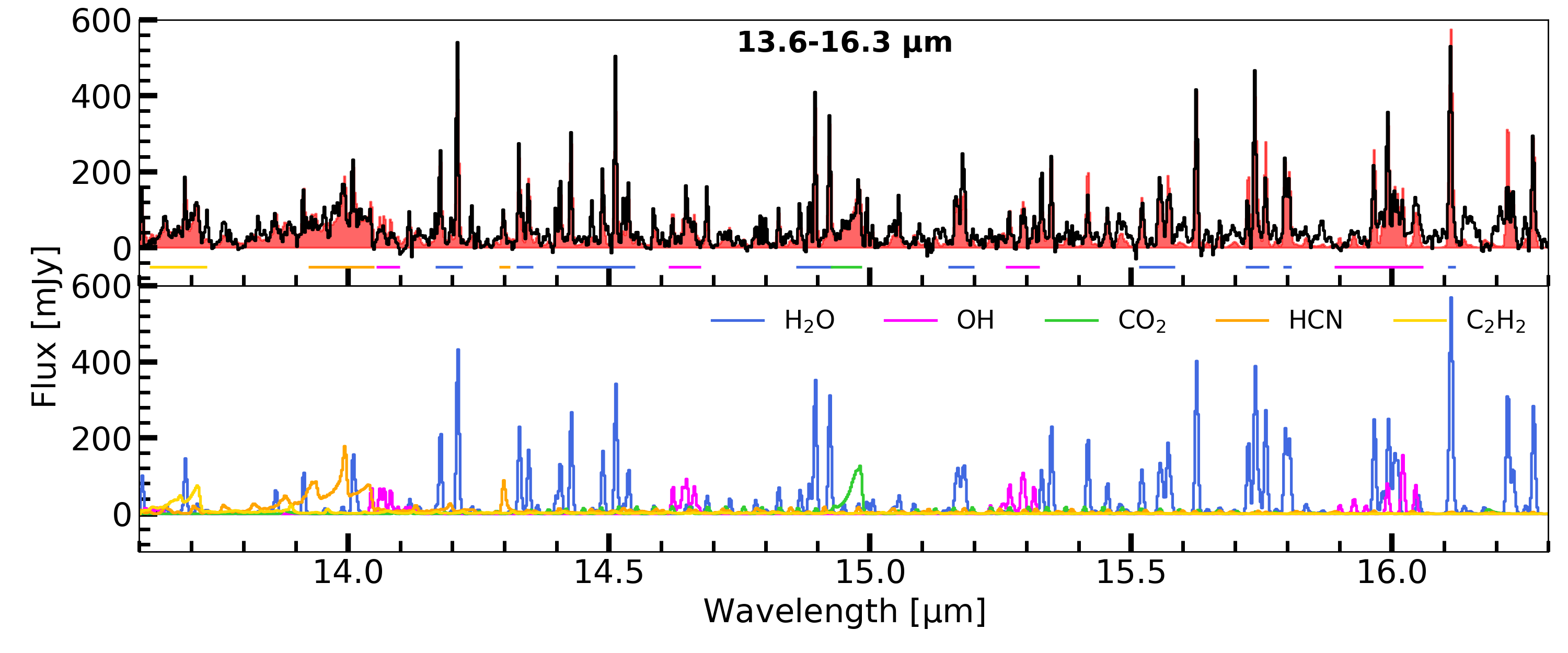}
    \caption{Best fitting slab models of \ce{^{12}CO_2}, \ce{HCN}, and \ce{C_2H_2} in the 13.6-16.3 $\mu$m wavelength region. The top panel displays the continuum subtracted JWST spectrum in a specific region, while the full model spectrum is shown in red. In addition, the horizontal bars show for each molecule the spectral windows used in the $\chi^2$ fits. The bottom panel shows the models for the individually detected molecules. The \ce{H_2O} and \ce{OH} slab models are shown for completeness. Their model properties will be presented in Temmink et al. (in prep.)}
    \label{fig:RegionSpectra}
\end{figure*}

\subsection{JWST spectrum and detected molecules} \label{sec:RD-JWST}
\begin{figure*}[h!]
    \centering
    \includegraphics[width=\textwidth]{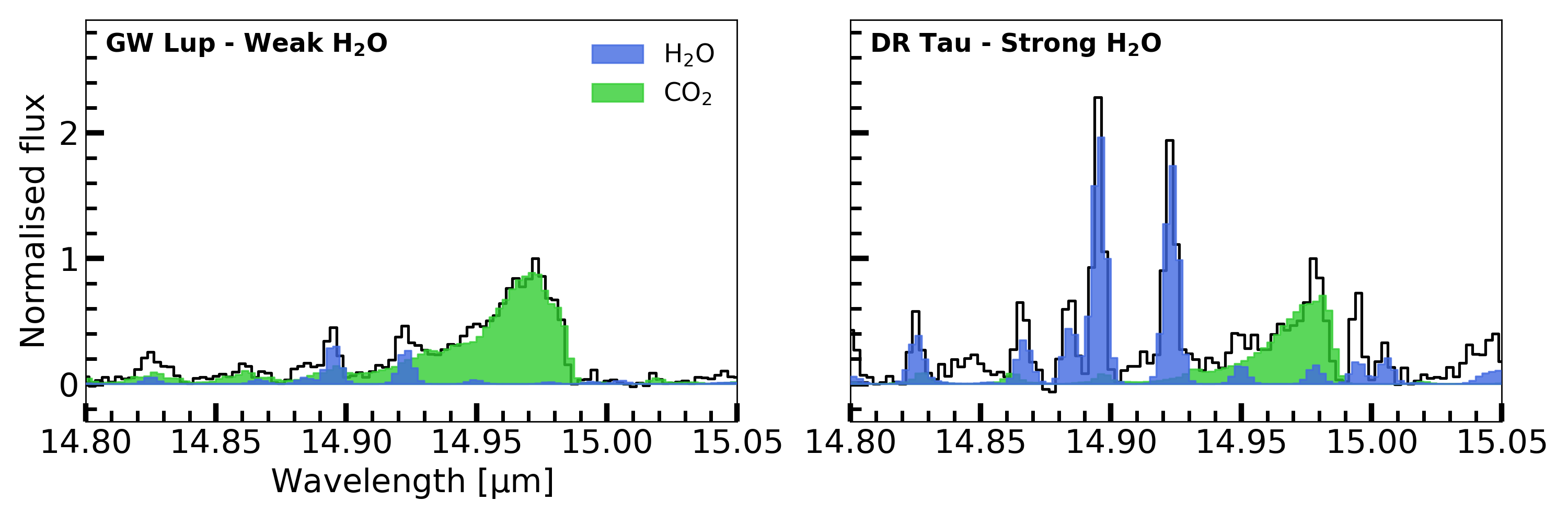}
    \caption{A zoom-in on the \ce{^{12}CO_2} $Q$-branch (14.80-15.05 $\mu$m) of GW~Lup (left; \citealt{GrantEA23}) and DR~Tau (right). Both spectra have been scaled to a common distance of 150 pc and normalised to the peak of the \ce{^{12}CO_2} $Q$-branch.}
    \label{fig:H2O-CO2}
\end{figure*}
The JWST-MIRI spectrum of DR~Tau presented in Figure \ref{fig:FullSpectrum} is very line-rich. Most of the visible lines correspond to \ce{H_2O} transitions, which will be analysed in Temmink et al. (in prep.). Here, we focus on the molecules, besides \ce{H_2O} and \ce{OH}, for which transitions are covered by JWST-MIRI. The bottom panel of Figure \ref{fig:FullSpectrum} also contains a zoom-in on the \ce{CO} transitions ($J\geq$25), highlighting the strong emission features. \\
\indent We detect emission from \ce{CO}, \ce{CO_2}, \ce{HCN}, and \ce{C_2H_2}. Table \ref{tab:FitParameters} displays the parameters of the best fitting slab models for \ce{CO_2}, \ce{HCN}, and \ce{C_2H_2}. The uncertainties listed have been derived from the minimum and maximum value for each parameter that falls within the 1$\sigma$-confidence interval. We also list the total number of molecules ($\mathcal{N}$), which is well constrained for optically thin emission. The total number of molecules has been calculated by multiplying the column density with the emitting area: $\mathcal{N}$=$N\pi R_\textnormal{em}^2$. The best fitting slab models are displayed in Figure \ref{fig:RegionSpectra}, whereas the $\chi^2$-maps are shown in Figure \ref{fig:MIRI-Chi2}. 
\begin{table}[ht!]
    \centering
    \caption{Parameters of the best-fitting slab models for the different molecules observed with JWST-MIRI.}
    \begin{tabular}{c c c c c}
        \hline
        \hline
        \multirow{2}{4em}{Molecule} & $\log_{10}\left(N\right)$ & $T$ & $R_\textnormal{em}$ & $\mathcal{N}$ \\
        & [cm$^{-2}$] & [K] & [au] & \\
        \hline
        \multicolumn{5}{c}{13.6-16.3 $\mu$m ($\sigma$=2.19-2.23 mJy)} \\
        \ce{^{12}CO_2} & 17.4$^{+0.7}_{-0.2}$ & 325$^{+50}_{-100}$ & 0.53$^{+0.36}_{-0.06}$ & 4.96$\times$10$^{43}$ \\
        \ce{HCN} & 14.7$^{+2.1}_{-0.6}$ & 900$^{+50}_{-50}$ & 4.83$^{+4.96}_{-4.38}$ & 8.22$\times$10$^{42}$ \\
        \ce{C_2H_2} & 15.0$^{+5.6}_{-1.7}$ & 775$^{+150}_{-625}$ & 1.45$^{+8.56}_{-1.32}$ & 1.48$\times$10$^{42}$ \\
        \hline
    \end{tabular}
    \label{tab:FitParameters}
\end{table}

\subsubsection{\ce{^{12}CO_2}}
The $Q$-branch of \ce{^{12}CO_2} is well detected in DR~Tau, whereas individual $P$- and $R$-branch, and the hot-bands are not strongly detected. Our fit only extends over the $Q$-branch, which yields a column density of $N$=10$^{17.4}$ cm$^{-2}$, a gas temperature of $T$=325 K and an emitting radius of $R$=0.53 au. Compared to \ce{H_2O}, our \ce{^{12}CO_2} $Q$-branch is significantly weaker. This is similar to Sz~98 \citep{GasmanEA23Subm} and EX~Lup \citep{KospalEA23}, but totally the opposite for GW~Lup \citep{GrantEA23} and PDS~70 \citep{PerottiEA23}. The difference with GW~Lup is also visualised in Figure \ref{fig:H2O-CO2}, which displays a zoom-in on the normalised \ce{^{12}CO_2} $Q$-branch. As opposed to GW~Lup, we do not detect any emission signature from \ce{^{13}CO_2}.

\subsubsection{\ce{HCN} \& \ce{C_2H_2}}
Similar to GW~Lup \citep{GrantEA23}, but contrarily to PDS~70 \citep{PerottiEA23}, we detect \ce{HCN} and \ce{C_2H2} in the wavelength region of 13.6-16.3 $\mu$m. The \ce{HCN} $Q$-branch ($\sim$14 $\mu$m) and hot-band ($\sim$14.3 $\mu$m) are both strongly detected. The emission is likely optically thin, which causes the column densities and emitting radius to be degenerate (see also Figure \ref{fig:MIRI-Chi2}). Nonetheless, our best-fitting slab model suggests a column density of $N$=10$^{14.7}$ cm$^{-2}$, a temperature of $T$=900 K, and an emitting radius of $R$=4.83 au. \\
\indent The parameters of \ce{C_2H_2} ($\sim$13.7 $\mu$m) are much less constrained compared to those of \ce{HCN}. Similar to GW~Lup, the inferred column density is on the border between optically thin and thick, which adds another degeneracy: for high column densities, there is a degeneracy between the column density and the gas temperature. For \ce{C_2H_2} we find a best-fitting column density of $N$=10$^{15.0}$ cm$^{-2}$, a temperature of $T$=775 K, and a emitting radius of $R$=1.45 au. 

\subsection{\ce{CO} analysis} \label{sec:iSHELL_Results}
As the main focus of this paper is the \ce{CO} emission and a comparison between the JWST-MIRI and IRTF-iSHELL observations, this following section presents the results of the extensive \ce{CO} analysis.

\subsubsection{\ce{^{12}}CO as seen by JWST-MIRI} \label{sec:12CO-MIRI}
\begin{figure}[ht!]
    \centering
    \includegraphics[width=\columnwidth]{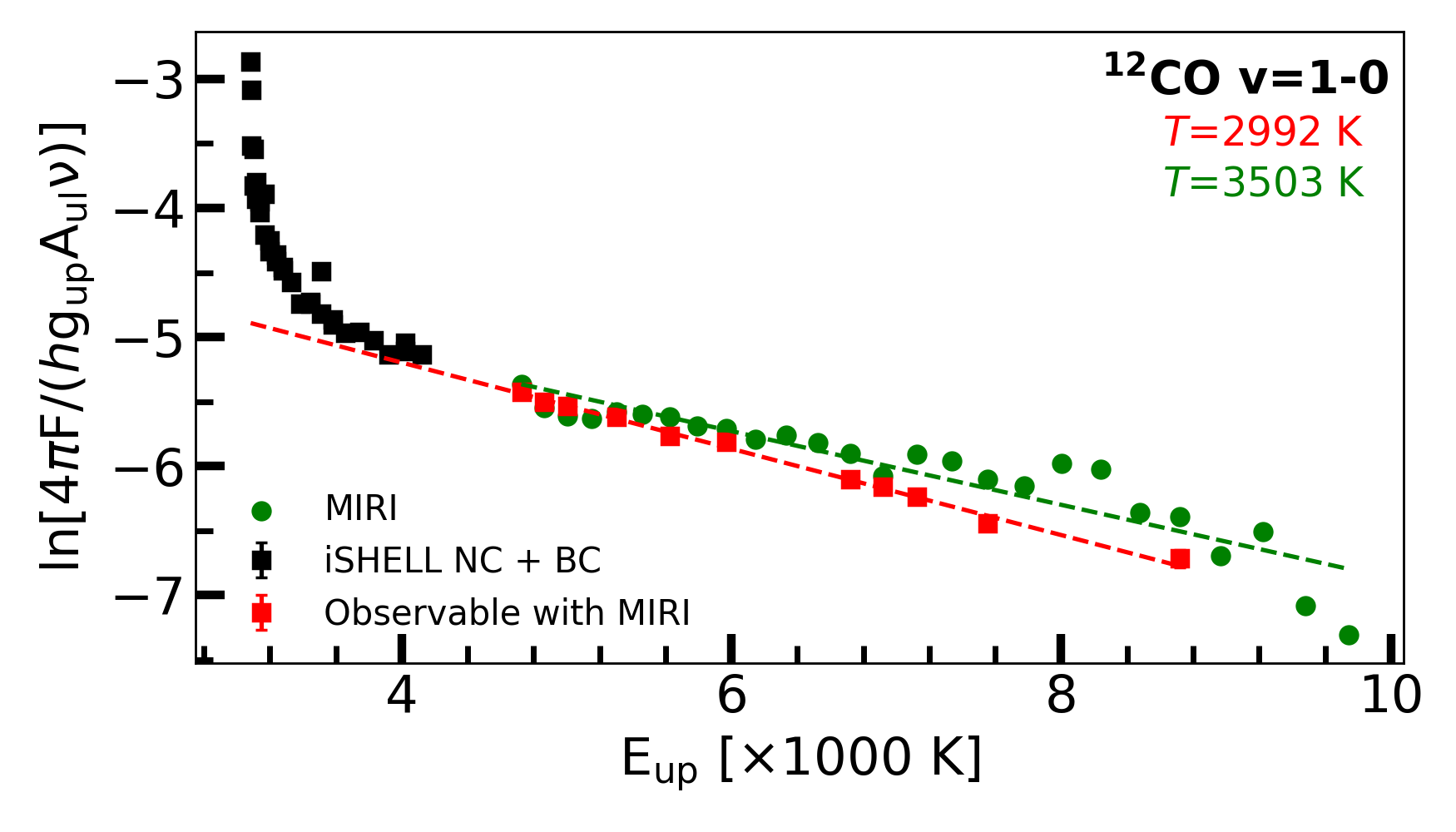}
    \caption{A rotational diagram of the \ce{^{12}CO} v=1-0 transtions. As JWST-MIRI is unable to distinguish between the narrow (NC) and broad (BC) components, their contributions have been added together. The black squares indicate the fluxes inferred from the iSHELL observations, while the ones marked denote the transitions that are also observable with JWST-MIRI. Errorbars on the iSHELL observations are plotted, however, these are too small with respect to the size of the squares. The circles in green are the integrated fluxes inferred from the JWST-MIRI observations using a pseudo-Voigt profile. The red, dashed line is a fit to the red points, used to derive what excitation temperature JWST-MIRI would probe as seen with the iSHELL observations ($T$=2992$\pm$40 K). The green dashed line indicates the excitation temperature derived from these observations, $T$=3503$^{+318}_{-269}$ K.}
    \label{fig:RD-MIRI}
\end{figure}

For molecules, such as \ce{CO}, with well identified individual ro-vibrational lines, the excitation is commonly analysed by plotting the integrated fluxes in a rotational diagram. Figure \ref{fig:RD-MIRI} shows in green the measured JWST-MIRI fluxes of \ce{^{12}CO} using a pseudo-Voigt profile \citep{LabianoEA21}, assuming that the observable flux originates in the v=1-0 lines, i.e. we ignore any blending with the \ce{^{12}CO} v=2-1, v=3-2, and \ce{^{13}CO} v=1-0 transitions. Recall that JWST-MIRI only probes the transitions with the largest $J$ values ($J\geq$25) of the \ce{^{12}CO} $P$-branch. \\
\indent The temperature, $T_\textnormal{ex}$, derived from such a rotational is given by the inverse of the slope, whereas the column density can be obtained from the intercept. For \ce{^12CO}, we acquire a temperature as high as $T_\textnormal{ex}$=3503$^{+318}_{-269}$ K and a column density as low as 1.3$\times$10$^{17}$ cm$^{-2}$ assuming an emitting area with a radius of 0.4 au (see Table \ref{tab:iSHELL-Parameters}). However, this simple analysis assumes that the line emission is optically thin. If the emission is optically thick, the inferred temperature can be much higher than the actual kinetic temperature. As demonstrated by, for example, \citet{HerczegEA11} (see their Appendix D) and Francis et al. (subm.), optically thick line emission introduces a curvature at medium $J$ values. This curvature causes the high-$J$ transitions (probed by JWST-MIRI) to mimic much higher excitation temperatures than the kinetic one when only those transitions are fitted in a diagram. \\
\indent We emphasise this point by including the integrated iSHELL fluxes of the \ce{^{12}CO} v=1-0 transitions in Figure \ref{fig:RD-MIRI}, with the black and red squares obtained by taking the sum over the narrow and broad components. The red squares mark the transitions that are observable with JWST-MIRI and, as seen in the Figure, some small differences are visible in the measured fluxes with respect to JWST-MIRI, as we have not accounted for the blending of lines in the JWST-MIRI spectrum. As shown in Figure \ref{fig:iSHELL-MIRI-SelLines}, many of the \ce{^{12}CO} v=1-0 transitions are blended with other lines, while only a handful are unblended, some of which are highlighted in Figure \ref{fig:iSHELL-MIRI-SelLines}, whereas a few more can be found at longer wavelengths ($\geq$5.18 $\mu$m). Such blending will be worse for sources with higher intrinsic line wings (e.g. sources with outflows or galaxies), where one may not be able to detect any unblended line at all. \\
\indent The iSHELL fluxes clearly show a curvature in the rotational diagrams. For comparison, we fit a straight line to the iSHELL integrated fluxes that cover the JWST-MIRI wavelength range (the red squares). The fit yields an excitation temperature of $T_\textnormal{ex}$=2992$\pm$40 K. As will be demonstrated in Section \ref{sec:iSHELL-RDAs} (see Table \ref{tab:iSHELL-Parameters}), the inferred temperatures are much lower when the iSHELL observations are fitted over the full range of $J$ values with a LTE slab model that accounts for optical depth. This demonstrates the importance of having information on the fluxes of the lower-$J$ lines. \\
\indent As mentioned before, JWST-MIRI's resolution is not sufficient to distinguish between the two components present in the complex line profile of \ce{^{12}CO}. The total \ce{^{12}CO} line profile can, on the other hand, only be marginally resolved with JWST-MIRI. For unblended \ce{^{12}CO} v=1-0 transitions (see Section \ref{sec:ConvMIRI}), we infer values for the FWHM in the range of $\sim$84-100 km s$^{-1}$, which are slightly larger than JWST-MIRI's velocity resolution of $\sim$80 km s$^{-1}$ \citep{LabianoEA21}. The narrow component, tracing the disk wind, could potentially be observed with JWST-MIRI as extended emission. However, no extended emission is observed in the \ce{CO} lines.

\subsubsection{iSHELL CO line profiles and rotational diagrams} \label{sec:iSHELL-RDAs}
\begin{figure*}[ht!]
    \centering
    \includegraphics[width=14cm]{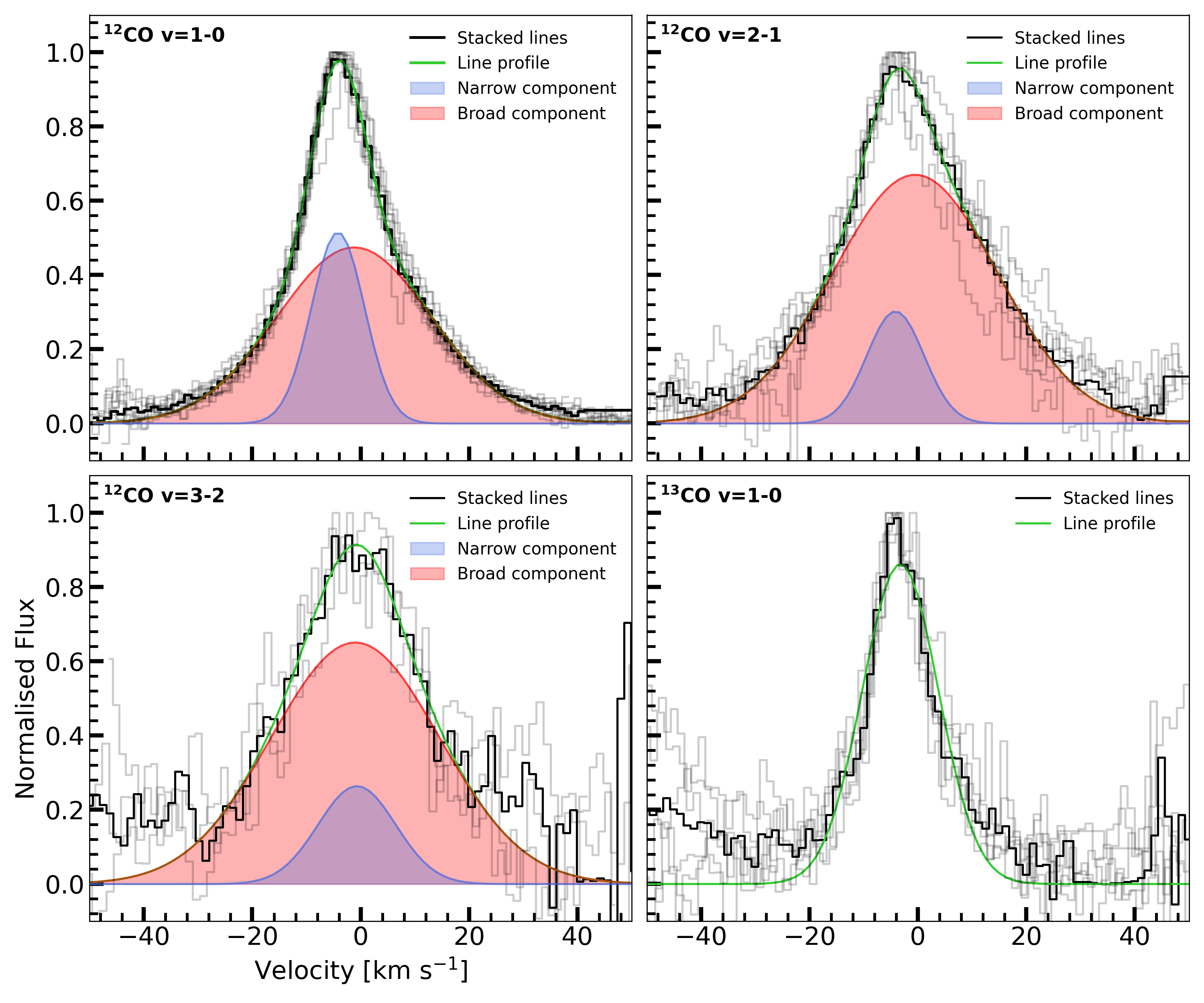}
    \caption{The fitted iSHELL line profiles for the \ce{^{12}CO} $v$=1-0, $v$=2-1, $v$=3-2, and the \ce{^{13}CO} $v$=1-0 transitions. The median line profile is shown in black, whereas the fitted line profile is shown in green. For the \ce{^{12}CO} transitions, we have indicated the narrow and broad components in, respectively, blue and red. The faint, grey line profiles visible in the background correspond to the individual line profiles used in creating the weighted-averaged, normalised line profile.}
    \label{fig:iSHELL-LineProfiles}
\end{figure*}
\begin{figure*}[ht!]
    \centering
    \includegraphics[width=14cm]{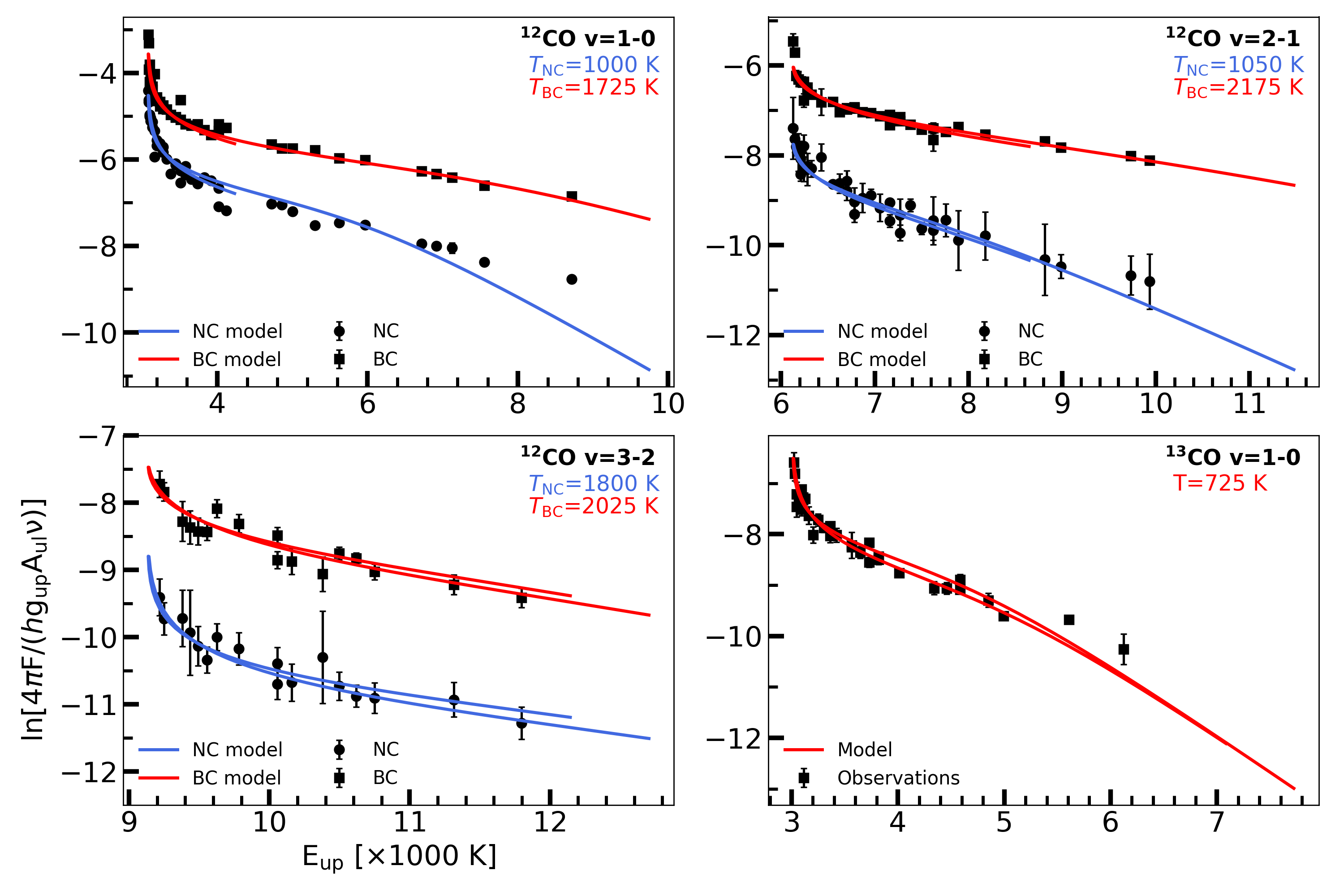}
    \caption{The rotational diagrams for \ce{^12}CO v=1-0 (top left), v=2-1 (top right), v=3-2 (bottom left), and \ce{^{13}CO} v=1-0 (bottom right). In each panel, the slab models are given by the solid line. For the \ce{^{12}CO} transitions, the narrow and broad components (NC and BC, respectively) are shown separately. The integrated fluxes for the narrow component are given by the dots, while those for the broad component are indicated by the squares. The two lines in each model correspond the $R$- and $P$-branches.}
    \label{fig:iSHELL-RDs}
\end{figure*}
\begin{figure*}[ht!]
    \centering
    \includegraphics[width=14cm]{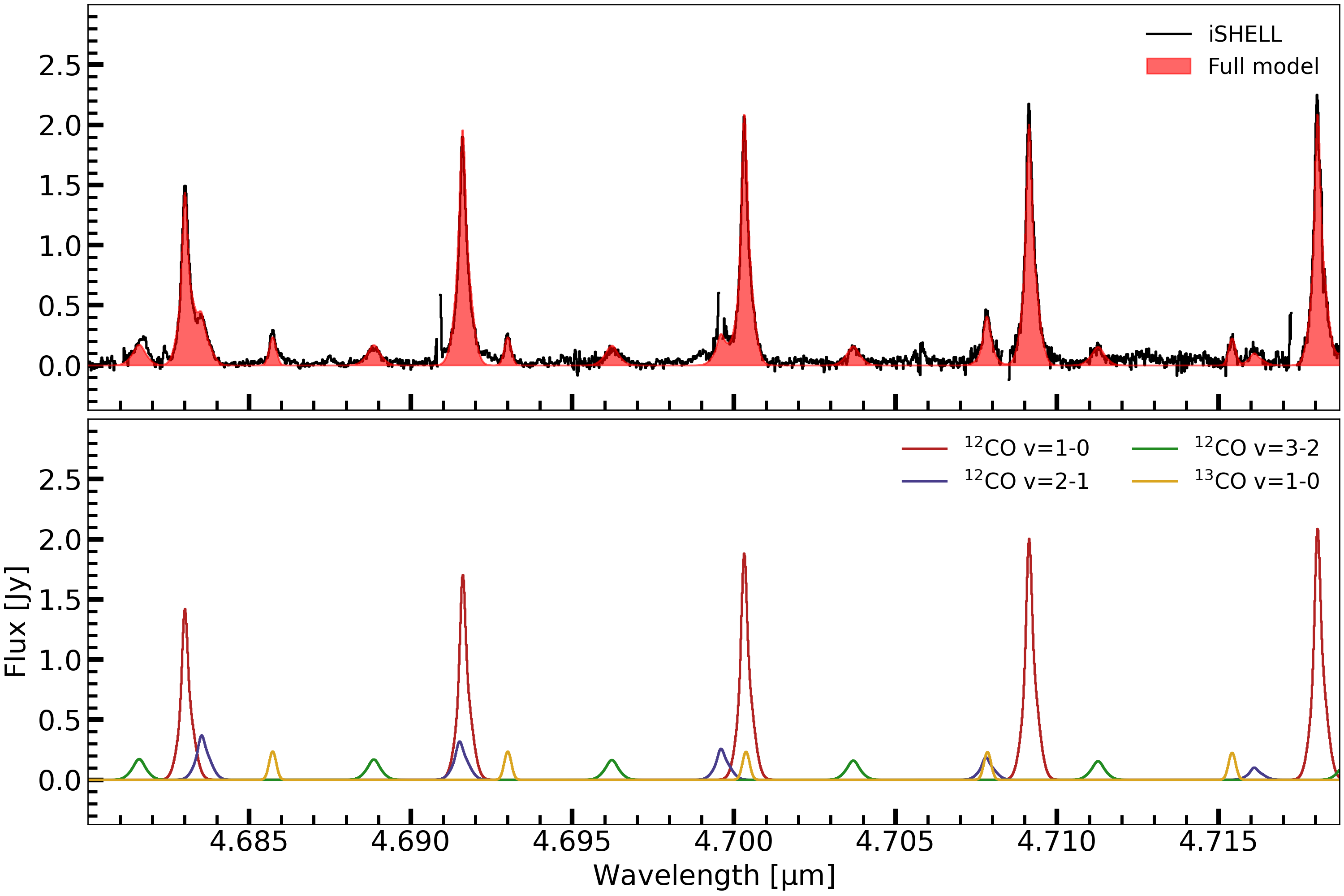}
    \caption{A zoom-in (4.68-4.72 $\mu$m) of the continuum-subtracted iSHELL spectrum of DR~Tau with a model based on the best-fitting parameters obtained through the rotational diagram analysis. A full version can be found in Figure \ref{fig:iSHELL-Model}.}
    \label{fig:iSHELL-Model-ZoomIn}
\end{figure*}

A proper analysis of the high-resolution, ground-based \ce{CO} observations starts with a thorough analysis of the line profiles. The line profiles are displayed in Figure \ref{fig:iSHELL-LineProfiles}, whereas the acquired values for $\Delta V$ and the FWHM are listed in Table \ref{tab:iSHELL-Parameters}. The rotational diagrams themselves are displayed in Figure \ref{fig:iSHELL-RDs}, while the best-fit parameters, for the broad and narrow components, are summarised, together with the values for the average FWHM and $\Delta V$, in Table \ref{tab:iSHELL-Parameters}. The rotational diagrams (Figure \ref{fig:iSHELL-RDs}) of the \ce{^{12}CO} vibrational ladders and \ce{^{13}CO} v=1-0 transitions are presented separately. In addition, we separated out the narrow and broad components for \ce{^{12}CO}. For the broad component we find consistently higher integrated fluxes and excitation temperatures compared to those of the narrow one. All diagrams display an upward curvature at the lower end of the $E_\textnormal{up}$ range, suggesting optically thick emission. Optical depth is further discussed in Section \ref{sec:OD}. We do not list any uncertainties on the best fitting parameters ($T$ and $N$), but they can be viewed in Figure \ref{fig:iSHELL-RD-Chi2}. Models, based on the rotational diagrams, for the transitions of \ce{^{12}CO} $v=$1-0 (red), $v=$2-1 (blue), $v=$3-2 (green), and \ce{^{13}CO} $v=$1-0 (yellow) are displayed in Figure \ref{fig:iSHELL-Model}, while a zoom-in on the 4.68-4.72 $\mu$m region is shown in Figure \ref{fig:iSHELL-Model-ZoomIn}, to provide a better view of how well the model fits the data. \\
\begin{table*}[ht!]
    \caption{Best fit parameters acquired through the rotational diagram analysis for the iSHELL data.}
    \centering
    \begin{tabular}{c c c c c c c c}
        \hline
        \hline
        Molecule & Vibr. Trans. & $\Delta V^{(\alpha)}$ & FWHM & $N$ & $T$ & $R_\textnormal{em}$ & $\mathcal{N}^{(\beta)}$ \\
         & & [km s$^{-1}$] & [km s$^{-1}$] & [cm$^{-2}$] & [K] & [au] & \\
        \hline
        \ce{^{12}CO} & v=1-0 (NC) & -4.2 & 11.6 & 3.2$\times$10$^{17}$ & 1000 & 0.46 & $>$4.6$\times$10$^{43}$ \\
                     & v=1-0 (BC) & -1.2 & 33.5 & 6.3$\times$10$^{17}$ & 1725 & 0.38 & $>$6.3$\times$10$^{44}$ \\ 
                     & v=2-1 (NC) & -4.1 & 12.5 & 7.9$\times$10$^{17}$ & 1050 & 0.21 & $>$2.3$\times$10$^{43}$ \\ 
                     & v=2-1 (BC) & -0.5 & 35.8 & 5.0$\times$10$^{17}$ & 2175 & 0.22 & $>$1.6$\times$10$^{43}$ \\ 
                     & v=3-2 (NC) & -0.7 & 16.9 & 5.0$\times$10$^{18}$ & 1800 & 0.05 & $>$8.8$\times$10$^{42}$ \\
                     & v=3-2 (BC) & -1.0 & 36.6 & 1.6$\times$10$^{18}$ & 2025 & 0.13 & $>$1.9$\times$10$^{43}$ \\
        \hline
        \ce{^{13}CO} & v=1-0 & -3.2 & 16.5 & 1.6$\times$10$^{17}$ & 725 & 0.47 & $>$2.4$\times$10$^{43}$ \\
        \ce{C^{18}O} & v=1-0 & -4.2 & 10.3 & 2.0$\times$10$^{16}$ & 975 & 0.23 & 7.4$\times$10$^{41}$ \\
        \hline
    \end{tabular}
    \label{tab:iSHELL-Parameters}
    \tablefoot{($\alpha$): $\Delta V$ is the offset of the line center with respect to the heliocentric velocity of DR~Tau, $\sim$27.6 km s$^{-1}$. \\
    ($\beta$): As both \ce{^{12}CO} and \ce{^{13}CO} are found to be optically thick (see Section \ref{sec:OD}), the values listed for the total number of molecules must be treated as lower limits.}
\end{table*}
\indent We find column densities in the range of 1.6$\times$10$^{17}$-5.0$\times$10$^{18}$ cm$^{-2}$ for the different transitions and temperatures of 725-2175 K. We find that the narrow components of the various \ce{^{12}CO} ro-vibrational transitions trace colder temperatures (respectively, 1000, 1050, and 1800 K) compared to the broad components (respectively, 1725, 2175, and 2025 K), confirming the different origins of the two components and suggesting that the broad component does trace the hot inner regions of the disk. For \ce{^{13}CO} we find a significantly lower temperature of 725 K, which is slightly higher than the value reported by \citet{BastEA11} and \citet{BrownEA13} of $T$=510-570 K. However, both \citet{BastEA11} and \citet{BrownEA13} used only transitions with $E_\textnormal{up}\sim$3000-4000 K, whereas our data include transitions with values of $E_\textnormal{up}$ up to $\sim$6200 K.  \\

\subsubsection{Optical depth: \ce{^{12}CO} and \ce{^{13}CO}} \label{sec:OD}
\begin{figure*}[ht!]
    \centering
    \includegraphics[width=\textwidth]{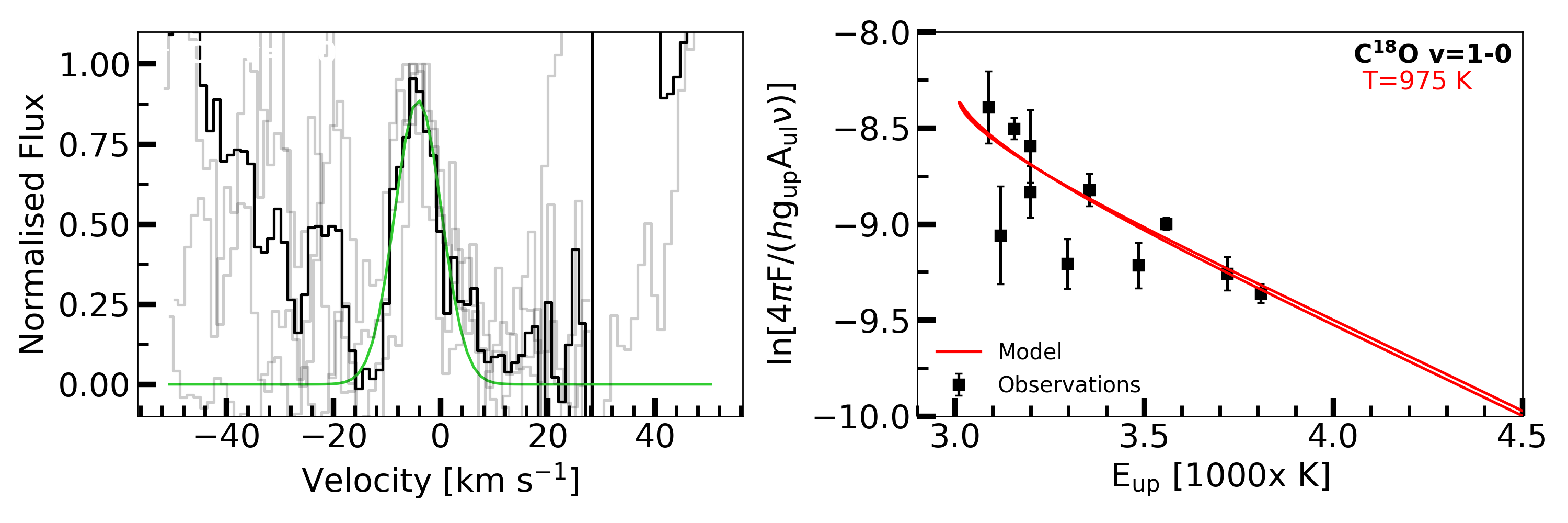}
    \caption{The left panel shows the acquired line Gaussian line profile for the \ce{C^{18}O} v=1-0 transitions, whereas the right plot displays the rotational diagram. The rotational diagram yields an excitation temperature of $T$=975 K.}
    \label{fig:RD-C18O}
\end{figure*}

As previously found by \citet{BrownEA13}, the \ce{^{12}CO} emission of DR~Tau is optically thick. Similarly as their Figure 11, Figure \ref{fig:RD-OD} displays the rotational diagrams of the \ce{^{12}CO} v=1-0, where we have combined the narrow and broad components, and \ce{^{13}CO} v=1-0 transitions together. In addition, we have multiplied the best fitting model of \ce{^{13}CO} by a factor of 68 to account for the isotopologue ratio as in the solar neighbourhood (\citealt{MilamEA05}; \ce{^{12}C}/\ce{^{13}C}=68$\pm$15). If \ce{^{12}CO} were optically thin, the scaled \ce{^{13}CO} model would have coincided with the \ce{^{12}CO} fluxes. However, we see that the scaled \ce{^{13}CO} lies above the \ce{^{12}CO} fluxes, confirming in a similar manner as \citet{BrownEA13} that \ce{^{12}CO} is optically thick. Furthermore, \ce{^{12}CO} being optically thick is also confirmed by the shape of the line fluxes. As shown by \citet{HerczegEA11} (see their Figure D.1), for optically thin emission one would expect a straight line, whereas if the emission becomes optically thick the fluxes of the lines with higher upper level energies flatten off. Furthermore, as seen in Figure \ref{fig:RD-OD}, the \ce{^{13}CO} v=1-0 model is also not completely a straight line, suggesting that the \ce{^{13}CO} is also (moderately) optically thick. However, as the \ce{^{13}CO} emission is less optically thick, a better lower limit on the total number of molecules can be obtained by multiplying the total number of molecules found for \ce{^{13}CO} by the isotopologue ratio of 68. This multiplication yields a total number of molecules of $\mathcal{N_{\ce{CO}}}\simeq$1.6$\times$10$^{45}$.  \\
\indent As the \ce{^{13}CO} emission is also (moderately) optically thick, rarer, less optically thick isotopologues are needed to infer the total amount of molecules. While \ce{C^{18}O} is not detected in the iSHELL observations, the v=1-0 transitions are detected in the VLT-CRIRES data of \citet{BastEA11}. For a description of the data, we refer to \citet{BastEA11} and \citet{BrownEA13}. We also scale the VLT-CRIRES data to match the JWST-MIRI flux at $\sim$5 $\mu$m. Using the same approach as described in Section \ref{sec:HighRes-CO}, we carry out a rotational diagram analysis for the \ce{C^{18}O} v=1-0 transitions to obtain a better estimate for the total number of molecules. \\
\indent First, for the \ce{C^{18}O} line profile we find a velocity offset of $\Delta V$=-4.2 km s$^{-1}$ with respect to DR~Tau's heliocentric velocity and a FWHM of 10.3 km s$^{-1}$. Second, the rotational diagram yields an excitation temperature of $T$=975 K, a column density of $N$=2.0$\times$10$^{16}$ cm$^{-2}$, and an emitting radius of $R_\textnormal{em}$=0.23 au. The acquired line profile and rotational diagram are displayed in, respectively, the left and right panel of Figure \ref{fig:RD-C18O}. The $\chi^2$-map, visualising the uncertainties of the fit, is shown in the bottom right panel of Figure \ref{fig:iSHELL-RD-Chi2}. The inferred excitation temperature for \ce{C^{18}O} is found to be higher than that of \ce{^{13}CO}. However, the $\chi^2$-maps of \ce{C^{18}O} show a large contour for the 1$\sigma$ uncertainty, suggesting that the temperatures agree within the uncertainties. Finally, multiplying the column density by the corresponding emitting area yields a total number of molecules of $\mathcal{N}$=7.4$\times$10$^{41}$, which multiplied by the isotopologue ratio of $^{16}$O/$^{18}$O=557$\pm$30 \citep{Wilson99} yields $\mathcal{N}_\textnormal{CO}$=4.1$\times$10$^{44}$. \\

\subsubsection{Estimated emitting radii of \ce{CO}}
Previous studies (e.g. \citealt{SalykEA11b,BP15,BanzattiEA22}) have shown that the emitting radius where the line flux peaks can be estimated from the line profiles using Kepler's law, as the emission is broadened by the Keplerian rotation:
\begin{align} \label{eq:KL}
    R = GM_*\left(\frac{\sin(i)}{\Delta v}\right)^2,
\end{align}
where $i$ is the inclination and $\Delta v$ is the line width, often taken to be equal to the half width at half maximum (HWHM). The inner emitting radius can, subsequently, also be estimated by taking $\Delta v$ equal to the half width at 10\% (HW10\%; \citealt{BanzattiEA22}). Both the HWHM and the HW10\% have been derived from the averaged line profiles displayed in Figure \ref{fig:iSHELL-LineProfiles}, for simplicity we have taken the HWHM equal to half the FWHM listed in Table \ref{tab:iSHELL-Parameters}. Assuming the inclination of the inner disk of DR~Tau to be equal to the inclination of the outer disk, we use $i$=5.4\degree \citep{LongEA19}. Furthermore, as this method only applies to the lines broadened by the Keplerian rotation, we calculate the (inner) emitting radius only for the broad components of the different \ce{^{12}CO} line profiles and the \ce{^{13}CO} v=1-0 line profile. The calculated values are listed in Table \ref{tab:EmittingRadii}. \\
\begin{table}[ht!]
    \centering
    \caption{The derived (inner) emitting radii for the ro-vibrational transitions of \ce{^{12}CO} and \ce{^{13}CO} using an inclination of $i$=5.4\degree.}
    \begin{tabular}{c c c c}
        \hline\hline
        Molecule & Vibr. Trans. & HWHM & $R_\textnormal{\ce{CO}}$ \\
        & & [km s$^{-1}$] & [au] \\
        \hline
        \ce{^{12}CO} & v=1-0 & 16.8 & 0.03 \\
        & v=2-1 & 17.9 & 0.02 \\
        & v=3-2 & 18.3 & 0.02 \\
        \ce{^{13}CO} & v=1-0 & 8.3 & 0.11 \\
        \hline 
        Molecule & Vibr. Trans. & HW10\% & $R_\textnormal{\ce{CO},in}$ \\
        & & [km s$^{-1}$] & [au] \\
        \hline
        \ce{^{12}CO} & v=1-0 & 29.5 & 0.01 \\
        & v=2-1 & 32.0 & 0.01 \\
        & v=3-2 & 32.7 & 0.01 \\
        \ce{^{13}CO} & v=1-0 & 14.4 & 0.04 \\
        \hline
    \end{tabular}
    \label{tab:EmittingRadii}
\end{table} 
\indent All the calculated emitting radii ($R_\textnormal{\ce{CO}}$) differ significantly from those inferred through the rotational diagrams. One potential explanation for the differences can be found in the chosen value for the inclination. The used inclination has been derived from the continuum emission as probed by ALMA. However, inner disks can be misaligned with respect to the outer disk, for instance caused by perturbations induced by companions (see, for example, \citealt{BohnEA22}). By testing a wide variety of larger inclinations, we infer that inclinations of, respectively, 20.1\degree, 22.4\degree, 23.0\degree, and, 10.2\degree are required to match the corresponding emitting radii for the broad components of the various \ce{^{12}CO} vibrational transitions and the \ce{^{13}CO} v=1-0 transitions inferred from the rotational diagrams. This holds for the assumption that these emitting radii correspond to circular area enclosing the host star. As the high inferred excitation temperatures suggest that the emission of the broad component must originate from inside the dust sublimation radius, this assumption may be valid for DR~Tau. Such a higher inferred inclination of the inner disk agrees with the value of $i_\textnormal{inner}\sim$18$^{+10}_{-18}$\degree\ derived by \citet{GravityEA23} using VLTI-GRAVITY observations. It must be noted, however, that the uncertainties on this inclination also encompass the inclination of the outer disk ($i_\textnormal{outer}$=5.4$^{+2.1}_{-2.6}$\degree). Although this cannot yet be confirmed, the VLTI-GRAVITY observations and the \ce{CO} line profiles both hint at a misalignment between the inner and outer disk. 

\subsubsection{LTE versus non-LTE} \label{sec:nLTE}
When using slab models or creating rotational diagrams, LTE is (almost) always assumed, but the question remains whether this is a valid assumption. A simple test to see if all \ce{^{12}CO} emission lines can be characterised by a single temperature can be carried out by using the \ce{^{12}CO} v=1-0 best-fit parameters for the v=2-1 transitions. If the v=1-0 parameters reproduce the observed v=2-1 fluxes, the emission can be considered in LTE. Otherwise, the vibrational excitation is in non-LTE and the various transitions need to be considered separately, as we have done for the high-resolution observations. The right panel of Figure \ref{fig:12CO-nonLTE} shows the model prediction of the \ce{^{12}CO} v=2-1 transitions using the best-fit parameters of the v=1-0 transitions (see Table \ref{tab:iSHELL-Parameters}). As can be seen, the model overpredicts the emission of the \ce{^{12}CO} v=2-1 transitions, confirming that the vibrational excitation is not in LTE. The various vibrational transitions thus need to be considered separately, while the rotational populations can still be in LTE. \\
\indent Our findings are not surprising, considering the work by \citet{BastEA11}. They find a higher vibrational temperature ($T\sim$1700 K) for DR Tau, while only a temperature of $T\sim$500 K for the rotational \ce{^{13}CO} lines. Higher vibrational temperatures are more commonly found for Herbig Ae/Be stars (e.g. \citealt{BrittainEA07} and \citealt{vdPlasEA15}), which may be attributed to their higher UV fields and subsequent UV-pumping into the higher vibrational states. For T-Tauri stars, UV-pumping into the higher vibrational states can be achieved through higher UV fluxes following accretion by the host star \citep{ValentiEA00}. As DR~Tau is known to be an actively accreting system, UV-pumping provides a possible explanation for the higher vibrational temperatures found by \citet{BastEA11}.

\subsubsection{Implication for JWST-MIRI analysis} \label{sec:ConvMIRI}
While the LTE slab model of \ce{^{12}CO} fitted to the JWST-MIRI observations (see Table \ref{tab:FitParameters}) roughly agrees with the excitation parameters found for the \ce{^{12}CO} v=1-0 broad component (see Table \ref{tab:iSHELL-Parameters}), these results must be treated with caution. As mentioned before, the JWST-MIRI observations do not have the spectral resolution to distinguish between the broad and narrow components observed for \ce{^{12}CO} and, as discussed in Section \ref{sec:nLTE}, the \ce{CO} emission is not in LTE, suggesting that the observed emission cannot be described by a single temperature. In the following section, we explore how the slab models fitted to the rotational diagrams of the iSHELL observations can be used to describe the \ce{CO} emission as observed by JWST-MIRI. \\
\indent The main difference between the JWST-MIRI and iSHELL observations is the spectral resolution of JWST-MIRI, which does not allow to distinguish between the narrow and broad components (this also holds for JWST-NIRSpec), as displayed in Figure \ref{fig:iSHELL-LineProfiles}. In addition, as mentioned before, the JWST-MIRI observations only cover the high-$J$ transitions of the $P$-branch. The excitation temperature as probed by JWST-MIRI would, subsequently, not only be affected by the combination of both components (likely approaching the weighted average of the two), but would also be biased to higher excitation temperatures following the inclusion of only transitions with high upper level energies. As discussed in Section \ref{sec:12CO-MIRI} and displayed in Figure \ref{fig:RD-MIRI}, a temperature of $\sim$2992 K is acquired from the lines covered by JWST-MIRI, which is significantly higher than probed by the high-resolution iSHELL observations. Additionally, as shown in Figure \ref{fig:iSHELL-MIRI}, various transitions that are unblended in the iSHELL spectrum are blended in the JWST-MIRI one. Subsequently, it is harder to identify the contributions from the various (vibrational) transitions in JWST-MIRI's spectrum. \\
\indent To be able to compare the \ce{CO} emission observed with iSHELL and JWST-MIRI, and, more importantly, to show that we probe the same emission with both instruments, we bin our iSHELL model to JWST-MIRI's lower spectral resolution and convolve the binned model with JWST-MIRI's line profile. The binning was carried out using the python-package \textsc{SpectRes}, whereas the convolution was implemented using \textsc{astropy} and a kernel based on JWST-MIRI's line profile. As discussed in \citet{LabianoEA21}, JWST-MIRI's line profile is best described by a pseudo-Voigt profile, which is a weighted sum between a Gaussian and a Lorentzian profile (see Equation \ref{eq:PdV}). In comparison, a normal Voigt profile is given by the convolution of a Gaussian and Lorentzian profile. To create the convolution kernel, we first identified the strongest, unblended \ce{^{12}CO} v=1-0 transitions in the JWST-MIRI spectrum. These lines are displayed in Figure \ref{fig:iSHELL-MIRI-SelLines}. Subsequently, we created an average, normalised line emission profile, which is displayed in Figure \ref{fig:PdV-LineProfile}, by considering the 7 resolution elements enclosing the line peaks (blue shaded areas in Figure \ref{fig:iSHELL-MIRI-SelLines}). Finally, we used \textsc{lmfit} to fit a pseudo-Voigt profile to the averaged emission, which is also shown in Figure \ref{fig:PdV-LineProfile}. The fitted pseudo-Voigt profile was, subsequently, adopted as our convolution kernel. \\
\indent The used pseudo-Voigt profile is described by the following equation:
\begin{align} \label{eq:PdV}
    f(x;A,\mu,\sigma_f,\alpha) = \frac{\left(1-\alpha\right)A}{\sigma_\textnormal{g}\sqrt{2\pi}}e^{-\left(x-\mu\right)^2/2\sigma_\textnormal{g}^2}+\frac{\alpha A}{\pi}\left[\frac{\sigma_f}{\left(x-\mu\right)^2+\sigma_f^2}\right].
\end{align}
Here, $A$ denotes the amplitude, $\mu$ the center of the line, $\sigma_f$ describes the width, and $\alpha$ gives the weights. Furthermore, $\sigma_\textnormal{g}$ is defined as $\sigma_\textnormal{g}=\sigma_f/\sqrt{2\ln(2)}$. Table \ref{tab:PdV-Parameters} gives the best fitting values and their confidence intervals for $A$, $\mu$, $\sigma_f$, and $\alpha$. \\
\begin{table}[ht!]
    \caption{The best-fitting values and the corresponding confidence intervals for the pseudo-Voigt profile used to characterise the JWST-MIRI line profile.}
    \centering
    \begin{tabular}{c c c}
        \hline
        \hline
        Parameter$^{(1)}$ & Value & Confidence interval \\
        \hline 
        $A$  & 2.99 & $\pm$0.22 \\
        $\mu$  & 2.84 & $\pm$0.06 \\
        $\sigma_f$ & 1.03 & $\pm$0.08 \\
        $\alpha$ & 0.76 & $\pm$0.20 \\
        \hline
    \end{tabular}
    \tablefoot{$(1)$: the units of the listed parameters are all arbitrary, see also Figure \ref{fig:PdV-LineProfile}.}
    \label{tab:PdV-Parameters}
\end{table}

\begin{figure*}[ht!]    
    \centering
    \includegraphics[width=0.95\textwidth]{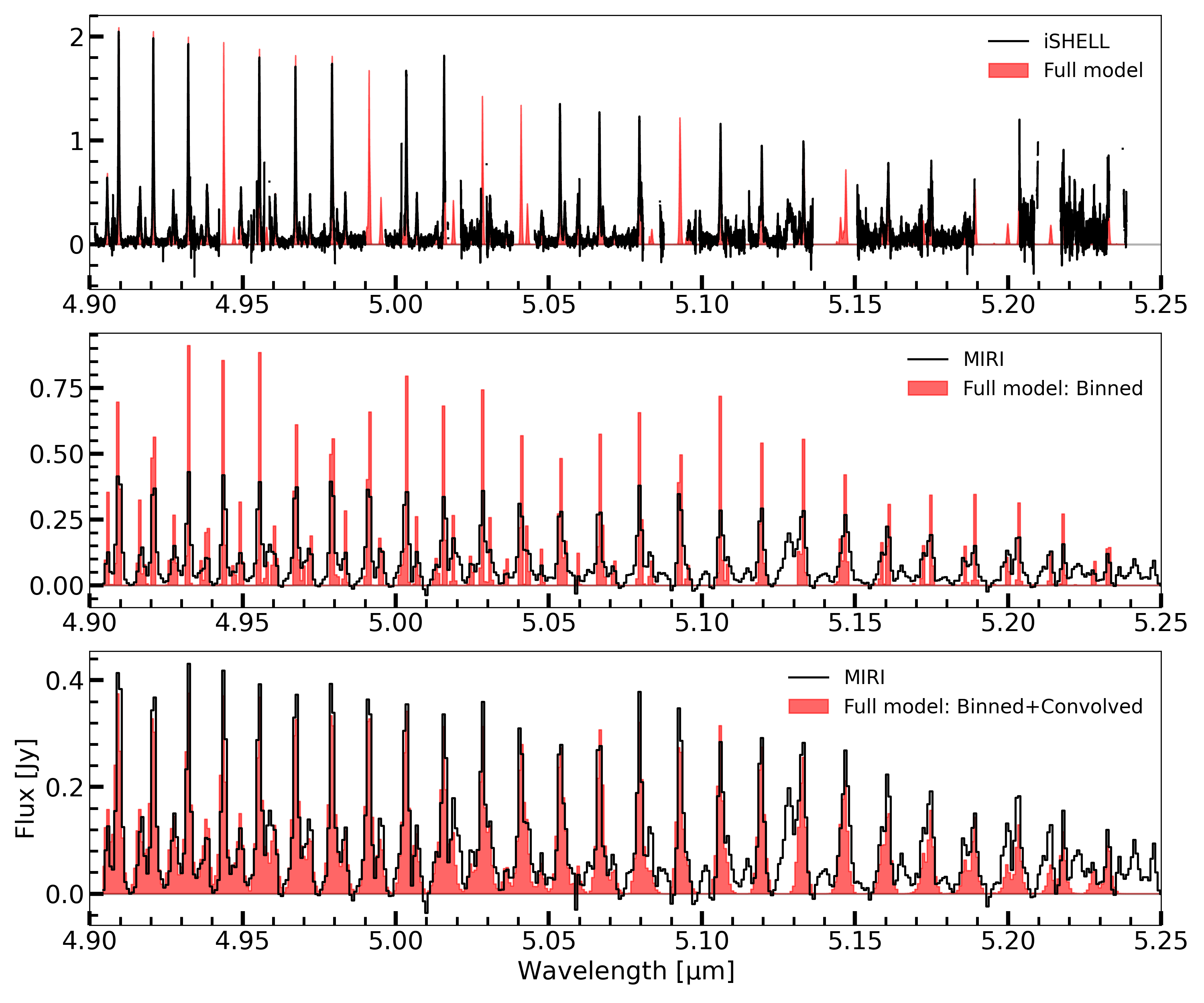}
    \caption{The top panel shows the continuum-subtracted iSHELL data (black) with the best fitting model (red) based on the various slab model fits to the rotational diagrams over the 4.90-5.25 $\mu$m wavelength region. The middle panel shows the best fitting model binned to the JWST-MIRI resolution together with the JWST-MIRI observations, whereas the bottom panel shows the model convolved with the pseudo-Voigt profile shown in Figure \ref{fig:PdV-LineProfile}.}
    \label{fig:iSHELL-MIRI-BC}
\end{figure*}

\indent Figure \ref{fig:iSHELL-MIRI-BC} shows our results: the top panel shows the full \ce{CO} model on top of the iSHELL observations, whereas the middle and bottom panel shows the model on top of the JWST-MIRI data after applying the binning and convolution, respectively. As clearly shown in the middle panel, solely binning the high-resolution model down to JWST-MIRI's resolution does not provide a good fit, as the peak flux is clearly overproduced. The subsequent convolution, on the other hand, yields a good description of the JWST-MIRI observations, as is shown in the bottom panel. To conclude, by binning a high-resolution model to JWST-MIRI's spectral resolution and by convolving the model with JWST-MIRI's line profile, we have shown that the high-resolution observations can be used as a good template for the JWST-MIRI observations and that a proper treatment of the line profile is crucial. \\
\indent We note that, even after convolution, the model does not provide a perfect fit to the JWST-MIRI observations. The small residual differences are potentially related to uncertainties of the flux calibration of the iSHELL observations or the different sensitivities of the instruments. In addition, stellar variability at the 20\% level cannot be excluded as an explanation. Nonetheless, we have shown and argue that the best way to characterise the \ce{CO} lines observed with JWST-MIRI is to combine them with high spectral resolution observations of ground-based instruments, such as iSHELL or VLT-CRIRES(+), that are publicly available for many sources through SpExoDisks.

\subsubsection{JWST-MIRI line profile: pseudo-Voigt versus Gaussian}
As line profiles are often considered to be of pure Gaussian nature, we discuss in the this subsection the different results one may acquire when treating JWST-MIRI's line profile as a Gaussian. When using a Gaussian line profile, we derive integrated fluxes that are 10\% or less smaller than those acquired when using a pseudo-Voigt profile. Figure \ref{fig:RD-MIRI-LP} compares the JWST-MIRI rotational diagrams when considering the different line profiles. A straight line fitted to the Gaussian integrated fluxes yields an excitation temperature of $T$=3565$^{+368}_{-305}$ K, which is slightly higher than the one ($T$=3503$^{+318}_{-269}$ K) probed when using a pseudo-Voigt profile. The temperatures do, however, agree well within the uncertainties. \\
\indent In addition, we reproduce the analysis conducted in Section \ref{sec:ConvMIRI} for a Gaussian line profile, given by the first part of Equation \ref{eq:PdV} (setting $\alpha$=0). Figure \ref{fig:PdV-LineProfile} contains the fitted Gaussian line profile to the normalised, unblended \ce{^{12}CO} transitions, which is described by the following parameters: $A$=2.40, $\mu$=2.85, and $\sigma_g$=0.98. The bottom panel of Figure \ref{fig:iSHELL-MIRI-BC-Gauss} shows the convolution of the binned \ce{CO} model with the Gaussian line profile. A close visual comparison between Figures \ref{fig:iSHELL-MIRI-BC} and \ref{fig:iSHELL-MIRI-BC-Gauss} yields only small differences between the two convolutions: the main difference can, as expected, be found in the line wings of the convolved model. The Lorentzian contribution ensures that the pseudo-Voigt profile provides a better fit to the line wings compared to the Gaussian one. On the other hand, the Gaussian convolution \textnormal{can yield} a model with slightly higher peak values, as less flux is distributed over the line wings. \\
\indent Even though the Gaussian convolution yields similar results as the pseudo-Voigt one, the pseudo-Voigt profile, as found by \citet{LabianoEA21}, provides a significantly better description of the JWST-MIRI observations in subband 1A, as demonstrated by Figure \ref{fig:iSHELL-MIRI-BC}. Depending on the application, a Gaussian profile may provide a sufficient fit to the observations, but one must take into consideration that the flux in the line wings will be underproduced. \\
\indent \textnormal{An incorrect treatment of the line profile can also impact the analysis of the excitation conditions, as the underproduced line wings may result in an underestimation of the line fluxes. For example, the integral of the Gaussian line profile displayed in Figure \ref{fig:PdV-LineProfile} is $\sim$5\% lower than that of the Pseudo-Voigt line profile. Consequently, a rotational diagrams analysis can yield a lower value on the total column density and a different value for the excitation temperature, depending on how the slope is changed by these underestimated fluxes.}

\section{Discussion} \label{sec:Disc}
\subsection{\ce{CO} excitation temperature as probed by JWST-MIRI}
As shown in Figure \ref{fig:RD-MIRI}, the excitation temperature derived when only considering the high-$J$ transitions observed by JWST-MIRI, assuming optically thin emission, is significantly higher than that derived from a rotational diagram that includes also lower-$J$ values. As discussed in Section \ref{sec:12CO-MIRI}, the higher excitation temperature follows from the fact that the \ce{^{12}CO} emission is optically thick. We emphasise the need for complementary observations when deriving the excitation properties of \ce{CO}, either from space (e.g. with JWST-NIRSpec) or from the ground (e.g. with VLT-CRIRES(+), IRTF-iSHELL or Keck-NIRSPEC). If only space-based observations are used, we advise to treat the results with caution, since JWST-NIRSpec and JWST-MIRI do not have the spectral resolution to resolve the commonly observed, complex \ce{CO} line profiles of planet-forming disks (see also \citealt{BanzattiEA22,BanzattiEA23}).

\subsection{Emitting region of \ce{CO}}
The high inferred excitation temperatures of $T\geq$1725 K suggest that the broad component of the \ce{^{12}CO} emission must come from inside the dust sublimation radius ($T_\textnormal{dust}\simeq$1500 K) as also indicated in Figure 14 of \citealt{BanzattiEA22}). Alternatively, the emission may originate from an elevated layer of the disk's atmosphere, where the gas temperature can be well above that of the dust. The \ce{^{13}CO} emission, on the other hand, should trace a deeper layer of the disk, as it is less optically thick compared to \ce{^{12}CO}, consistent with the lower probed excitation temperature. Depending on the flaring of the inner disks, the emission originating from these elevated emitting layers (as seen from the disk's midplane) could also explain the larger inclinations, or viewing angles, required for the emitting radii derived from the \ce{CO} line profiles and slab models fitted to the rotational diagrams to match. Alternatively, a misalignment between the inner and outer disks, as hinted at by the VLTI-GRAVITY and ALMA observations, may also explain the required inclinations. \\
\indent Thermo-chemical models have shown that the gas temperature in the inner regions of T-Tauri disks, in particular the elevated layers, can reach temperatures of $\geq$1000 K \citep{ThiEA13,BrudererEA15,WalshEA15,WoitkeEA18}. These high temperatures further strengthen the notion that the \ce{CO} emission must originate from the region inside of the dust sublimation radius and/or elevated layers of the disk. 

\subsection{Comparing the emitting properties of \ce{CO} with the other molecules}
\begin{figure*}[ht!]    
    \centering
    \includegraphics[trim=0cm 1.25cm 0cm 1.5cm, clip=true,width=0.9\textwidth]{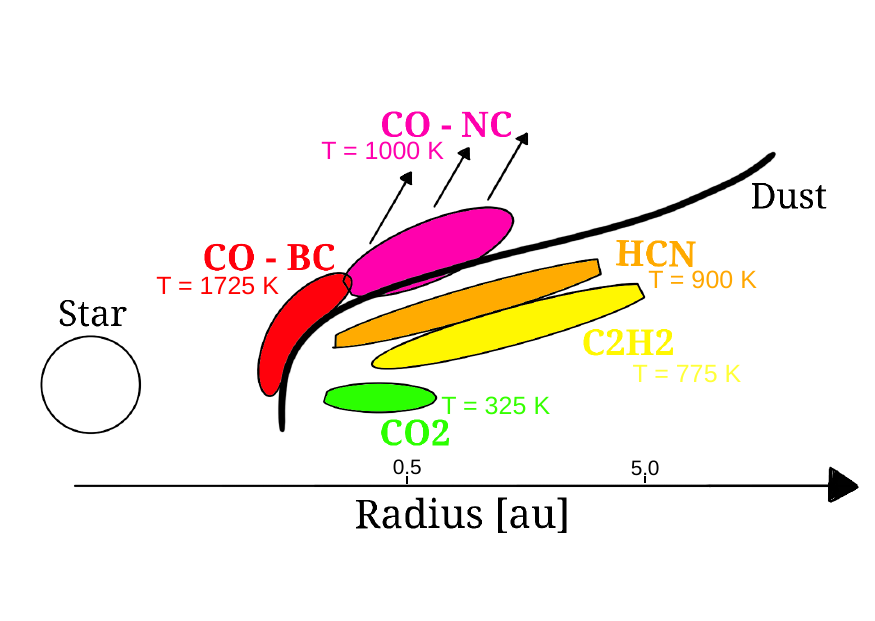}
    \caption{Cartoon visualising the expected emission regions, based on the slab models and interpreting the emitting radius as the disk radius, for our investigated molecules (\ce{CO}, \ce{CO_2}, \ce{HCN}, and \ce{C_2H_2}).}
    \label{fig:Sketch}
\end{figure*}

Compared with the other molecules studied here, \ce{CO_2}, \ce{HCN}, and \ce{C_2H_2}, the excitation temperatures probed by \ce{CO} are much higher. It must be noted that the temperatures inferred for \ce{CO_2}, \ce{HCN}, and \ce{C_2H_2} are similar to those in the other T-Tauri disks, GW~Lup and Sz~98 \citep{GrantEA23,GasmanEA23Subm}. Models by \citet{WoitkeEA18} suggest that the emission likely originates from an onion-like structure, where \ce{CO} has the highest emitting height and \ce{C_2H_2} the lowest. \ce{CO_2} and \ce{HCN} originate from layers in between, with \ce{CO_2} originating from a slightly higher layer than \ce{HCN}. Based on the derived excitation temperatures, this is in partial disagreement with our slab modelling efforts, which suggest that the \ce{CO_2} emission should originate from either a deeper emitting layer or from a larger radial distance compared to that of \ce{HCN} and \ce{C_2H_2}. The expected emission regions, based on our slab models, for the investigated molecules are visualised in Figure \ref{fig:Sketch}. Since the same findings hold for the disks around GW~Lup and Sz~98, additional (thermo)chemical modelling work is necessary to explain the observations and to infer from which layers these molecules originate from.

\section{Conclusions \& summary} \label{sec:CS}
In this work we have investigated the JWST-MIRI observations of the disk around the young star DR~Tau. In addition, we have used complementary, high resolution \ce{CO} ro-vibrational observations to fully investigate the emission properties of \ce{CO}. Below we summarise our main conclusions:
\begin{itemize}
    \item We have confirmed the detections of \ce{CO}, \ce{^{12}CO_2}, \ce{HCN}, and \ce{C_2H_2} with JWST-MIRI, all of which have been observed before with Spitzer and from the ground. With JWST-MIRI we are now able to better constrain the excitation conditions and find excitation conditions of \ce{^{12}CO_2}, \ce{HCN}, and \ce{C_2H_2} for DR~Tau that are similar to those derived for GW~Lup and Sz~98. The conditions suggest that \ce{HCN} and \ce{C_2H_2} originate from a layer higher up in the disk's atmosphere than \ce{^{12}CO_2}.
    \item Using high-resolution IRTF-iSHELL observations we have thoroughly analysed the \ce{CO} emission properties using slab model fits to rotational diagrams. These observations cover various transitions of \ce{^{12}CO} v=1-0, v=2-1, and v=3-2, and \ce{^{13}CO} v=1-0. Similar to previous studies, we note that the \ce{^{12}CO} transitions are comprised of two components: a narrow and a broad component. The narrow part is linked to a potential inner disk wind, whereas the broad component is thought to arise from the Keplerian rotation of the disk. The high inferred temperatures ($\geq$1725 K) for the broad component of the \ce{^{12}CO} transitions suggest that the emission likely originates from a high atmospheric layer and/or from inside the dust sublimation radius.
    \item In addition to the emitting radii inferred from the slab models, we have calculated the emitting radii based on the line profiles, which are found to be significantly smaller than those inferred from the rotational diagrams. To match the emitting radii, we require higher inclinations (or viewing angles if the emission comes from a flared elevated layer) of the inner disk of $i\sim$10-23\degree, suggesting, under the assumption that the inferred emitting radii correspond to a circular emitting area enclosing the host star, that the inner disk of DR~Tau may be slightly misaligned with the respect to the outer disk (i$_\textnormal{outer}\sim$5\degree).
    \item Based on the flux ratios and curvature visible in the rotational diagrams, we find that the \ce{^{12}CO} and \ce{^{13}CO} transitions are optically thick. Using \ce{C^{18}O} transitions covered in VLT-CRIRES observations and applying the same rotational diagram analysis, we have inferred a total number of molecules for \ce{CO} in the inner disk ($\leq$0.23 au) of DR~Tau: $\mathcal{N}_\textnormal{CO}\sim$4.1$\times$10$^{44}$.
    \item Finally, we show that, to properly compare the JWST-MIRI and high-resolution, ground-based \ce{CO} observations, one needs to bin the high-resolution model to JWST-MIRI's spectral resolution and convolve the binned spectrum with a pseudo-Voigt profile.
\end{itemize}
With this work we emphasise that, although JWST-MIRI is a powerful instrument and is able to confidently detect the high-$J$ transitions of \ce{CO}, complementary, high-resolution observations are necessary to properly investigate the physical properties of the emitting gas. For DR~Tau, the Keplerian \ce{CO} emission must originate from inside the dust sublimation radius at temperatures of $T\geq$1725 K, close to the host star. Other molecules, such as \ce{^{12}CO_2}, \ce{HCN}, and \ce{C_2H_2} are found to originate from farther out in the disk, with \ce{^{12}CO_2} likely originating from the deepest layer in the disk of DR~Tau. In addition, the slow disk wind, as traced by the narrow component of \ce{^{12}CO} ($T\geq$1000 K), is launched from a larger radial distance compared to the Keplerian emission.

%% ---
\begin{acknowledgements}
    The authors would like to thank Andrea Banzatti for the many very useful discussions. \\
    \indent This work is based on observations made with the NASA/ESA/CSA James Webb Space Telescope. The data were obtained from the Mikulski Archive for Space Telescopes at the Space Telescope Science Institute, which is operated by the Association of Universities for Research in Astronomy, Inc., under NASA contract NAS 5-03127 for JWST. These observations are associated with program \#1282. The following National and International Funding Agencies funded and supported the MIRI development: NASA; ESA; Belgian Science Policy Office (BELSPO); Centre Nationale d’Etudes Spatiales (CNES); Danish National Space Centre; Deutsches Zentrum fur Luft- und Raumfahrt (DLR); Enterprise Ireland; Ministerio De Econom\'ia y Competividad; Netherlands Research School for Astronomy (NOVA); Netherlands Organisation for Scientific Research (NWO); Science and Technology Facilities Council; Swiss Space Office; Swedish National Space Agency; and UK Space Agency. \\
    \indent This research used the SpExoDisks Database at www.spexodisks.com \\
    \indent M.T. and E.v.D. acknowledge support from the ERC grant 101019751 MOLDISK. E.v.D. acknowledges support from the Danish National Research Foundation through the Center of Excellence ``InterCat'' (DNRF150). B.T. is a Laureate of the Paris Region fellowship program, which is supported by the Ile-de-France Region and has received funding under the Horizon 2020 innovation framework program and Marie Sklodowska-Curie grant agreement No. 945298. D.G., V.C., I.A., A.O., and B.V. thank the Belgian Federal Science Policy Office (BELSPO) for the provision of financial support in the framework of the PRODEX Programme of the European Space Agency (ESA). V.C. and A.O. acknowledge funding from the Belgian F.R.S.-FNRS. G.P. gratefully acknowledges support from the Max Planck Society. T.H. and K.S. acknowledge support from the European Research Council under the Horizon 2020 Framework Program via the ERC Advanced Grant Origins 83 24 28. D.B. and M.MC have been funded by Spanish MCIN/AEI/10.13039/501100011033 grants PID2019-107061GB-C61 and No. MDM-2017-0737. A.C.G. acknowledges from PRIN-MUR 2022 20228JPA3A “The path to star and planet formation in the JWST era (PATH)” and by INAF-GoG 2022 “NIR-dark Accretion Outbursts in Massive Young stellar objects (NAOMY)” and Large Grant INAF 2022 “YSOs Outflows, Disks and Accretion: towards a global framework for the evolution of planet forming systems (YODA)”. I.K., A.M.A., and E.v.D. acknowledge support from grant TOP-1 614.001.751 from the Dutch Research Council (NWO). I.K. and J.K. acknowledge funding from H2020-MSCA-ITN-2019, grant no. 860470 (CHAMELEON). T.P.R acknowledges support from ERC grant 743029 EASY. L.C. acknowledges support by grant PIB2021-127718NB-I00,  from the Spanish Ministry of Science and Innovation/State Agency of Research MCIN/AEI/10.13039/501100011033.
    
\end{acknowledgements}

\bibliographystyle{aa}
\bibliography{Bibliography}

\begin{thebibliography}{72}
\expandafter\ifx\csname natexlab\endcsname\relax\def\natexlab#1{#1}\fi

\bibitem[{{Andrews} {et~al.}(2018){Andrews}, {Huang}, {P{\'e}rez}, {Isella}, {Dullemond}, {Kurtovic}, {Guzm{\'a}n}, {Carpenter}, {Wilner}, {Zhang}, {Zhu}, {Birnstiel}, {Bai}, {Benisty}, {Hughes}, {{\"O}berg}, \& {Ricci}}]{AndrewsEA18}
{Andrews}, S.~M., {Huang}, J., {P{\'e}rez}, L.~M., {et~al.} 2018, \apjl, 869, L41

\bibitem[{{Ardila} {et~al.}(2002){Ardila}, {Basri}, {Walter}, {Valenti}, \& {Johns-Krull}}]{ArdilaEA02}
{Ardila}, D.~R., {Basri}, G., {Walter}, F.~M., {Valenti}, J.~A., \& {Johns-Krull}, C.~M. 2002, \apj, 567, 1013

\bibitem[{{Argyriou} {et~al.}(2023){Argyriou}, {Glasse}, {Law}, {Labiano}, {{\'A}lvarez-M{\'a}rquez}, {Patapis}, {Kavanagh}, {Gasman}, {Mueller}, {Larson}, {Vandenbussche}, {Glauser}, {Royer}, {Dicken}, {Harkett}, {Sargent}, {Engesser}, {Jones}, {Kendrew}, {Noriega-Crespo}, {Brandl}, {Rieke}, {Wright}, {Lee}, \& {Wells}}]{ArgyriouEA23}
{Argyriou}, I., {Glasse}, A., {Law}, D.~R., {et~al.} 2023, \aap, 675, A111

\bibitem[{{Avni}(1976)}]{Avni76}
{Avni}, Y. 1976, \apj, 210, 642

\bibitem[{{Banzatti} {et~al.}(2022){Banzatti}, {Abernathy}, {Brittain}, {Bosman}, {Pontoppidan}, {Boogert}, {Jensen}, {Carr}, {Najita}, {Grant}, {Sigler}, {Sanchez}, {Kern}, \& {Rayner}}]{BanzattiEA22}
{Banzatti}, A., {Abernathy}, K.~M., {Brittain}, S., {et~al.} 2022, \aj, 163, 174

\bibitem[{{Banzatti} {et~al.}(2012){Banzatti}, {Meyer}, {Bruderer}, {Geers}, {Pascucci}, {Lahuis}, {Juh{\'a}sz}, {Henning}, \& {{\'A}brah{\'a}m}}]{BanzattiEA12}
{Banzatti}, A., {Meyer}, M.~R., {Bruderer}, S., {et~al.} 2012, \apj, 745, 90

\bibitem[{{Banzatti} {et~al.}(2020){Banzatti}, {Pascucci}, {Bosman}, {Pinilla}, {Salyk}, {Herczeg}, {Pontoppidan}, {Vazquez}, {Watkins}, {Krijt}, {Hendler}, \& {Long}}]{BanzattiEA20}
{Banzatti}, A., {Pascucci}, I., {Bosman}, A.~D., {et~al.} 2020, \apj, 903, 124

\bibitem[{{Banzatti} \& {Pontoppidan}(2015)}]{BP15}
{Banzatti}, A. \& {Pontoppidan}, K.~M. 2015, \apj, 809, 167

\bibitem[{{Banzatti} {et~al.}(2023){Banzatti}, {Pontoppidan}, {P{\'e}re Ch{\'a}vez}, {Salyk}, {Diehl}, {Bruderer}, {Herczeg}, {Carmona}, {Pascucci}, {Brittain}, {Jensen}, {Grant}, {van Dishoeck}, {Kamp}, {Bosman}, {{\"O}berg}, {Blake}, {Meyer}, {Gaidos}, {Boogert}, {Rayner}, \& {Wheeler}}]{BanzattiEA23}
{Banzatti}, A., {Pontoppidan}, K.~M., {P{\'e}re Ch{\'a}vez}, J., {et~al.} 2023, \aj, 165, 72

\bibitem[{{Bast} {et~al.}(2011){Bast}, {Brown}, {Herczeg}, {van Dishoeck}, \& {Pontoppidan}}]{BastEA11}
{Bast}, J.~E., {Brown}, J.~M., {Herczeg}, G.~J., {van Dishoeck}, E.~F., \& {Pontoppidan}, K.~M. 2011, \aap, 527, A119

\bibitem[{{Bohn} {et~al.}(2022){Bohn}, {Benisty}, {Perraut}, {van der Marel}, {W{\"o}lfer}, {van Dishoeck}, {Facchini}, {Manara}, {Teague}, {Francis}, {Berger}, {Garcia-Lopez}, {Ginski}, {Henning}, {Kenworthy}, {Kraus}, {M{\'e}nard}, {M{\'e}rand}, \& {P{\'e}rez}}]{BohnEA22}
{Bohn}, A.~J., {Benisty}, M., {Perraut}, K., {et~al.} 2022, \aap, 658, A183

\bibitem[{{Booth} {et~al.}(2024){Booth}, {Temmink}, {van Dishoeck}, {Evans}, {Ilee}, {Kama}, {Keyte}, {Law}, {Leemker}, {van der Marel}, {Nomura}, {Notsu}, {{\"O}berg}, \& {Walsh}}]{BoothEA24}
{Booth}, A.~S., {Temmink}, M., {van Dishoeck}, E.~F., {et~al.} 2024, arXiv e-prints, arXiv:2402.04002

\bibitem[{{Booth} {et~al.}(2021){Booth}, {Walsh}, {Terwisscha van Scheltinga}, {van Dishoeck}, {Ilee}, {Hogerheijde}, {Kama}, \& {Nomura}}]{BoothEA21}
{Booth}, A.~S., {Walsh}, C., {Terwisscha van Scheltinga}, J., {et~al.} 2021, Nature Astronomy, 5, 684

\bibitem[{{Bosman} {et~al.}(2022){Bosman}, {Bergin}, {Calahan}, \& {Duval}}]{BosmanEA22}
{Bosman}, A.~D., {Bergin}, E.~A., {Calahan}, J.~K., \& {Duval}, S.~E. 2022, \apjl, 933, L40

\bibitem[{{Braun} {et~al.}(2021){Braun}, {Yen}, {Koch}, {Manara}, {Miotello}, \& {Testi}}]{BraunEA21}
{Braun}, T. A.~M., {Yen}, H.-W., {Koch}, P.~M., {et~al.} 2021, \apj, 908, 46

\bibitem[{{Brittain} {et~al.}(2007){Brittain}, {Simon}, {Najita}, \& {Rettig}}]{BrittainEA07}
{Brittain}, S.~D., {Simon}, T., {Najita}, J.~R., \& {Rettig}, T.~W. 2007, \apj, 659, 685

\bibitem[{{Brown} {et~al.}(2013){Brown}, {Pontoppidan}, {van Dishoeck}, {Herczeg}, {Blake}, \& {Smette}}]{BrownEA13}
{Brown}, J.~M., {Pontoppidan}, K.~M., {van Dishoeck}, E.~F., {et~al.} 2013, \apj, 770, 94

\bibitem[{{Bruderer} {et~al.}(2015){Bruderer}, {Harsono}, \& {van Dishoeck}}]{BrudererEA15}
{Bruderer}, S., {Harsono}, D., \& {van Dishoeck}, E.~F. 2015, \aap, 575, A94

\bibitem[{{Brunken} {et~al.}(2022){Brunken}, {Booth}, {Leemker}, {Nazari}, {van der Marel}, \& {van Dishoeck}}]{BrunkenEA22}
{Brunken}, N. G.~C., {Booth}, A.~S., {Leemker}, M., {et~al.} 2022, \aap, 659, A29

\bibitem[{{Bushouse} {et~al.}(2023){Bushouse}, {Eisenhamer}, {Dencheva}, {Davies}, {Greenfield}, {Morrison}, {Hodge}, {Simon}, {Grumm}, {Droettboom}, {Slavich}, {Sosey}, {Pauly}, {Miller}, {Jedrzejewski}, {Hack}, {Davis}, {Crawford}, {Law}, {Gordon}, {Regan}, {Cara}, {MacDonald}, {Bradley}, {Shanahan}, {Jamieson}, {Teodoro}, \& {Williams}}]{BushouseEA23}
{Bushouse}, H., {Eisenhamer}, J., {Dencheva}, N., {et~al.} 2023, {JWST Calibration Pipeline}, Zenodo

\bibitem[{{Carnall}(2017)}]{SpectRes}
{Carnall}, A.~C. 2017, arXiv e-prints, arXiv:1705.05165

\bibitem[{{Christiaens} {et~al.}(2023){Christiaens}, {Gonzalez}, {Farkas}, {Dahlqvist}, {Nasedkin}, {Milli}, {Absil}, {Ngo}, {Cantero}, {Rainot}, {Hammond}, {Bonse}, {Cantalloube}, {Vigan}, {Kompella}, \& {Hancock}}]{ChristiaensEA23}
{Christiaens}, V., {Gonzalez}, C., {Farkas}, R., {et~al.} 2023, The Journal of Open Source Software, 8, 4774

\bibitem[{{Cutri} {et~al.}(2013){Cutri}, {Wright}, {Conrow}, {Fowler}, {Eisenhardt}, {Grillmair}, {Kirkpatrick}, {Masci}, {McCallon}, {Wheelock}, {Fajardo-Acosta}, {Yan}, {Benford}, {Harbut}, {Jarrett}, {Lake}, {Leisawitz}, {Ressler}, {Stanford}, {Tsai}, {Liu}, {Helou}, {Mainzer}, {Gettings}, {Gonzalez}, {Hoffman}, {Marsh}, {Padgett}, {Skrutskie}, {Beck}, {Papin}, \& {Wittman}}]{AllWISE}
{Cutri}, R.~M., {Wright}, E.~L., {Conrow}, T., {et~al.} 2013, {Explanatory Supplement to the AllWISE Data Release Products}, Explanatory Supplement to the AllWISE Data Release Products, by R. M. Cutri et al.

\bibitem[{{Dawson} \& {Johnson}(2018)}]{DJ18}
{Dawson}, R.~I. \& {Johnson}, J.~A. 2018, \araa, 56, 175

\bibitem[{{Erb}(2022)}]{PyBaselines}
{Erb}, D. 2022, {pybaselines: A Python library of algorithms for the baseline correction of experimental data}, Zenodo

\bibitem[{{Furuya} {et~al.}(2022){Furuya}, {Tsukagoshi}, {Qi}, {Nomura}, {Cleeves}, {Lee}, \& {Yoshida}}]{FuruyaEA22}
{Furuya}, K., {Tsukagoshi}, T., {Qi}, C., {et~al.} 2022, \apj, 926, 148

\bibitem[{{Gaia Collaboration} {et~al.}(2018){Gaia Collaboration}, {Brown}, {Vallenari}, {Prusti}, {de Bruijne}, {Babusiaux}, {Bailer-Jones}, {Biermann}, {Evans}, {Eyer}, {Jansen}, {Jordi}, {Klioner}, {Lammers}, {Lindegren}, {Luri}, {Mignard}, {Panem}, {Pourbaix}, {Randich}, {Sartoretti}, {Siddiqui}, {Soubiran}, {van Leeuwen}, {Walton}, {Arenou}, {Bastian}, {Cropper}, {Drimmel}, {Katz}, {Lattanzi}, {Bakker}, {Cacciari}, {Casta{\~n}eda}, {Chaoul}, {Cheek}, {De Angeli}, {Fabricius}, {Guerra}, {Holl}, {Masana}, {Messineo}, {Mowlavi}, {Nienartowicz}, {Panuzzo}, {Portell}, {Riello}, {Seabroke}, {Tanga}, {Th{\'e}venin}, {Gracia-Abril}, {Comoretto}, {Garcia-Reinaldos}, {Teyssier}, {Altmann}, {Andrae}, {Audard}, {Bellas-Velidis}, {Benson}, {Berthier}, {Blomme}, {Burgess}, {Busso}, {Carry}, {Cellino}, {Clementini}, {Clotet}, {Creevey}, {Davidson}, {De Ridder}, {Delchambre}, {Dell'Oro}, {Ducourant}, {Fern{\'a}ndez-Hern{\'a}ndez}, {Fouesneau}, {Fr{\'e}mat}, {Galluccio}, {Garc{\'\i}a-Torres},
  {Gonz{\'a}lez-N{\'u}{\~n}ez}, {Gonz{\'a}lez-Vidal}, {Gosset}, {Guy}, {Halbwachs}, {Hambly}, {Harrison}, {Hern{\'a}ndez}, {Hestroffer}, {Hodgkin}, {Hutton}, {Jasniewicz}, {Jean-Antoine-Piccolo}, {Jordan}, {Korn}, {Krone-Martins}, {Lanzafame}, {Lebzelter}, {L{\"o}ffler}, {Manteiga}, {Marrese}, {Mart{\'\i}n-Fleitas}, {Moitinho}, {Mora}, {Muinonen}, {Osinde}, {Pancino}, {Pauwels}, {Petit}, {Recio-Blanco}, {Richards}, {Rimoldini}, {Robin}, {Sarro}, {Siopis}, {Smith}, {Sozzetti}, {S{\"u}veges}, {Torra}, {van Reeven}, {Abbas}, {Abreu Aramburu}, {Accart}, {Aerts}, {Altavilla}, {{\'A}lvarez}, {Alvarez}, {Alves}, {Anderson}, {Andrei}, {Anglada Varela}, {Antiche}, {Antoja}, {Arcay}, {Astraatmadja}, {Bach}, {Baker}, {Balaguer-N{\'u}{\~n}ez}, {Balm}, {Barache}, {Barata}, {Barbato}, {Barblan}, {Barklem}, {Barrado}, {Barros}, {Barstow}, {Bartholom{\'e} Mu{\~n}oz}, {Bassilana}, {Becciani}, {Bellazzini}, {Berihuete}, {Bertone}, {Bianchi}, {Bienaym{\'e}}, {Blanco-Cuaresma}, {Boch}, {Boeche}, {Bombrun}, {Borrachero},
  {Bossini}, {Bouquillon}, {Bourda}, {Bragaglia}, {Bramante}, {Breddels}, {Bressan}, {Brouillet}, {Br{\"u}semeister}, {Brugaletta}, {Bucciarelli}, {Burlacu}, {Busonero}, {Butkevich}, {Buzzi}, {Caffau}, {Cancelliere}, {Cannizzaro}, {Cantat-Gaudin}, {Carballo}, {Carlucci}, {Carrasco}, {Casamiquela}, {Castellani}, {Castro-Ginard}, {Charlot}, {Chemin}, {Chiavassa}, {Cocozza}, {Costigan}, {Cowell}, {Crifo}, {Crosta}, {Crowley}, {Cuypers}, {Dafonte}, {Damerdji}, {Dapergolas}, {David}, {David}, {de Laverny}, {De Luise}, {De March}, {de Martino}, {de Souza}, {de Torres}, {Debosscher}, {del Pozo}, {Delbo}, {Delgado}, {Delgado}, {Di Matteo}, {Diakite}, {Diener}, {Distefano}, {Dolding}, {Drazinos}, {Dur{\'a}n}, {Edvardsson}, {Enke}, {Eriksson}, {Esquej}, {Eynard Bontemps}, {Fabre}, {Fabrizio}, {Faigler}, {Falc{\~a}o}, {Farr{\`a}s Casas}, {Federici}, {Fedorets}, {Fernique}, {Figueras}, {Filippi}, {Findeisen}, {Fonti}, {Fraile}, {Fraser}, {Fr{\'e}zouls}, {Gai}, {Galleti}, {Garabato}, {Garc{\'\i}a-Sedano}, {Garofalo},
  {Garralda}, {Gavel}, {Gavras}, {Gerssen}, {Geyer}, {Giacobbe}, {Gilmore}, {Girona}, {Giuffrida}, {Glass}, {Gomes}, {Granvik}, {Gueguen}, {Guerrier}, {Guiraud}, {Guti{\'e}rrez-S{\'a}nchez}, {Haigron}, {Hatzidimitriou}, {Hauser}, {Haywood}, {Heiter}, {Helmi}, {Heu}, {Hilger}, {Hobbs}, {Hofmann}, {Holland}, {Huckle}, {Hypki}, {Icardi}, {Jan{\ss}en}, {Jevardat de Fombelle}, {Jonker}, {Juh{\'a}sz}, {Julbe}, {Karampelas}, {Kewley}, {Klar}, {Kochoska}, {Kohley}, {Kolenberg}, {Kontizas}, {Kontizas}, {Koposov}, {Kordopatis}, {Kostrzewa-Rutkowska}, {Koubsky}, {Lambert}, {Lanza}, {Lasne}, {Lavigne}, {Le Fustec}, {Le Poncin-Lafitte}, {Lebreton}, {Leccia}, {Leclerc}, {Lecoeur-Taibi}, {Lenhardt}, {Leroux}, {Liao}, {Licata}, {Lindstr{\o}m}, {Lister}, {Livanou}, {Lobel}, {L{\'o}pez}, {Managau}, {Mann}, {Mantelet}, {Marchal}, {Marchant}, {Marconi}, {Marinoni}, {Marschalk{\'o}}, {Marshall}, {Martino}, {Marton}, {Mary}, {Massari}, {Matijevi{\v{c}}}, {Mazeh}, {McMillan}, {Messina}, {Michalik}, {Millar}, {Molina}, {Molinaro},
  {Moln{\'a}r}, {Montegriffo}, {Mor}, {Morbidelli}, {Morel}, {Morris}, {Mulone}, {Muraveva}, {Musella}, {Nelemans}, {Nicastro}, {Noval}, {O'Mullane}, {Ord{\'e}novic}, {Ord{\'o}{\~n}ez-Blanco}, {Osborne}, {Pagani}, {Pagano}, {Pailler}, {Palacin}, {Palaversa}, {Panahi}, {Pawlak}, {Piersimoni}, {Pineau}, {Plachy}, {Plum}, {Poggio}, {Poujoulet}, {Pr{\v{s}}a}, {Pulone}, {Racero}, {Ragaini}, {Rambaux}, {Ramos-Lerate}, {Regibo}, {Reyl{\'e}}, {Riclet}, {Ripepi}, {Riva}, {Rivard}, {Rixon}, {Roegiers}, {Roelens}, {Romero-G{\'o}mez}, {Rowell}, {Royer}, {Ruiz-Dern}, {Sadowski}, {Sagrist{\`a} Sell{\'e}s}, {Sahlmann}, {Salgado}, {Salguero}, {Sanna}, {Santana-Ros}, {Sarasso}, {Savietto}, {Schultheis}, {Sciacca}, {Segol}, {Segovia}, {S{\'e}gransan}, {Shih}, {Siltala}, {Silva}, {Smart}, {Smith}, {Solano}, {Solitro}, {Sordo}, {Soria Nieto}, {Souchay}, {Spagna}, {Spoto}, {Stampa}, {Steele}, {Steidelm{\"u}ller}, {Stephenson}, {Stoev}, {Suess}, {Surdej}, {Szabados}, {Szegedi-Elek}, {Tapiador}, {Taris}, {Tauran}, {Taylor},
  {Teixeira}, {Terrett}, {Teyssandier}, {Thuillot}, {Titarenko}, {Torra Clotet}, {Turon}, {Ulla}, {Utrilla}, {Uzzi}, {Vaillant}, {Valentini}, {Valette}, {van Elteren}, {Van Hemelryck}, {van Leeuwen}, {Vaschetto}, {Vecchiato}, {Veljanoski}, {Viala}, {Vicente}, {Vogt}, {von Essen}, {Voss}, {Votruba}, {Voutsinas}, {Walmsley}, {Weiler}, {Wertz}, {Wevers}, {Wyrzykowski}, {Yoldas}, {{\v{Z}}erjal}, {Ziaeepour}, {Zorec}, {Zschocke}, {Zucker}, {Zurbach}, \& {Zwitter}}]{GC18}
{Gaia Collaboration}, {Brown}, A.~G.~A., {Vallenari}, A., {et~al.} 2018, \aap, 616, A1

\bibitem[{{Gasman} {et~al.}(2023){Gasman}, {van Dishoeck}, {Grant}, {Temmink}, {Tabone}, {Henning}, {Kamp}, {G{\"u}del}, {Lagage}, {Perotti}, {Christiaens}, {Samland}, {Arabhavi}, {Argyriou}, {Abergel}, {Absil}, {Barrado}, {Boccaletti}, {Bouwman}, {Garatti}, {Geers}, {Glauser}, {Guadarrama}, {Jang}, {Kanwar}, {Lahuis}, {Morales-Calder{\'o}n}, {Mueller}, {Nehm{\'e}}, {Olofsson}, {Pantin}, {Pawellek}, {Ray}, {Rodgers-Lee}, {Scheithauer}, {Schreiber}, {Schwarz}, {Vandenbussche}, {Vlasblom}, {Waters}, {Wright}, {Colina}, {Greve}, \& {{\"O}stlin}}]{GasmanEA23Subm}
{Gasman}, D., {van Dishoeck}, E.~F., {Grant}, S.~L., {et~al.} 2023, arXiv e-prints, arXiv:2307.09301

\bibitem[{{Gomez Gonzalez} {et~al.}(2017){Gomez Gonzalez}, {Wertz}, {Absil}, {Christiaens}, {Defr{\`e}re}, {Mawet}, {Milli}, {Absil}, {Van Droogenbroeck}, {Cantalloube}, {Hinz}, {Skemer}, {Karlsson}, \& {Surdej}}]{GomezGonzalezEA17}
{Gomez Gonzalez}, C.~A., {Wertz}, O., {Absil}, O., {et~al.} 2017, \aj, 154, 7

\bibitem[{{Gordon} {et~al.}(2022){Gordon}, {Rothman}, {Hargreaves}, {Hashemi}, {Karlovets}, {Skinner}, {Conway}, {Hill}, {Kochanov}, {Tan}, {Wcis{\l}o}, {Finenko}, {Nelson}, {Bernath}, {Birk}, {Boudon}, {Campargue}, {Chance}, {Coustenis}, {Drouin}, {Flaud}, {Gamache}, {Hodges}, {Jacquemart}, {Mlawer}, {Nikitin}, {Perevalov}, {Rotger}, {Tennyson}, {Toon}, {Tran}, {Tyuterev}, {Adkins}, {Baker}, {Barbe}, {Can{\`e}}, {Cs{\'a}sz{\'a}r}, {Dudaryonok}, {Egorov}, {Fleisher}, {Fleurbaey}, {Foltynowicz}, {Furtenbacher}, {Harrison}, {Hartmann}, {Horneman}, {Huang}, {Karman}, {Karns}, {Kassi}, {Kleiner}, {Kofman}, {Kwabia-Tchana}, {Lavrentieva}, {Lee}, {Long}, {Lukashevskaya}, {Lyulin}, {Makhnev}, {Matt}, {Massie}, {Melosso}, {Mikhailenko}, {Mondelain}, {M{\"u}ller}, {Naumenko}, {Perrin}, {Polyansky}, {Raddaoui}, {Raston}, {Reed}, {Rey}, {Richard}, {T{\'o}bi{\'a}s}, {Sadiek}, {Schwenke}, {Starikova}, {Sung}, {Tamassia}, {Tashkun}, {Vander Auwera}, {Vasilenko}, {Vigasin}, {Villanueva}, {Vispoel}, {Wagner}, {Yachmenev}, \&
  {Yurchenko}}]{HITRAN}
{Gordon}, I.~E., {Rothman}, L.~S., {Hargreaves}, R.~J., {et~al.} 2022, \jqsrt, 277, 107949

\bibitem[{{Grant} {et~al.}(2023{\natexlab{a}}){Grant}, {Bettoni}, {Banzatti}, {van Dishoeck}, {Brittain}, {Fedele}, {Henning}, {Manara}, {Semonov}, \& {Whelan}}]{GrantEA23b}
{Grant}, S.~L., {Bettoni}, G., {Banzatti}, A., {et~al.} 2023{\natexlab{a}}, arXiv e-prints, arXiv:2309.03888

\bibitem[{{Grant} {et~al.}(2023{\natexlab{b}}){Grant}, {van Dishoeck}, {Tabone}, {Gasman}, {Henning}, {Kamp}, {G{\"u}del}, {Lagage}, {Bettoni}, {Perotti}, {Christiaens}, {Samland}, {Arabhavi}, {Argyriou}, {Abergel}, {Absil}, {Barrado}, {Boccaletti}, {Bouwman}, {o Garatti}, {Geers}, {Glauser}, {Guadarrama}, {Jang}, {Kanwar}, {Lahuis}, {Morales-Calder{\'o}n}, {Mueller}, {Nehm{\'e}}, {Olofsson}, {Pantin}, {Pawellek}, {Ray}, {Rodgers-Lee}, {Scheithauer}, {Schreiber}, {Schwarz}, {Temmink}, {Vandenbussche}, {Vlasblom}, {Waters}, {Wright}, {Colina}, {Greve}, {Justannont}, \& {{\"O}stlin}}]{GrantEA23}
{Grant}, S.~L., {van Dishoeck}, E.~F., {Tabone}, B., {et~al.} 2023{\natexlab{b}}, \apjl, 947, L6

\bibitem[{{GRAVITY Collaboration} {et~al.}(2021){GRAVITY Collaboration}, {Perraut}, {Labadie}, {Bouvier}, {M{\'e}nard}, {Klarmann}, {Dougados}, {Benisty}, {Berger}, {Bouarour}, {Brandner}, {Caratti O Garatti}, {Caselli}, {de Zeeuw}, {Garcia-Lopez}, {Henning}, {Sanchez-Bermudez}, {Sousa}, {van Dishoeck}, {Al{\'e}cian}, {Amorim}, {Cl{\'e}net}, {Davies}, {Drescher}, {Duvert}, {Eckart}, {Eisenhauer}, {F{\"o}rster-Schreiber}, {Garcia}, {Gendron}, {Genzel}, {Gillessen}, {Grellmann}, {Hei{\ss}el}, {Hippler}, {Horrobin}, {Hubert}, {Jocou}, {Kervella}, {Lacour}, {Lapeyr{\`e}re}, {Le Bouquin}, {L{\'e}na}, {Lutz}, {Ott}, {Paumard}, {Perrin}, {Scheithauer}, {Shangguan}, {Shimizu}, {Stadler}, {Straub}, {Straubmeier}, {Sturm}, {Tacconi}, {Vincent}, {von Fellenberg}, \& {Widmann}}]{GravityEA23}
{GRAVITY Collaboration}, {Perraut}, K., {Labadie}, L., {et~al.} 2021, \aap, 655, A73

\bibitem[{{Herczeg} {et~al.}(2011){Herczeg}, {Brown}, {van Dishoeck}, \& {Pontoppidan}}]{HerczegEA11}
{Herczeg}, G.~J., {Brown}, J.~M., {van Dishoeck}, E.~F., \& {Pontoppidan}, K.~M. 2011, \aap, 533, A112

\bibitem[{{Jennings} {et~al.}(2020){Jennings}, {Booth}, {Tazzari}, {Rosotti}, \& {Clarke}}]{JenningsEA20}
{Jennings}, J., {Booth}, R.~A., {Tazzari}, M., {Rosotti}, G.~P., \& {Clarke}, C.~J. 2020, \mnras, 495, 3209

\bibitem[{{Jennings} {et~al.}(2022){Jennings}, {Tazzari}, {Clarke}, {Booth}, \& {Rosotti}}]{JenningsEA22}
{Jennings}, J., {Tazzari}, M., {Clarke}, C.~J., {Booth}, R.~A., \& {Rosotti}, G.~P. 2022, \mnras, 514, 6053

\bibitem[{{K{\'o}sp{\'a}l} {et~al.}(2023){K{\'o}sp{\'a}l}, {{\'A}brah{\'a}m}, {Diehl}, {Banzatti}, {Bouwman}, {Chen}, {Cruz-S{\'a}enz de Miera}, {Green}, {Henning}, \& {Rab}}]{KospalEA23}
{K{\'o}sp{\'a}l}, {\'A}., {{\'A}brah{\'a}m}, P., {Diehl}, L., {et~al.} 2023, \apjl, 945, L7

\bibitem[{{Labiano} {et~al.}(2021){Labiano}, {Argyriou}, {{\'A}lvarez-M{\'a}rquez}, {Glasse}, {Glauser}, {Patapis}, {Law}, {Brandl}, {Justtanont}, {Lahuis}, {Mart{\'\i}nez-Galarza}, {Mueller}, {Noriega-Crespo}, {Royer}, {Shaughnessy}, \& {Vandenbussche}}]{LabianoEA21}
{Labiano}, A., {Argyriou}, I., {{\'A}lvarez-M{\'a}rquez}, J., {et~al.} 2021, \aap, 656, A57

\bibitem[{{Long} {et~al.}(2019){Long}, {Herczeg}, {Harsono}, {Pinilla}, {Tazzari}, {Manara}, {Pascucci}, {Cabrit}, {Nisini}, {Johnstone}, {Edwards}, {Salyk}, {Menard}, {Lodato}, {Boehler}, {Mace}, {Liu}, {Mulders}, {Hendler}, {Ragusa}, {Fischer}, {Banzatti}, {Rigliaco}, {van de Plas}, {Dipierro}, {Gully-Santiago}, \& {Lopez-Valdivia}}]{LongEA19}
{Long}, F., {Herczeg}, G.~J., {Harsono}, D., {et~al.} 2019, \apj, 882, 49

\bibitem[{{Manara} {et~al.}(2022){Manara}, {Ansdell}, {Rosotti}, {Hughes}, {Armitage}, {Lodato}, \& {Williams}}]{ManaraEA22}
{Manara}, C.~F., {Ansdell}, M., {Rosotti}, G.~P., {et~al.} 2022, arXiv e-prints, arXiv:2203.09930

\bibitem[{{Mandell} {et~al.}(2012){Mandell}, {Bast}, {van Dishoeck}, {Blake}, {Salyk}, {Mumma}, \& {Villanueva}}]{MandellEA12}
{Mandell}, A.~M., {Bast}, J., {van Dishoeck}, E.~F., {et~al.} 2012, \apj, 747, 92

\bibitem[{{McGuire}(2022)}]{McGuire22}
{McGuire}, B.~A. 2022, \apjs, 259, 30

\bibitem[{{Milam} {et~al.}(2005){Milam}, {Savage}, {Brewster}, {Ziurys}, \& {Wyckoff}}]{MilamEA05}
{Milam}, S.~N., {Savage}, C., {Brewster}, M.~A., {Ziurys}, L.~M., \& {Wyckoff}, S. 2005, \apj, 634, 1126

\bibitem[{{Molli{\`e}re} {et~al.}(2022){Molli{\`e}re}, {Molyarova}, {Bitsch}, {Henning}, {Schneider}, {Kreidberg}, {Eistrup}, {Burn}, {Nasedkin}, {Semenov}, {Mordasini}, {Schlecker}, {Schwarz}, {Lacour}, {Nowak}, \& {Schulik}}]{MolliereEA22}
{Molli{\`e}re}, P., {Molyarova}, T., {Bitsch}, B., {et~al.} 2022, \apj, 934, 74

\bibitem[{{Morbidelli} {et~al.}(2012){Morbidelli}, {Lunine}, {O'Brien}, {Raymond}, \& {Walsh}}]{MorbidelliEA12}
{Morbidelli}, A., {Lunine}, J.~I., {O'Brien}, D.~P., {Raymond}, S.~N., \& {Walsh}, K.~J. 2012, Annual Review of Earth and Planetary Sciences, 40, 251

\bibitem[{{Najita} {et~al.}(2010){Najita}, {Carr}, {Strom}, {Watson}, {Pascucci}, {Hollenbach}, {Gorti}, \& {Keller}}]{NajitaEA10}
{Najita}, J.~R., {Carr}, J.~S., {Strom}, S.~E., {et~al.} 2010, \apj, 712, 274

\bibitem[{Newville {et~al.}(2014)Newville, Stensitzki, Allen, \& Ingargiola}]{lmfit}
Newville, M., Stensitzki, T., Allen, D.~B., \& Ingargiola, A. 2014, {LMFIT: Non-Linear Least-Square Minimization and Curve-Fitting for Python}

\bibitem[{{{\"O}berg} \& {Bergin}(2021)}]{OB21}
{{\"O}berg}, K.~I. \& {Bergin}, E.~A. 2021, \physrep, 893, 1

\bibitem[{{{\"O}berg} {et~al.}(2023){{\"O}berg}, {Facchini}, \& {Anderson}}]{ObergEA23}
{{\"O}berg}, K.~I., {Facchini}, S., \& {Anderson}, D.~E. 2023, \araa, 61, 287

\bibitem[{{Pascucci} {et~al.}(2008){Pascucci}, {Apai}, {Hardegree-Ullman}, {Kim}, {Meyer}, \& {Bouwman}}]{PascucciEA08}
{Pascucci}, I., {Apai}, D., {Hardegree-Ullman}, E.~E., {et~al.} 2008, \apj, 673, 477

\bibitem[{{Perotti} {et~al.}(2023){Perotti}, {Christiaens}, {Henning}, {Tabone}, {Waters}, {Kamp}, {Olofsson}, {Grant}, {Gasman}, {Bouwman}, {Samland}, {Franceschi}, {van Dishoeck}, {Schwarz}, {G{\"u}del}, {Lagage}, {Ray}, {Vandenbussche}, {Abergel}, {Absil}, {Arabhavi}, {Argyriou}, {Barrado}, {Boccaletti}, {Caratti o Garatti}, {Geers}, {Glauser}, {Justannont}, {Lahuis}, {Mueller}, {Nehm{\'e}}, {Pantin}, {Scheithauer}, {Waelkens}, {Guadarrama}, {Jang}, {Kanwar}, {Morales-Calder{\'o}n}, {Pawellek}, {Rodgers-Lee}, {Schreiber}, {Colina}, {Greve}, {{\"O}stlin}, \& {Wright}}]{PerottiEA23}
{Perotti}, G., {Christiaens}, V., {Henning}, T., {et~al.} 2023, \nat, 620, 516

\bibitem[{{Pontoppidan} {et~al.}(2014){Pontoppidan}, {Salyk}, {Bergin}, {Brittain}, {Marty}, {Mousis}, \& {{\"O}berg}}]{PontoppidanEA14}
{Pontoppidan}, K.~M., {Salyk}, C., {Bergin}, E.~A., {et~al.} 2014, in Protostars and Planets VI, ed. H.~{Beuther}, R.~S. {Klessen}, C.~P. {Dullemond}, \& T.~{Henning}, 363--385

\bibitem[{{Pontoppidan} {et~al.}(2010){Pontoppidan}, {Salyk}, {Blake}, {Meijerink}, {Carr}, \& {Najita}}]{PontoppidanEA10}
{Pontoppidan}, K.~M., {Salyk}, C., {Blake}, G.~A., {et~al.} 2010, \apj, 720, 887

\bibitem[{{Press} {et~al.}(1992){Press}, {Teukolsky}, {Vetterling}, \& {Flannery}}]{PressEA92}
{Press}, W.~H., {Teukolsky}, S.~A., {Vetterling}, W.~T., \& {Flannery}, B.~P. 1992, {Numerical recipes in C. The art of scientific computing}

\bibitem[{{Rieke} {et~al.}(2015){Rieke}, {Wright}, {B{\"o}ker}, {Bouwman}, {Colina}, {Glasse}, {Gordon}, {Greene}, {G{\"u}del}, {Henning}, {Justtanont}, {Lagage}, {Meixner}, {N{\o}rgaard-Nielsen}, {Ray}, {Ressler}, {van Dishoeck}, \& {Waelkens}}]{RiekeEA15}
{Rieke}, G.~H., {Wright}, G.~S., {B{\"o}ker}, T., {et~al.} 2015, \pasp, 127, 584

\bibitem[{{Rigby} {et~al.}(2023){Rigby}, {Perrin}, {McElwain}, {Kimble}, {Friedman}, {Lallo}, {Doyon}, {Feinberg}, {Ferruit}, {Glasse}, {Rieke}, {Rieke}, {Wright}, {Willott}, {Colon}, {Milam}, {Neff}, {Stark}, {Valenti}, {Abell}, {Abney}, {Abul-Huda}, {Scott Acton}, {Adams}, {Adler}, {Aguilar}, {Ahmed}, {Albert}, {Alberts}, {Aldridge}, {Allen}, {Altenburg}, {{\'A}lvarez-M{\'a}rquez}, {Alves de Oliveira}, {Andersen}, {Anderson}, {Anderson}, {Argyriou}, {Armstrong}, {Arribas}, {Artigau}, {Arvai}, {Atkinson}, {Bacon}, {Bair}, {Banks}, {Barrientes}, {Barringer}, {Bartosik}, {Bast}, {Baudoz}, {Beatty}, {Bechtold}, {Beck}, {Bergeron}, {Bergkoetter}, {Bhatawdekar}, {Birkmann}, {Blazek}, {Blome}, {Boccaletti}, {B{\"o}ker}, {Boia}, {Bonaventura}, {Bond}, {Bosley}, {Boucarut}, {Bourque}, {Bouwman}, {Bower}, {Bowers}, {Boyer}, {Bradley}, {Brady}, {Braun}, {Breda}, {Bresnahan}, {Bright}, {Britt}, {Bromenschenkel}, {Brooks}, {Brooks}, {Brown}, {Brown}, {Brown}, {Bunker}, {Burger}, {Bushouse}, {Cale}, {Cameron}, {Cameron},
  {Canipe}, {Caplinger}, {Caputo}, {Cara}, {Carey}, {Carniani}, {Carrasquilla}, {Carruthers}, {Case}, {Catherine}, {Chance}, {Chapman}, {Charlot}, {Charlow}, {Chayer}, {Chen}, {Cherinka}, {Chichester}, {Chilton}, {Chonis}, {Clampin}, {Clark}, {Clark}, {Coe}, {Coleman}, {Comber}, {Comeau}, {Connolly}, {Cooper}, {Cooper}, {Coppock}, {Correnti}, {Cossou}, {Coulais}, {Coyle}, {Cracraft}, {Curti}, {Cuturic}, {Davis}, {Davis}, {Dean}, {DeLisa}, {deMeester}, {Dencheva}, {Dencheva}, {DePasquale}, {Deschenes}, {Hunor Detre}, {Diaz}, {Dicken}, {DiFelice}, {Dillman}, {Dixon}, {Doggett}, {Donaldson}, {Douglas}, {DuPrie}, {Dupuis}, {Durning}, {Easmin}, {Eck}, {Edeani}, {Egami}, {Ehrenwinkler}, {Eisenhamer}, {Eisenhower}, {Elie}, {Elliott}, {Elliott}, {Ellis}, {Engesser}, {Espinoza}, {Etienne}, {Etxaluze}, {Falini}, {Feeney}, {Ferry}, {Filippazzo}, {Fincham}, {Fix}, {Flagey}, {Florian}, {Flynn}, {Fontanella}, {Ford}, {Forshay}, {Fox}, {Franz}, {Fu}, {Fullerton}, {Galkin}, {Galyer}, {Garc{\'\i}a Mar{\'\i}n}, {Gardner},
  {Gardner}, {Garland}, {Garrett}, {Gasman}, {Gaspar}, {Gaudreau}, {Gauthier}, {Geers}, {Geithner}, {Gennaro}, {Giardino}, {Girard}, {Giuliano}, {Glassmire}, {Glauser}, {Glazer}, {Godfrey}, {Golimowski}, {Gollnitz}, {Gong}, {Gonzaga}, {Gordon}, {Gordon}, {Goudfrooij}, {Greene}, {Greenhouse}, {Grimaldi}, {Groebner}, {Grundy}, {Guillard}, {Gutman}, {Ha}, {Haderlein}, {Hagedorn}, {Hainline}, {Haley}, {Hami}, {Hamilton}, {Hammel}, {Hansen}, {Harkins}, {Harr}, {Hart}, {Hart}, {Hartig}, {Hashimoto}, {Haskins}, {Hathaway}, {Havey}, {Hayden}, {Hecht}, {Heller-Boyer}, {Henriques}, {Henry}, {Hermann}, {Hernandez}, {Hesman}, {Hicks}, {Hilbert}, {Hines}, {Hoffman}, {Holfeltz}, {Holler}, {Hoppa}, {Hott}, {Howard}, {Howard}, {Hunter}, {Hunter}, {Hurst}, {Husemann}, {Hustak}, {Ilinca Ignat}, {Illingworth}, {Irish}, {Jackson}, {Jahromi}, {Jakobsen}, {James}, {James}, {Januszewski}, {Jenkins}, {Jirdeh}, {Johnson}, {Johnson}, {Jones}, {Jones}, {Jones}, {Jones}, {Jordan}, {Jordan}, {Jurczyk}, {Jurling}, {Kaleida}, {Kalmanson},
  {Kammerer}, {Kang}, {Kao}, {Karakla}, {Kavanagh}, {Kelly}, {Kendrew}, {Kennedy}, {Kenny}, {Keski-kuha}, {Keyes}, {Kidwell}, {Kinzel}, {Kirk}, {Kirkpatrick}, {Kirshenblat}, {Klaassen}, {Knapp}, {Scott Knight}, {Knollenberg}, {Koehler}, {Koekemoer}, {Kovacs}, {Kulp}, {Kumari}, {Kyprianou}, {La Massa}, {Labador}, {Labiano}, {Lagage}, {Lajoie}, {Lallo}, {Lam}, {Lamb}, {Lambros}, {Lampenfield}, {Langston}, {Larson}, {Law}, {Lawrence}, {Lee}, {Leisenring}, {Lepo}, {Leveille}, {Levenson}, {Levine}, {Levy}, {Lewis}, {Lewis}, {Libralato}, {Lightsey}, {Link}, {Liu}, {Lo}, {Lockwood}, {Logue}, {Long}, {Long}, {Loomis}, {Lopez-Caniego}, {Lorenzo Alvarez}, {Love-Pruitt}, {Lucy}, {Luetzgendorf}, {Maghami}, {Maiolino}, {Major}, {Malla}, {Malumuth}, {Manjavacas}, {Mannfolk}, {Marrione}, {Marston}, {Martel}, {Maschmann}, {Masci}, {Masciarelli}, {Maszkiewicz}, {Mather}, {McKenzie}, {McLean}, {McMaster}, {Melbourne}, {Mel{\'e}ndez}, {Menzel}, {Merz}, {Meyett}, {Meza}, {Miskey}, {Misselt}, {Moller}, {Morrison}, {Morse},
  {Moseley}, {Mosier}, {Mountain}, {Mueckay}, {Mueller}, {Mullally}, {Murphy}, {Murray}, {Murray}, {Mustelier}, {Muzerolle}, {Mycroft}, {Myers}, {Myrick}, {Nanavati}, {Nance}, {Nayak}, {Naylor}, {Nelan}, {Nickson}, {Nielson}, {Nieto-Santisteban}, {Nikolov}, {Noriega-Crespo}, {O'Shaughnessy}, {O'Sullivan}, {Ochs}, {Ogle}, {Oleszczuk}, {Olmsted}, {Osborne}, {Ottens}, {Owens}, {Pacifici}, {Pagan}, {Page}, {Park}, {Parrish}, {Patapis}, {Paul}, {Pauly}, {Pavlovsky}, {Pedder}, {Peek}, {Pena-Guerrero}, {Penanen}, {Perez}, {Perna}, {Perriello}, {Phillips}, {Pietraszkiewicz}, {Pinaud}, {Pirzkal}, {Pitman}, {Piwowar}, {Platais}, {Player}, {Plesha}, {Pollizi}, {Polster}, {Pontoppidan}, {Porterfield}, {Proffitt}, {Pueyo}, {Pulliam}, {Quirt}, {Quispe Neira}, {Ramos Alarcon}, {Ramsay}, {Rapp}, {Rapp}, {Rauscher}, {Ravindranath}, {Rawle}, {Regan}, {Reichard}, {Reis}, {Ressler}, {Rest}, {Reynolds}, {Rhue}, {Richon}, {Rickman}, {Ridgaway}, {Ritchie}, {Rix}, {Robberto}, {Robinson}, {Robinson}, {Robinson}, {Rock}, {Rodriguez},
  {Rodriguez Del Pino}, {Roellig}, {Rohrbach}, {Roman}, {Romelfanger}, {Rose}, {Roteliuk}, {Roth}, {Rothwell}, {Rowlands}, {Roy}, {Royer}, {Royle}, {Rui}, {Rumler}, {Runnels}, {Russ}, {Rustamkulov}, {Ryden}, {Ryer}, {Sabata}, {Sabatke}, {Sabbi}, {Samuelson}, {Sapp}, {Sappington}, {Sargent}, {Sauer}, {Scheithauer}, {Schlawin}, {Schlitz}, {Schmitz}, {Schneider}, {Schreiber}, {Schulze}, {Schwab}, {Scott}, {Sembach}, {Shanahan}, {Shaughnessy}, {Shaw}, {Shawger}, {Shay}, {Sheehan}, {Shen}, {Sherman}, {Shiao}, {Shih}, {Shivaei}, {Sienkiewicz}, {Sing}, {Sirianni}, {Sivaramakrishnan}, {Skipper}, {Sloan}, {Slocum}, {Slowinski}, {Smith}, {Smith}, {Smith}, {Smith}, {Snyder}, {Soh}, {Tony Sohn}, {Soto}, {Spencer}, {Stallcup}, {Stansberry}, {Starr}, {Starr}, {Stewart}, {Stiavelli}, {Straughn}, {Strickland}, {Stys}, {Summers}, {Sun}, {Sunnquist}, {Swade}, {Swam}, {Swaters}, {Swoish}, {Taylor}, {Taylor}, {Te Plate}, {Tea}, {Teague}, {Telfer}, {Temim}, {Thatte}, {Thompson}, {Thompson}, {Thomson}, {Tikkanen}, {Tippet},
  {Todd}, {Toolan}, {Tran}, {Trejo}, {Truong}, {Tsukamoto}, {Tustain}, {Tyra}, {Ubeda}, {Underwood}, {Uzzo}, {Van Campen}, {Vandal}, {Vandenbussche}, {Vila}, {Volk}, {Wahlgren}, {Waldman}, {Walker}, {Wander}, {Warfield}, {Warner}, {Wasiak}, {Watkins}, {Weaver}, {Weilert}, {Weiser}, {Weiss}, {Weissman}, {Welty}, {West}, {Wheate}, {Wheatley}, {Wheeler}, {White}, {Whiteaker}, {Whitehouse}, {Whiteleather}, {Whitman}, {Williams}, {Willmer}, {Willoughby}, {Wilson}, {Wirth}, {Wislowski}, {Wolf}, {Wolfe}, {Wolff}, {Workman}, {Wright}, {Wu}, {Wu}, {Wymer}, {Yates}, {Yeager}, {Yeates}, {Yerger}, {Yoon}, {Young}, {Yu}, {Zak}, {Zeidler}, {Zhou}, {Zielinski}, {Zincke}, \& {Zonak}}]{RigbyEA23}
{Rigby}, J., {Perrin}, M., {McElwain}, M., {et~al.} 2023, \pasp, 135, 048001

\bibitem[{{Salyk} {et~al.}(2011{\natexlab{a}}){Salyk}, {Blake}, {Boogert}, \& {Brown}}]{SalykEA11b}
{Salyk}, C., {Blake}, G.~A., {Boogert}, A.~C.~A., \& {Brown}, J.~M. 2011{\natexlab{a}}, \apj, 743, 112

\bibitem[{{Salyk} {et~al.}(2011{\natexlab{b}}){Salyk}, {Pontoppidan}, {Blake}, {Najita}, \& {Carr}}]{SalykEA11}
{Salyk}, C., {Pontoppidan}, K.~M., {Blake}, G.~A., {Najita}, J.~R., \& {Carr}, J.~S. 2011{\natexlab{b}}, \apj, 731, 130

\bibitem[{{Tabone} {et~al.}(2023){Tabone}, {Bettoni}, {van Dishoeck}, {Arabhavi}, {Grant}, {Gasman}, {Henning}, {Kamp}, {G{\"u}del}, {Lagage}, {Ray}, {Vandenbussche}, {Abergel}, {Absil}, {Argyriou}, {Barrado}, {Boccaletti}, {Bouwman}, {Caratti o Garatti}, {Geers}, {Glauser}, {Justannont}, {Lahuis}, {Mueller}, {Nehm{\'e}}, {Olofsson}, {Pantin}, {Scheithauer}, {Waelkens}, {Waters}, {Black}, {Christiaens}, {Guadarrama}, {Morales-Calder{\'o}n}, {Jang}, {Kanwar}, {Pawellek}, {Perotti}, {Perrin}, {Rodgers-Lee}, {Samland}, {Schreiber}, {Schwarz}, {Colina}, {{\"O}stlin}, \& {Wright}}]{TaboneEA23}
{Tabone}, B., {Bettoni}, G., {van Dishoeck}, E.~F., {et~al.} 2023, Nature Astronomy [\eprint[arXiv]{2304.05954}]

\bibitem[{{Thi} {et~al.}(2013){Thi}, {Kamp}, {Woitke}, {van der Plas}, {Bertelsen}, \& {Wiesenfeld}}]{ThiEA13}
{Thi}, W.~F., {Kamp}, I., {Woitke}, P., {et~al.} 2013, \aap, 551, A49

\bibitem[{{Trapman} {et~al.}(2019){Trapman}, {Facchini}, {Hogerheijde}, {van Dishoeck}, \& {Bruderer}}]{TrapmanEA19}
{Trapman}, L., {Facchini}, S., {Hogerheijde}, M.~R., {van Dishoeck}, E.~F., \& {Bruderer}, S. 2019, \aap, 629, A79

\bibitem[{{Valenti} {et~al.}(2000){Valenti}, {Johns-Krull}, \& {Linsky}}]{ValentiEA00}
{Valenti}, J.~A., {Johns-Krull}, C.~M., \& {Linsky}, J.~L. 2000, \apjs, 129, 399

\bibitem[{{van der Marel} {et~al.}(2021){van der Marel}, {Booth}, {Leemker}, {van Dishoeck}, \& {Ohashi}}]{MarelEA21}
{van der Marel}, N., {Booth}, A.~S., {Leemker}, M., {van Dishoeck}, E.~F., \& {Ohashi}, S. 2021, \aap, 651, L5

\bibitem[{{van der Plas} {et~al.}(2015){van der Plas}, {van den Ancker}, {Waters}, \& {Dominik}}]{vdPlasEA15}
{van der Plas}, G., {van den Ancker}, M.~E., {Waters}, L.~B.~F.~M., \& {Dominik}, C. 2015, \aap, 574, A75

\bibitem[{{Walsh} {et~al.}(2015){Walsh}, {Nomura}, \& {van Dishoeck}}]{WalshEA15}
{Walsh}, C., {Nomura}, H., \& {van Dishoeck}, E. 2015, \aap, 582, A88

\bibitem[{{Wells} {et~al.}(2015){Wells}, {Pel}, {Glasse}, {Wright}, {Aitink-Kroes}, {Azzollini}, {Beard}, {Brandl}, {Gallie}, {Geers}, {Glauser}, {Hastings}, {Henning}, {Jager}, {Justtanont}, {Kruizinga}, {Lahuis}, {Lee}, {Martinez-Delgado}, {Mart{\'\i}nez-Galarza}, {Meijers}, {Morrison}, {M{\"u}ller}, {Nakos}, {O'Sullivan}, {Oudenhuysen}, {Parr-Burman}, {Pauwels}, {Rohloff}, {Schmalzl}, {Sykes}, {Thelen}, {van Dishoeck}, {Vandenbussche}, {Venema}, {Visser}, {Waters}, \& {Wright}}]{WellsEA15}
{Wells}, M., {Pel}, J.~W., {Glasse}, A., {et~al.} 2015, \pasp, 127, 646

\bibitem[{{Wilson}(1999)}]{Wilson99}
{Wilson}, T.~L. 1999, Reports on Progress in Physics, 62, 143

\bibitem[{{Woitke} {et~al.}(2018){Woitke}, {Min}, {Thi}, {Roberts}, {Carmona}, {Kamp}, {M{\'e}nard}, \& {Pinte}}]{WoitkeEA18}
{Woitke}, P., {Min}, M., {Thi}, W.~F., {et~al.} 2018, \aap, 618, A57

\bibitem[{{Wright} {et~al.}(2010){Wright}, {Eisenhardt}, {Mainzer}, {Ressler}, {Cutri}, {Jarrett}, {Kirkpatrick}, {Padgett}, {McMillan}, {Skrutskie}, {Stanford}, {Cohen}, {Walker}, {Mather}, {Leisawitz}, {Gautier}, {McLean}, {Benford}, {Lonsdale}, {Blain}, {Mendez}, {Irace}, {Duval}, {Liu}, {Royer}, {Heinrichsen}, {Howard}, {Shannon}, {Kendall}, {Walsh}, {Larsen}, {Cardon}, {Schick}, {Schwalm}, {Abid}, {Fabinsky}, {Naes}, \& {Tsai}}]{WISE}
{Wright}, E.~L., {Eisenhardt}, P. R.~M., {Mainzer}, A.~K., {et~al.} 2010, \aj, 140, 1868

\bibitem[{{Wright} {et~al.}(2023){Wright}, {Rieke}, {Glasse}, {Ressler}, {Garc{\'\i}a Mar{\'\i}n}, {Aguilar}, {Alberts}, {{\'A}lvarez-M{\'a}rquez}, {Argyriou}, {Banks}, {Baudoz}, {Boccaletti}, {Bouchet}, {Bouwman}, {Brandl}, {Breda}, {Bright}, {Cale}, {Colina}, {Cossou}, {Coulais}, {Cracraft}, {De Meester}, {Dicken}, {Engesser}, {Etxaluze}, {Fox}, {Friedman}, {Fu}, {Gasman}, {G{\'a}sp{\'a}r}, {Gastaud}, {Geers}, {Glauser}, {Gordon}, {Greene}, {Greve}, {Grundy}, {G{\"u}del}, {Guillard}, {Haderlein}, {Hashimoto}, {Henning}, {Hines}, {Holler}, {Detre}, {Jahromi}, {James}, {Jones}, {Justtanont}, {Kavanagh}, {Kendrew}, {Klaassen}, {Krause}, {Labiano}, {Lagage}, {Lambros}, {Larson}, {Law}, {Lee}, {Libralato}, {Lorenzo Alverez}, {Meixner}, {Morrison}, {Mueller}, {Murray}, {Mycroft}, {Myers}, {Nayak}, {Naylor}, {Nickson}, {Noriega-Crespo}, {{\"O}stlin}, {O'Sullivan}, {Ottens}, {Patapis}, {Penanen}, {Pietraszkiewicz}, {Ray}, {Regan}, {Roteliuk}, {Royer}, {Samara-Ratna}, {Samuelson}, {Sargent}, {Scheithauer},
  {Schneider}, {Schreiber}, {Shaughnessy}, {Sheehan}, {Shivaei}, {Sloan}, {Tamas}, {Teague}, {Temim}, {Tikkanen}, {Tustain}, {van Dishoeck}, {Vandenbussche}, {Weilert}, {Whitehouse}, \& {Wolff}}]{WrightEA23}
{Wright}, G.~S., {Rieke}, G.~H., {Glasse}, A., {et~al.} 2023, \pasp, 135, 048003

\bibitem[{{Wright} {et~al.}(2015){Wright}, {Wright}, {Goodson}, {Rieke}, {Aitink-Kroes}, {Amiaux}, {Aricha-Yanguas}, {Azzollini}, {Banks}, {Barrado-Navascues}, {Belenguer-Davila}, {Blommaert}, {Bouchet}, {Brandl}, {Colina}, {Detre}, {Diaz-Catala}, {Eccleston}, {Friedman}, {Garc{\'\i}a-Mar{\'\i}n}, {G{\"u}del}, {Glasse}, {Glauser}, {Greene}, {Groezinger}, {Grundy}, {Hastings}, {Henning}, {Hofferbert}, {Hunter}, {Jessen}, {Justtanont}, {Karnik}, {Khorrami}, {Krause}, {Labiano}, {Lagage}, {Langer}, {Lemke}, {Lim}, {Lorenzo-Alvarez}, {Mazy}, {McGowan}, {Meixner}, {Morris}, {Morrison}, {M{\"u}ller}, {rgaard-Nielson}, {Olofsson}, {O'Sullivan}, {Pel}, {Penanen}, {Petach}, {Pye}, {Ray}, {Renotte}, {Renouf}, {Ressler}, {Samara-Ratna}, {Scheithauer}, {Schneider}, {Shaughnessy}, {Stevenson}, {Sukhatme}, {Swinyard}, {Sykes}, {Thatcher}, {Tikkanen}, {van Dishoeck}, {Waelkens}, {Walker}, {Wells}, \& {Zhender}}]{WrightEA15}
{Wright}, G.~S., {Wright}, D., {Goodson}, G.~B., {et~al.} 2015, \pasp, 127, 595

\bibitem[{{Zhang} {et~al.}(2023){Zhang}, {Kalscheur}, {Long}, {Zhang}, {Long}, {Bergin}, {Zhu}, \& {Trapman}}]{ZhangEA23}
{Zhang}, S., {Kalscheur}, M., {Long}, F., {et~al.} 2023, \apj, 952, 108

\end{thebibliography}

%% ---
\newpage
\onecolumn

\begin{appendix}
\section{Continuum subtraction} \label{app:CS}
\begin{figure*}[ht!]
    \centering
    \includegraphics[width=0.95\textwidth]{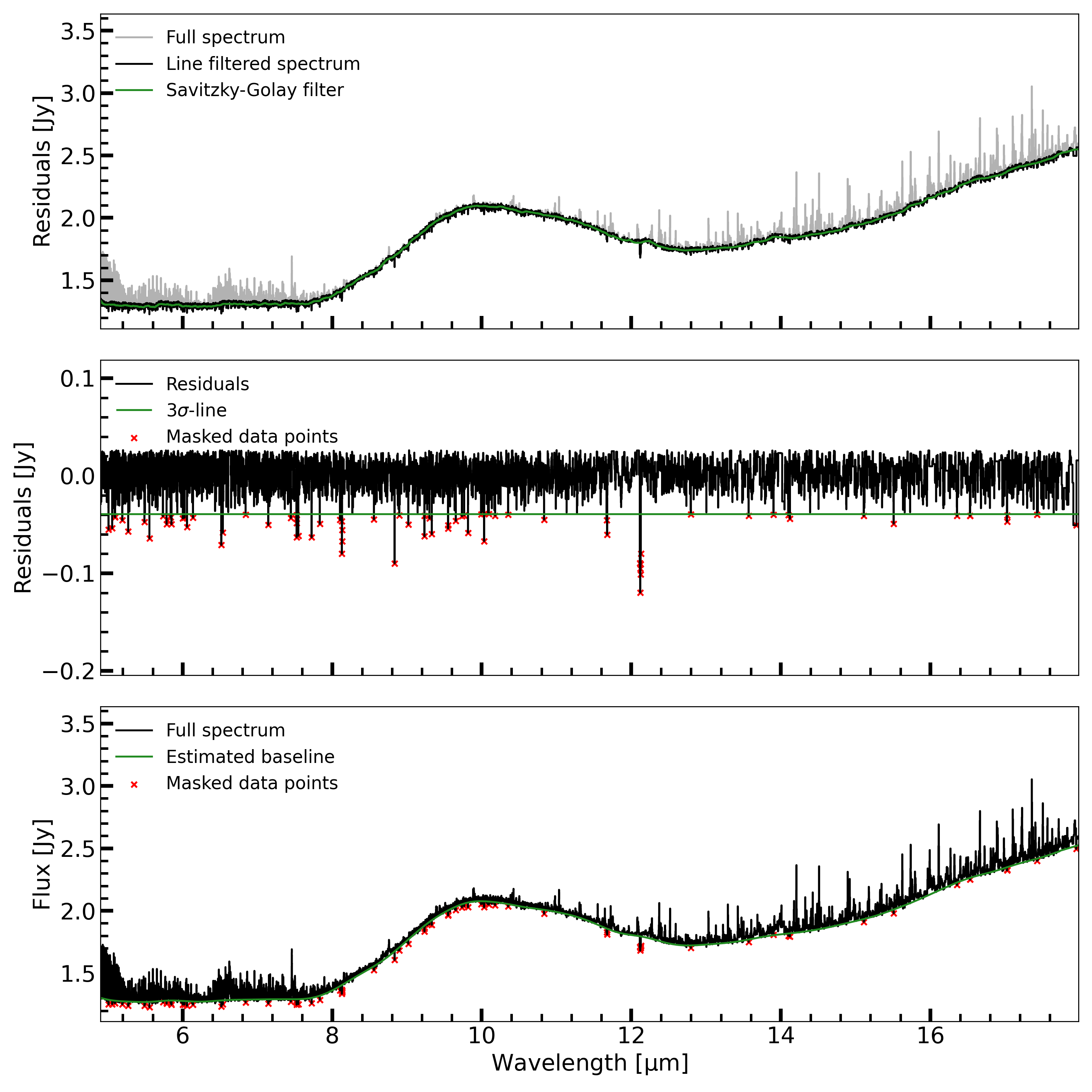}
    \caption{The intermediate steps of the continuum subtraction method described in Section \ref{sec:ContinuumSubtraction}. The top panel shows the final Savitzky-Golay filter (green) on the filtered line spectrum (black) and the filtered lines (light grey). The middle panel displays the residuals (black; line-filtered spectrum after subtracting the Savitzky-Golay filter), together with their -3$\times$STD level (green), and the data points that fall below the -3$\times$ level (red crosses). The bottom panel shows the full JWST-MIRI spectrum (black) together with the estimated baseline (green) and the data points that have been masked throughout the estimation of the baseline (red crosses).}
    \label{fig:CS_Intermediate}
\end{figure*}

This section provides the immediate steps of the continuum subtraction described in Section \ref{sec:ContinuumSubtraction}. The necessary images are shown in Figure \ref{fig:CS_Intermediate}. The top panel of this figure shows the final Savitzky-Golay filter (green) after all $>$2$\sigma$ emission lines (light grey) have been filtered, together with the line-filtered spectrum. The middle panel displays the residuals after subtracting the Savitzky-Golay filter from the line-filtered spectrum. The green lines indicate the -3$\times$ level of the STD of the residuals. All the red crosses indicate the data points that lie below the -3$\times$STD lines and have, subsequently, been masked throughout the baseline estimation. Finally, the bottom panel shows the full JWST-MIRI spectrum, together with the estimated baseline, and the masked data points.

\clearpage
\section{Uncertainties for the JWST-MIRI subbands}
\begin{table}[ht!]
    \centering
    \caption{The median continuum flux, estimated $S/N$ from the ETC, and the acquired uncertainty $\sigma$ for each subband.}
    \begin{tabular}{c c c c}
    \hline\hline
    Subband ([$\mu$m]) & Median flux [Jy] & ETC S/N & $\sigma$ [mJy] \\
    \hline
    1A (4.90-5.74) & 1.31 & 482.3 & 2.7 \\
    1B (5.66-6.63) & 1.31 & 539.2 & 2.4 \\
    1C (6.53-7.65) & 1.32 & 653.2 & 2.0 \\
    2A (7.51-8.77) & 1.43 & 677.7 & 2.1 \\
    2B (8.67-10.13) & 1.98 & 895.9 & 2.2 \\
    2C (10.02-11.70) & 2.03 & 982.5 & 2.1 \\
    3A (11.55-13.47) & 1.79 & 845.3 & 2.1 \\
    3B (13.34-15.57) & 1.89 & 865.5 & 2.2 \\
    3C (15.41-17.98) & 2.34 & 1030.3 & 2.3 \\
    4A (17.70-20.95) & 2.74 & 425.0 & 6.4 \\
    4B (20.69-24.48) & 2.85 & 246.3 & 11.6 \\
    4C (24.19-27.90) & 2.69 & 76.6 & 35.2 \\
    \hline
    \end{tabular}
    \label{tab:Subband-Sigma}
\end{table}

\section{$\chi^2$-maps of \ce{CO_2}, \ce{HCN}, and \ce{C_2H_2}}
\begin{figure*}[ht!]
    \centering
    \includegraphics[width=0.95\textwidth]{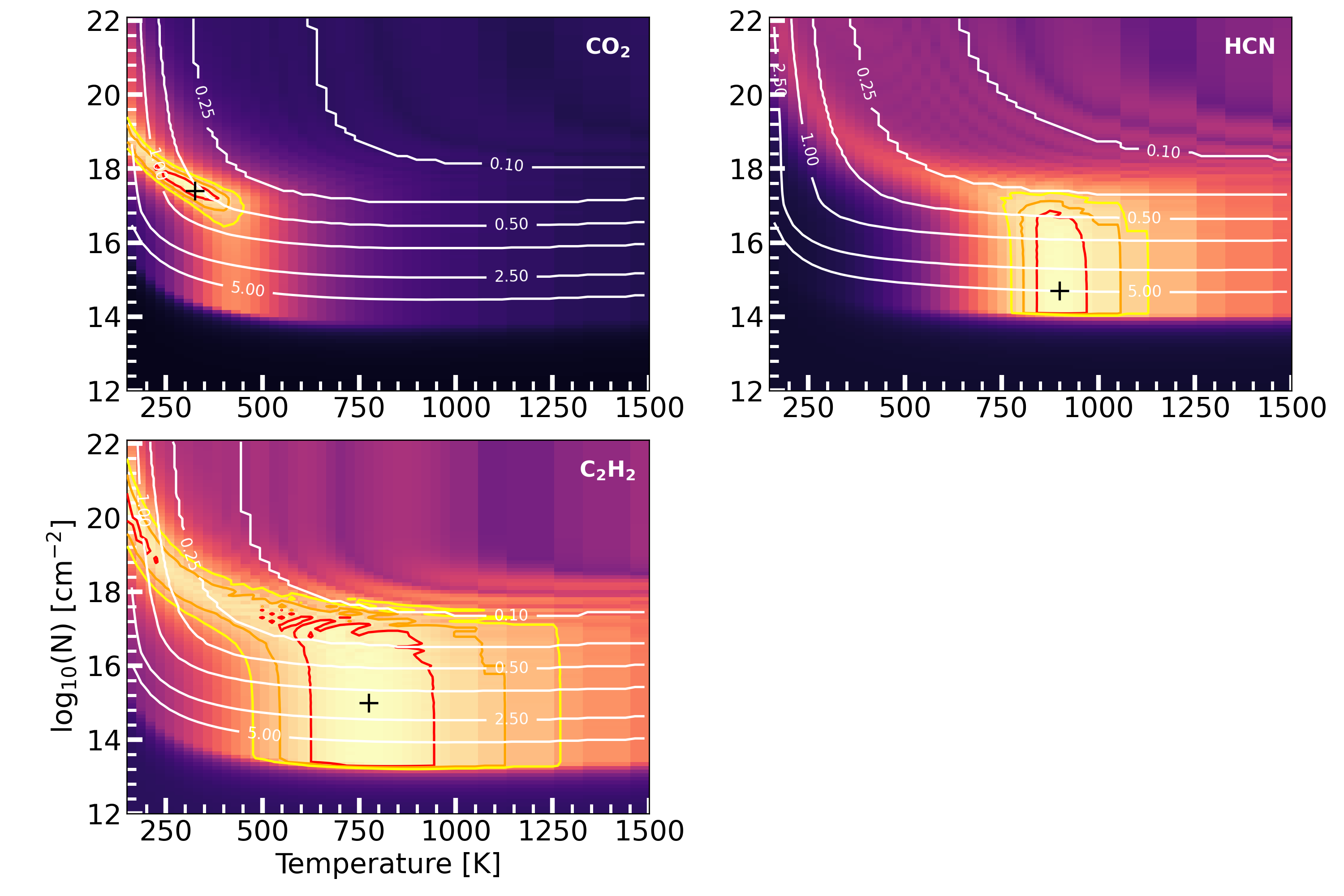}
    \caption{The normalised $\chi^2$-maps ($\chi^2_\textnormal{min}/\chi^2$) for \ce{CO_2} (top left), \ce{HCN} (top right), and \ce{C_2H_2} (bottom left). The dark plus indicates the best fitting parameters, while red, orange, and yellow contours indicate the 1$\sigma$-, 2$\sigma$-, and 3$\sigma$-confidence intervals, respectively. The white contours indicate the emitting radii in au.}
    \label{fig:MIRI-Chi2}
\end{figure*}

\clearpage
\section{Complementary \ce{CO} ro-vibrational observations}
\subsection{Line information and integrated fluxes}
\begin{table}[ht!]
    \centering
    \caption{Information on the \ce{CO} transitions used for the rotational diagrams and corresponding integrated fluxes.}
    \begin{tabular}{c c c c c c c c c c}
        \hline\hline
        Molecule & Rot. Trans. & Wavelength & $E_\textnormal{up}$ & $A_\textnormal{ul}$ & $g_\textnormal{up}$ & \multicolumn{2}{c}{Int. Flux$^{(\alpha)}$} & Line prof.$^{(\beta)}$ & Blend$^{(\gamma)}$ \\
        & & & & & & Narrow comp. & Broad comp. & \\
        & & [$\mu$m] & [K] & [s$^{-1}$] & & \multicolumn{2}{c}{[erg s$^{-1}$ cm$^{-2}$]} & \\
        \hline
        \ce{^{12}CO} v=1-0 & R-18 & 4.52589 & 4123.9 & 19.15 & 39 & 1.98$\pm$0.05$\times$10$^{-14}$ & 1.34$\pm$0.03$\times$10$^{-13}$ & N & N \\
        & R-17 & 4.53237 & 4020.0 & 19.03 & 37 & 2.04$\pm$0.05$\times$10$^{-14}$ & 1.38$\pm$0.03$\times$10$^{-13}$ & N & N \\
        & R-14 & 4.55236 & 3741.0 & 18.66 & 31 & 2.84$\pm$0.17$\times$10$^{-14}$ & 1.12$\pm$0.04$\times$10$^{-13}$ & N & Y \\
        & R-13 & 4.55922 & 3658.9 & 18.52 & 29 & 2.91$\pm$0.06$\times$10$^{-14}$ & 1.01$\pm$0.01$\times$10$^{-13}$ & N & Y \\
        & R-12 & 4.56617 & 3582.2 & 18.38 & 27 & 3.00$\pm$0.09$\times$10$^{-14}$ & 9.77$\pm$0.17$\times$10$^{-14}$ & N & N \\
        & R-11 & 4.57323 & 3511.0 & 18.24 & 25 & 2.99$\pm$0.06$\times$10$^{-14}$ & 9.75$\pm$0.20$\times$10$^{-14}$ & N & Y \\
        & R-10 & 4.58038 & 3445.3 & 18.08 & 23 & 3.12$\pm$0.07$\times$10$^{-14}$ & 9.36$\pm$0.14$\times$10$^{-14}$ & N & N \\
        & R-9 & 4.58764 & 3385.1 & 17.91 & 21 & 2.30$\pm$0.19$\times$10$^{-14}$ & 8.99$\pm$0.20$\times$10$^{-14}$ & N & Y \\
        & R-7 & 4.60244 & 3281.0 & 17.52 & 17 & 3.00$\pm$0.07$\times$10$^{-14}$ & 8.86$\pm$0.12$\times$10$^{-14}$ & N & Y \\
        & R-6 & 4.61000 & 3237.1 & 17.28 & 15 & 3.00$\pm$0.31$\times$10$^{-14}$ & 8.33$\pm$0.07$\times$10$^{-14}$ & N & Y \\
        & R-5 & 4.61766 & 3198.8 & 17.00 & 13 & 2.92$\pm$0.06$\times$10$^{-14}$ & 7.86$\pm$0.12$\times$10$^{-14}$ & N & Y \\
        & R-4 & 4.62541 & 3165.9 & 16.65 & 11 & 1.65$\pm$0.04$\times$10$^{-14}$ & 1.12$\pm$0.02$\times$10$^{-13}$ & N & N \\
        & R-3 & 4.63328 & 3138.5 & 16.20 & 9 & 2.94$\pm$0.06$\times$10$^{-14}$ & 6.64$\pm$0.09$\times$10$^{-14}$ & Y & N \\
        & R-2 & 4.64124 & 3116.6 & 15.54 & 7 & 2.39$\pm$0.04$\times$10$^{-14}$ & 5.88$\pm$0.07$\times$10$^{-14}$ & Y & N \\
        & R-1 & 4.64931 & 3100.1 & 14.42 & 5 & 1.63$\pm$0.03$\times$10$^{-14}$ & 5.45$\pm$0.06$\times$10$^{-14}$ & Y & N \\
        & R-0 & 4.65749 & 3089.2 & 11.95 & 3 & 1.13$\pm$0.04$\times$10$^{-14}$ & 4.43$\pm$0.06$\times$10$^{-14}$ & N & Y \\
        & P-1 & 4.67415 & 3083.7 & 35.46 & 1 & 1.47$\pm$0.05$\times$10$^{-14}$ & 5.35$\pm$0.09$\times$10$^{-14}$ & N & N \\
        & P-2 & 4.68264 & 3089.2 & 23.51 & 3 & 2.34$\pm$0.04$\times$10$^{-14}$ & 4.71$\pm$0.07$\times$10$^{-14}$ & N & Y \\
        & P-3 & 4.69124 & 3100.1 & 21.04 & 5 & 2.46$\pm$0.04$\times$10$^{-14}$ & 5.27$\pm$0.16$\times$10$^{-14}$ & N & Y \\
        & P-4 & 4.69995 & 3116.6 & 19.92 & 7 & 2.83$\pm$0.05$\times$10$^{-14}$ & 6.42$\pm$0.10$\times$10$^{-14}$ & N & Y \\
        & P-5 & 4.70877 & 3138.5 & 19.26 & 9 & 3.03$\pm$0.05$\times$10$^{-14}$ & 7.26$\pm$0.09$\times$10$^{-14}$ & Y & N \\
        & P-6 & 4.71769 & 3165.9 & 18.80 & 11 & 3.32$\pm$0.09$\times$10$^{-14}$ & 7.01$\pm$0.15$\times$10$^{-14}$ & Y & N \\
        & P-7 & 4.72673 & 3198.8 & 18.45 & 13 & 2.75$\pm$0.08$\times$10$^{-14}$ & 7.73$\pm$0.13$\times$10$^{-14}$ & N & Y \\
        & P-8 & 4.73587 & 3237.1 & 18.17 & 15 & 3.28$\pm$0.04$\times$10$^{-14}$ & 7.71$\pm$0.06$\times$10$^{-14}$ & Y & N \\
        & P-9 & 4.74513 & 3281.0 & 17.93 & 17 & 3.35$\pm$0.07$\times$10$^{-14}$ & 8.10$\pm$0.10$\times$10$^{-14}$ & Y & N \\
        & P-10 & 4.75450 & 3330.3 & 17.71 & 19 & 2.80$\pm$0.06$\times$10$^{-14}$ & 8.78$\pm$0.09$\times$10$^{-14}$ & N & Y \\
        & P-12 & 4.77358 & 3445.3 & 17.35 & 23 & 2.98$\pm$0.06$\times$10$^{-14}$ & 8.70$\pm$0.08$\times$10$^{-14}$ & Y & N \\
        & P-13 & 4.78330 & 3511.0 & 17.19 & 25 & 2.05$\pm$0.09$\times$10$^{-14}$ & 1.38$\pm$0.06$\times$10$^{-13}$ & N & N \\
        & P-14 & 4.79312 & 3582.2 & 17.03 & 27 & 3.21$\pm$0.05$\times$10$^{-14}$ & 8.50$\pm$0.07$\times$10$^{-14}$ & N & N \\
        & P-17 & 4.82331 & 3828.5 & 16.61 & 33 & 2.95$\pm$0.06$\times$10$^{-14}$ & 8.83$\pm$0.08$\times$10$^{-14}$ & N & Y \\
        & P-18 & 4.83361 & 3921.6 & 16.48 & 35 & 2.85$\pm$0.06$\times$10$^{-14}$ & 8.26$\pm$0.22$\times$10$^{-14}$ & N & Y \\
        & P-19 & 4.84403 & 4020.0 & 16.35 & 37 & 2.50$\pm$0.08$\times$10$^{-14}$ & 9.45$\pm$0.14$\times$10$^{-14}$ & N & Y \\
        & P-25 & 4.90912 & 4725.0 & 15.62 & 49 & 2.19$\pm$0.04$\times$10$^{-14}$ & 8.66$\pm$0.07$\times$10$^{-14}$ & Y & N \\
        & P-26 & 4.92041 & 4861.5 & 15.50 & 51 & 2.20$\pm$0.04$\times$10$^{-14}$ & 8.16$\pm$0.06$\times$10$^{-14}$ & Y & N \\
        & P-27 & 4.93182 & 5003.4 & 15.39 & 53 & 1.94$\pm$0.07$\times$10$^{-14}$ & 8.36$\pm$0.12$\times$10$^{-14}$ & N & N \\
        & P-29 & 4.95504 & 5303.5 & 15.16 & 57 & 1.48$\pm$0.10$\times$10$^{-14}$ & 8.50$\pm$0.16$\times$10$^{-14}$ & N & N \\
        & P-31 & 4.97878 & 5625.1 & 14.93 & 61 & 1.65$\pm$0.07$\times$10$^{-14}$ & 7.38$\pm$0.16$\times$10$^{-14}$ & N & Y \\
        & P-33 & 5.00305 & 5968.2 & 14.71 & 65 & 1.65$\pm$0.09$\times$10$^{-14}$ & 7.36$\pm$0.14$\times$10$^{-14}$ & N & N \\
        & P-37 & 5.05324 & 6718.9 & 14.28 & 73 & 1.15$\pm$0.04$\times$10$^{-14}$ & 6.13$\pm$0.07$\times$10$^{-14}$ & N & N \\
        & P-38 & 5.06613 & 6920.0 & 14.17 & 75 & 1.11$\pm$0.05$\times$10$^{-14}$ & 5.90$\pm$0.08$\times$10$^{-14}$ & N & N \\
        & P-39 & 5.07917 & 7126.4 & 14.06 & 77 & 1.08$\pm$0.12$\times$10$^{-14}$ & 5.50$\pm$0.22$\times$10$^{-14}$ & N & Y \\
        & P-41 & 5.10568 & 7555.1 & 13.85 & 81 & 8.02$\pm$0.58$\times$10$^{-15}$ & 4.73$\pm$0.11$\times$10$^{-14}$ & N & Y \\
        & P-46 & 5.17452 & 8719.7 & 13.32 & 91 & 5.76$\pm$0.43$\times$10$^{-15}$ & 3.90$\pm$0.29$\times$10$^{-14}$ & N & N \\
        \hline
        \ce{^{12}CO} v=2-1 & R-22 & 4.55692 & 7625.6 & 37.63 & 47 & 3.86$\pm$0.83$\times$10$^{-15}$ & 2.91$\pm$0.72$\times$10$^{-14}$ & N & Y \\
        & R-21 & 4.56309 & 7501.1 & 37.43 & 45 & 3.85$\pm$0.52$\times$10$^{-15}$ & 3.47$\pm$0.09$\times$10$^{-14}$ & N & Y \\
        & R-19 & 4.57573 & 7268.3 & 37.01 & 41 & 3.12$\pm$0.54$\times$10$^{-15}$ & 4.13$\pm$0.11$\times$10$^{-14}$ & N & Y \\
        & R-18 & 4.58219 & 7159.9 & 36.79 & 39 & 5.77$\pm$0.38$\times$10$^{-15}$ & 3.25$\pm$0.07$\times$10$^{-14}$ & N & Y \\
        & R-17 & 4.58875 & 7057.0 & 36.57 & 37 & 4.86$\pm$1.49$\times$10$^{-15}$ & 3.75$\pm$0.26$\times$10$^{-14}$ & N & N \\
        & R-14 & 4.60902 & 6780.5 & 35.86 & 31 & 3.46$\pm$0.63$\times$10$^{-15}$ & 3.62$\pm$0.12$\times$10$^{-14}$ & N & Y \\
        & R-13 & 4.61598 & 6699.2 & 35.60 & 29 & 5.34$\pm$1.10$\times$10$^{-15}$ & 3.29$\pm$0.18$\times$10$^{-14}$ & Y & N \\
        & R-12 & 4.62303 & 6623.2 & 35.33 & 27 & 5.86$\pm$1.25$\times$10$^{-15}$ & 2.86$\pm$0.22$\times$10$^{-14}$ & N & N \\
        \hline
    \end{tabular}
    \label{tab:CO_IntFluxes}
\end{table}
\clearpage
\begin{table}[ht!]
    \ContinuedFloat
    \centering
    \caption{Continuation of Table \ref{tab:CO_IntFluxes}}
    \begin{tabular}{c c c c c c c c c c}
        \hline\hline
        Molecule & Rot. Trans. & Wavelength & $E_\textnormal{up}$ & $A_\textnormal{ul}$ & $g_\textnormal{up}$ & \multicolumn{2}{c}{Int. Flux$^{(\alpha)}$} & Line prof.$^{(\beta)}$ & Blend$^{(\gamma)}$ \\
        & & & & & & Narrow comp. & Broad comp. & \\
        & & [$\mu$m] & [K] & [s$^{-1}$] & & \multicolumn{2}{c}{[erg s$^{-1}$ cm$^{-2}$]} & \\
        \hline
        & R-11 & 4.63019 & 6552.7 & 35.05 & 25 & 5.28$\pm$0.39$\times$10$^{-15}$ & 3.31$\pm$0.07$\times$10$^{-14}$ & N & Y \\
        & R-9 & 4.64481 & 6427.9 & 34.43 & 21 & 7.91$\pm$2.35$\times$10$^{-15}$ & 2.70$\pm$0.80$\times$10$^{-14}$ & N & N \\
        & R-6 & 4.66750 & 6281.3 & 33.22 & 15 & 4.41$\pm$0.37$\times$10$^{-15}$ & 2.30$\pm$0.06$\times$10$^{-14}$ & N & Y \\
        & R-5 & 4.67528 & 6243.3 & 32.69 & 13 & 5.93$\pm$1.45$\times$10$^{-15}$ & 1.63$\pm$0.25$\times$10$^{-14}$ & Y & N \\
        & R-4 & 4.68315 & 6210.7 & 32.02 & 11 & 2.60$\pm$0.37$\times$10$^{-15}$ & 2.07$\pm$0.07$\times$10$^{-14}$ & N & Y \\
        & R-1 & 4.70742 & 6145.5 & 27.73 & 5 & 2.25$\pm$0.66$\times$10$^{-15}$ & 1.54$\pm$0.12$\times$10$^{-14}$ & N & Y \\
        & P-1 & 4.73265 & 6129.2 & 68.19 & 1 & 1.40$\pm$0.96$\times$10$^{-15}$ & 9.67$\pm$1.64$\times$10$^{-15}$ & N & N \\
        & P-4 & 4.75886 & 6161.8 & 38.31 & 7 & 3.62$\pm$0.60$\times$10$^{-15}$ & 1.75$\pm$0.21$\times$10$^{-14}$ & N & Y \\
        & P-5 & 4.76782 & 6183.6 & 37.04 & 9 & 3.96$\pm$1.66$\times$10$^{-15}$ & 2.02$\pm$0.33$\times$10$^{-14}$ & N & N \\
        & P-6 & 4.77689 & 6210.7 & 36.15 & 11 & 4.10$\pm$1.59$\times$10$^{-15}$ & 2.31$\pm$0.32$\times$10$^{-14}$ & N & N \\
        & P-7 & 4.78608 & 6243.3 & 35.48 & 13 & 3.96$\pm$1.31$\times$10$^{-15}$ & 2.64$\pm$0.22$\times$10$^{-14}$ & N & N \\
        & P-8 & 4.79537 & 6281.3 & 34.94 & 15 & 4.24$\pm$1.52$\times$10$^{-15}$ & 2.61$\pm$0.27$\times$10$^{-14}$ & N & N \\
        & P-9 & 4.80479 & 6324.7 & 34.48 & 17 & 4.80$\pm$0.90$\times$10$^{-15}$ & 2.51$\pm$0.20$\times$10$^{-14}$ & N & Y \\
        & P-15 & 4.86370 & 6699.2 & 32.48 & 29 & 5.79$\pm$1.32$\times$10$^{-15}$ & 2.95$\pm$0.23$\times$10$^{-14}$ & Y & N \\
        & P-16 & 4.87394 & 6780.5 & 32.21 & 31 & 3.86$\pm$1.19$\times$10$^{-15}$ & 3.17$\pm$0.22$\times$10$^{-14}$ & Y & N \\
        & P-17 & 4.88429 & 6867.3 & 31.95 & 33 & 4.44$\pm$1.44$\times$10$^{-15}$ & 2.99$\pm$0.25$\times$10$^{-14}$ & Y & N \\
        & P-18 & 4.89477 & 6959.4 & 31.70 & 35 & 4.92$\pm$0.68$\times$10$^{-15}$ & 3.07$\pm$0.15$\times$10$^{-14}$ & N & Y \\
        & P-20 & 4.91609 & 7159.9 & 31.21 & 39 & 3.04$\pm$0.42$\times$10$^{-15}$ & 3.23$\pm$0.07$\times$10$^{-14}$ & N & Y \\
        & P-21 & 4.92694 & 7268.3 & 30.97 & 41 & 3.62$\pm$1.29$\times$10$^{-15}$ & 2.92$\pm$0.23$\times$10$^{-14}$ & N & N \\
        & P-22 & 4.93792 & 7382.0 & 30.73 & 43 & 4.68$\pm$0.65$\times$10$^{-15}$ & 2.82$\pm$0.13$\times$10$^{-14}$ & N & Y \\
        & P-24 & 4.96025 & 7625.6 & 30.27 & 47 & 3.55$\pm$1.90$\times$10$^{-15}$ & 2.78$\pm$0.33$\times$10$^{-14}$ & N & N \\
        & P-25 & 4.97161 & 7755.5 & 30.05 & 49 & 3.70$\pm$1.36$\times$10$^{-15}$ & 2.65$\pm$0.22$\times$10$^{-14}$ & Y & N \\
        & P-26 & 4.98310 & 7890.7 & 29.82 & 51 & 2.44$\pm$1.61$\times$10$^{-15}$ & 3.04$\pm$0.27$\times$10$^{-14}$ & Y & N \\
        & P-28 & 5.00647 & 8177.3 & 29.38 & 55 & 2.85$\pm$1.51$\times$10$^{-15}$ & 2.72$\pm$0.25$\times$10$^{-14}$ & N & N \\
        & P-32 & 5.05482 & 8814.6 & 28.52 & 63 & 1.85$\pm$1.47$\times$10$^{-15}$ & 2.57$\pm$0.24$\times$10$^{-14}$ & N & N \\
        & P-33 & 5.06725 & 8987.2 & 28.31 & 65 & 1.62$\pm$0.43$\times$10$^{-15}$ & 2.29$\pm$0.07$\times$10$^{-14}$ & N & Y \\
        & P-37 & 5.11838 & 9731.0 & 27.47 & 73 & 1.43$\pm$0.62$\times$10$^{-15}$ & 2.05$\pm$0.12$\times$10$^{-14}$ & N & Y \\
        & P-38 & 5.13151 & 9930.1 & 27.26 & 75 & 1.27$\pm$0.78$\times$10$^{-15}$ & 1.89$\pm$0.14$\times$10$^{-14}$ & N & Y \\
        \hline
        \ce{^{12}CO v=3-2} & R-30 & 4.56764 & 11797.2 & 56.33 & 63 & 1.55$\pm$0.38$\times$10$^{-15}$ & 1.00$\pm$0.14$\times$10$^{-14}$ & N & N \\
        & R-27 & 4.58436 & 11315.7 & 55.55 & 57 & 1.96$\pm$0.50$\times$10$^{-15}$ & 1.08$\pm$0.16$\times$10$^{-14}$ & N & N \\
        & R-23 & 4.60800 & 10748.0 & 54.46 & 49 & 1.68$\pm$0.38$\times$10$^{-15}$ & 1.10$\pm$0.13$\times$10$^{-14}$ & N & N \\
        & R-22 & 4.61415 & 10619.3 & 54.17 & 47 & 1.65$\pm$0.27$\times$10$^{-15}$ & 1.29$\pm$0.10$\times$10$^{-14}$ & N & N \\
        & R-21 & 4.62041 & 10496.0 & 53.88 & 45 & 1.82$\pm$0.38$\times$10$^{-15}$ & 1.31$\pm$0.12$\times$10$^{-14}$ & N & N \\
        & R-20 & 4.62676 & 10378.0 & 53.59 & 43 & 2.65$\pm$1.83$\times$10$^{-15}$ & 9.14$\pm$2.35$\times$10$^{-15}$ & Y & N \\
        & R-18 & 4.63976 & 10157.9 & 52.97 & 39 & 1.63$\pm$0.45$\times$10$^{-15}$ & 9.88$\pm$1.94$\times$10$^{-15}$ & Y & N \\
        & R-17 & 4.64641 & 10055.9 & 52.66 & 37 & 1.49$\pm$0.33$\times$10$^{-15}$ & 9.49$\pm$1.20$\times$10$^{-15}$ & N & N \\
        & R-12 & 4.68118 & 9626.2 & 50.89 & 27 & 2.11$\pm$0.42$\times$10$^{-15}$ & 1.43$\pm$0.19$\times$10$^{-14}$ & N & N \\
        & R-11 & 4.68844 & 9556.3 & 50.49 & 25 & 1.38$\pm$0.27$\times$10$^{-15}$ & 9.24$\pm$1.15$\times$10$^{-15}$ & Y & N \\
        & R-10 & 4.69581 & 9491.8 & 50.06 & 23 & 1.54$\pm$0.45$\times$10$^{-15}$ & 8.49$\pm$1.71$\times$10$^{-15}$ & N & N \\
        & R-9 & 4.70328 & 9432.7 & 49.59 & 21 & 1.70$\pm$1.07$\times$10$^{-15}$ & 8.16$\pm$2.01$\times$10$^{-15}$ & N & N \\
        & R-8 & 4.71085 & 9378.9 & 49.08 & 19 & 1.88$\pm$0.79$\times$10$^{-15}$ & 7.95$\pm$2.35$\times$10$^{-15}$ & N & N \\
        & R-5 & 4.73421 & 9249.8 & 47.09 & 13 & 1.22$\pm$0.29$\times$10$^{-15}$ & 8.07$\pm$1.09$\times$10$^{-15}$ & N & N \\
        & P-6 & 4.83745 & 9217.5 & 52.11 & 11 & 1.54$\pm$0.42$\times$10$^{-15}$ & 8.30$\pm$1.62$\times$10$^{-15}$ & N & N \\
        & P-16 & 4.93614 & 9782.0 & 46.44 & 31 & 1.76$\pm$0.43$\times$10$^{-15}$ & 1.13$\pm$0.15$\times$10$^{-14}$ & N & N \\
        & P-19 & 4.96812 & 10055.9 & 45.34 & 37 & 1.63$\pm$0.39$\times$10$^{-15}$ & 1.10$\pm$0.13$\times$10$^{-14}$ & N & N \\
        \hline
        \ce{^{13}CO} v=1-0 & R-30 & 4.55970 & 5607.8 & 18.60 & 126 & - & 5.10$\pm$0.40$\times$10$^{-15}$ & N & Y \\
        & R-26 & 4.58188 & 4992.5 & 18.25 & 110 & - & 4.65$\pm$0.30$\times$10$^{-15}$ & N & Y \\
        & R-23 & 4.59947 & 4585.4 & 17.97 & 98 & - & 8.28$\pm$0.93$\times$10$^{-15}$ & N & N \\
        & R-22 & 4.60551 & 4460.0 & 17.88 & 94 & - & 6.71$\pm$0.78$\times$10$^{-15}$ & N & N \\
        & R-21 & 4.61165 & 4339.9 & 17.78 & 90 & - & 6.39$\pm$0.81$\times$10$^{-15}$ & Y & N \\
        & R-18 & 4.63062 & 4010.5 & 17.48 & 78 & - & 7.30$\pm$0.31$\times$10$^{-15}$ & N & Y \\
        & R-16 & 4.64373 & 3817.0 & 17.26 & 70 & - & 8.36$\pm$0.80$\times$10$^{-15}$ & Y & N \\
        & R-15 & 4.65043 & 3728.0 & 17.15 & 66 & - & 7.45$\pm$0.72$\times$10$^{-15}$ & Y & N \\
        & R-13 & 4.66411 & 3565.8 & 16.91 & 58 & - & 8.69$\pm$0.73$\times$10$^{-15}$ & N & N \\
        & R-10 & 4.68535 & 3361.6 & 16.51 & 46 & - & 1.01$\pm$0.08$\times$10$^{-14}$ & N & N \\
        \hline
    \end{tabular}
    \label{tab:CO_IntFluxes}
\end{table}
\clearpage
\begin{table}[ht!]
    \ContinuedFloat
    \centering
    \caption{Continuation of Table \ref{tab:CO_IntFluxes}}
    \begin{tabular}{c c c c c c c c c c}
        \hline\hline
        Molecule & Rot. Trans. & Wavelength & $E_\textnormal{up}$ & $A_\textnormal{ul}$ & $g_\textnormal{up}$ & \multicolumn{2}{c}{Int. Flux$^{(\alpha)}$} & Line prof.$^{(\beta)}$ & Blend$^{(\gamma)}$ \\
        & & & & & & Narrow comp. & Broad comp. & \\
        & & [$\mu$m] & [K] & [s$^{-1}$] & & \multicolumn{2}{c}{[erg s$^{-1}$ cm$^{-2}$]} & \\
        \hline
        & R-9 & 4.69262 & 3304.0 & 16.36 & 42 & - & 8.79$\pm$0.74$\times$10$^{-15}$ & Y & N \\
        & R-7 & 4.70747 & 3204.4 & 16.01 & 34 & - & 6.05$\pm$0.95$\times$10$^{-15}$ & N & Y \\
        & R-6 & 4.71504 & 3162.5 & 15.79 & 30 & - & 7.69$\pm$1.33$\times$10$^{-15}$ & N & N \\
        & R-4 & 4.73047 & 3094.4 & 15.22 & 22 & - & 6.12$\pm$0.77$\times$10$^{-15}$ & N & N \\
        & R-3 & 4.73834 & 3068.2 & 14.81 & 18 & - & 6.06$\pm$0.92$\times$10$^{-15}$ & N & N \\
        & R-2 & 4.74631 & 3047.2 & 14.20 & 14 & - & 4.87$\pm$1.18$\times$10$^{-15}$ & N & N \\
        & P-2 & 4.78771 & 3021.0 & 21.50 & 6 & - & 5.84$\pm$1.17$\times$10$^{-15}$ & N & N \\
        & P-3 & 4.79630 & 3031.5 & 19.25 & 10 & - & 7.03$\pm$0.99$\times$10$^{-15}$ & N & N \\
        & P-4 & 4.80499 & 3047.2 & 18.23 & 14 & - & 4.86$\pm$1.01$\times$10$^{-15}$ & N & Y \\
        & P-6 & 4.82270 & 3094.4 & 17.20 & 22 & - & 1.01$\pm$0.05$\times$10$^{-14}$ & N & Y \\
        & P-7 & 4.83172 & 3125.8 & 16.89 & 26 & - & 9.66$\pm$1.11$\times$10$^{-15}$ & Y & N \\
        & P-10 & 4.85942 & 3251.6 & 16.22 & 38 & - & 8.91$\pm$1.05$\times$10$^{-15}$ & N & N \\
        & P-12 & 4.87844 & 3361.6 & 15.89 & 46 & - & 7.67$\pm$1.00$\times$10$^{-15}$ & Y & N \\
        & P-13 & 4.88812 & 3424.4 & 15.75 & 50 & - & 8.40$\pm$1.15$\times$10$^{-15}$ & N & N \\
        & P-15 & 4.90781 & 3565.8 & 15.47 & 58 & - & 7.79$\pm$2.00$\times$10$^{-15}$ & N & N \\
        & P-16 & 4.91783 & 3644.3 & 15.35 & 62 & - & 7.16$\pm$0.82$\times$10$^{-15}$ & N & N \\
        & P-17 & 4.92796 & 3728.0 & 15.23 & 66 & - & 9.16$\pm$0.26$\times$10$^{-15}$ & N & Y \\
        & P-18 & 4.93821 & 3817.0 & 15.11 & 70 & - & 7.35$\pm$0.56$\times$10$^{-15}$ & N & Y \\
        & P-25 & 5.01328 & 4585.4 & 14.34 & 98 & - & 5.03$\pm$0.39$\times$10$^{-15}$ & N & Y \\
        & P-27 & 5.03582 & 4851.7 & 14.13 & 106 & - & 4.32$\pm$0.58$\times$10$^{-15}$ & N & Y \\
        & P-35 & 5.13110 & 6123.2 & 13.33 & 138 & - & 1.98$\pm$0.59$\times$10$^{-15}$ & N & Y \\
        \hline
        \ce{C^{18}O v=1-0} & R-12 & 4.68007 & 3485.0 & 16.66 & 27 & - & 1.51$\pm$0.18$\times$10$^{-15}$ & Y & N \\
        & R-10 & 4.69434 & 3354.6 & 16.39 & 23 & - & 1.87$\pm$0.16$\times$10$^{-15}$ & N & N \\
        & R-7 & 4.71646 & 3198.0 & 15.88 & 17 & - & 1.68$\pm$0.32$\times$10$^{-15}$ & Y & N \\
        & R-5 & 4.73170 & 3119.7 & 15.42 & 13 & - & 7.80$\pm$1.98$\times$10$^{-16}$ & Y & N \\
        & R-4 & 4.73947 & 3088.4 & 15.11 & 11 & - & 1.26$\pm$0.24$\times$10$^{-15}$ & N & N \\
        & P-8 & 4.84983 & 3156.3 & 16.51 & 15 & - & 1.64$\pm$0.09$\times$10$^{-15}$ & N & Y \\
        & P-9 & 4.85907 & 3198.0 & 16.29 & 17 & - & 1.32$\pm$0.18$\times$10$^{-15}$ & N & N \\
        & P-11 & 4.87786 & 3297.2 & 15.93 & 21 & - & 1.09$\pm$0.14$\times$10$^{-15}$ & Y & N \\
        & P-15 & 4.91678 & 3558.0 & 15.36 & 29 & - & 1.77$\pm$0.06$\times$10$^{-15}$ & N & Y \\
        & P-17 & 4.93692 & 3719.7 & 15.11 & 33 & - & 1.52$\pm$0.13$\times$10$^{-15}$ & N & N \\
        & P-18 & 4.94717 & 3808.3 & 15.00 & 35 & - & 1.44$\pm$0.07$\times$10$^{-15}$ & N & Y \\
        \hline
    \end{tabular}
    \tablefoot{($\alpha$): for the \ce{^{13}CO} and \ce{C^{18}O} v=1-0 transitions, we list the integrated flux of the single component under the column of the broad component, assuming that the observed emission originates from the Keplerian disk. \\
    ($\beta$): this entry lists all the transitions that have been used in acquiring the median, normalised line profile (Y) and those that have only been used for the rotational diagram (N). \\
    ($\gamma$): this entry lists all the transitions that are blended with other transitions.}
    \label{tab:CO_IntFluxes}
\end{table}

\clearpage
\subsection{Full iSHELL model}
\begin{figure*}[ht!]
    \centering
    \includegraphics[width=\textwidth]{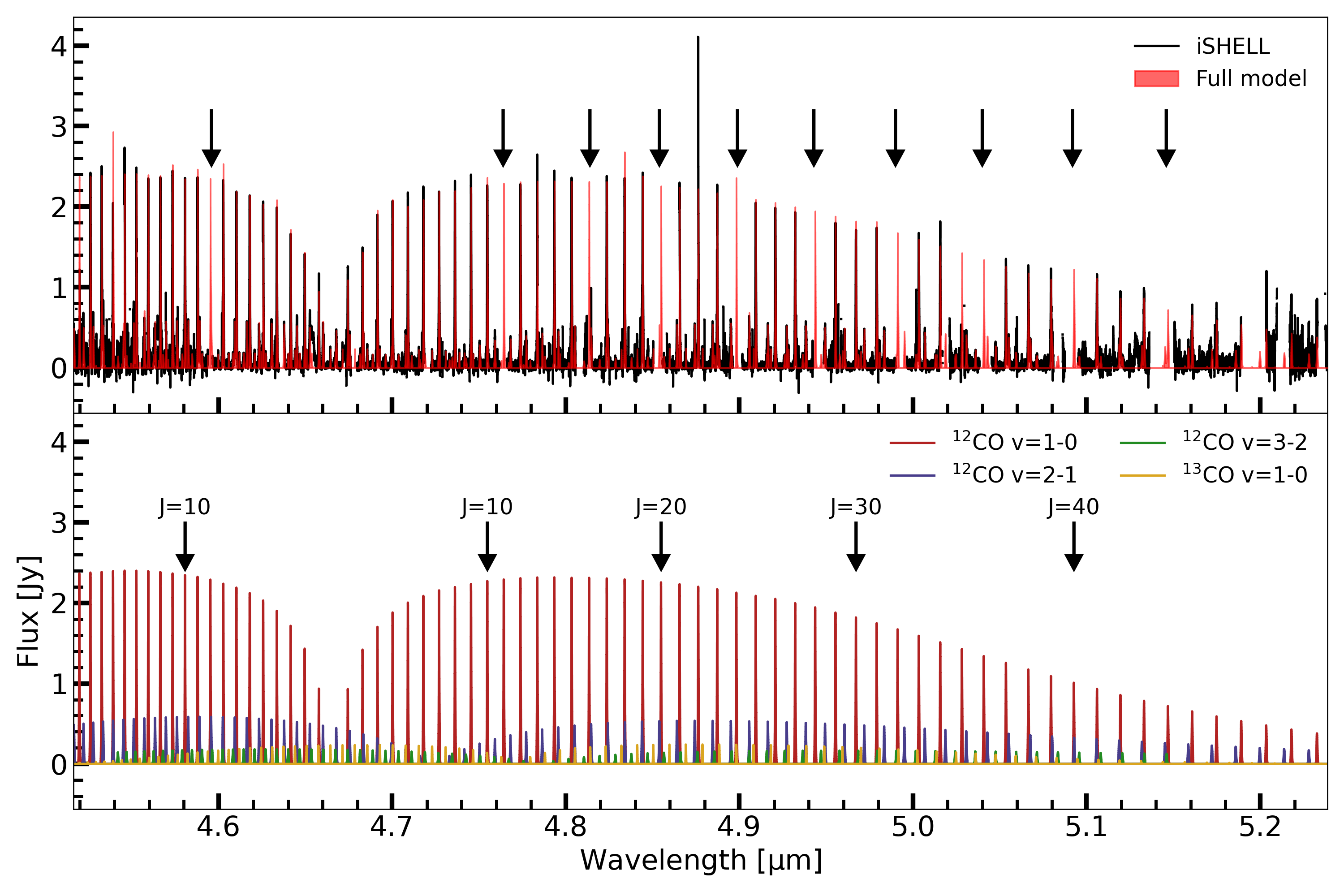}
    \caption{The continuum-subtracted iSHELL spectrum of DR~Tau with a model based on the best-fitting parameters obtained through the rotational diagram analysis. The gaps visible (highlighted by the black arrows in the top panel) in the iSHELL spectrum correspond to removed telluric features. The top panel shows the full model, whereas the bottom panel shows the contribution of each band separately. For \ce{^{12}CO}, the narrow and broad component are shown together. In addition, in the bottom panel we highlight the \ce{^{12}CO} transitions which have an upper level $J$-value of 10, 20, 30, or 40.}
    \label{fig:iSHELL-Model}
\end{figure*}

\clearpage
\subsection{$\chi^2$-maps of the \ce{CO} rotational diagrams}
\begin{figure*}[ht!]
    \centering
    \includegraphics[width=0.85\textwidth]{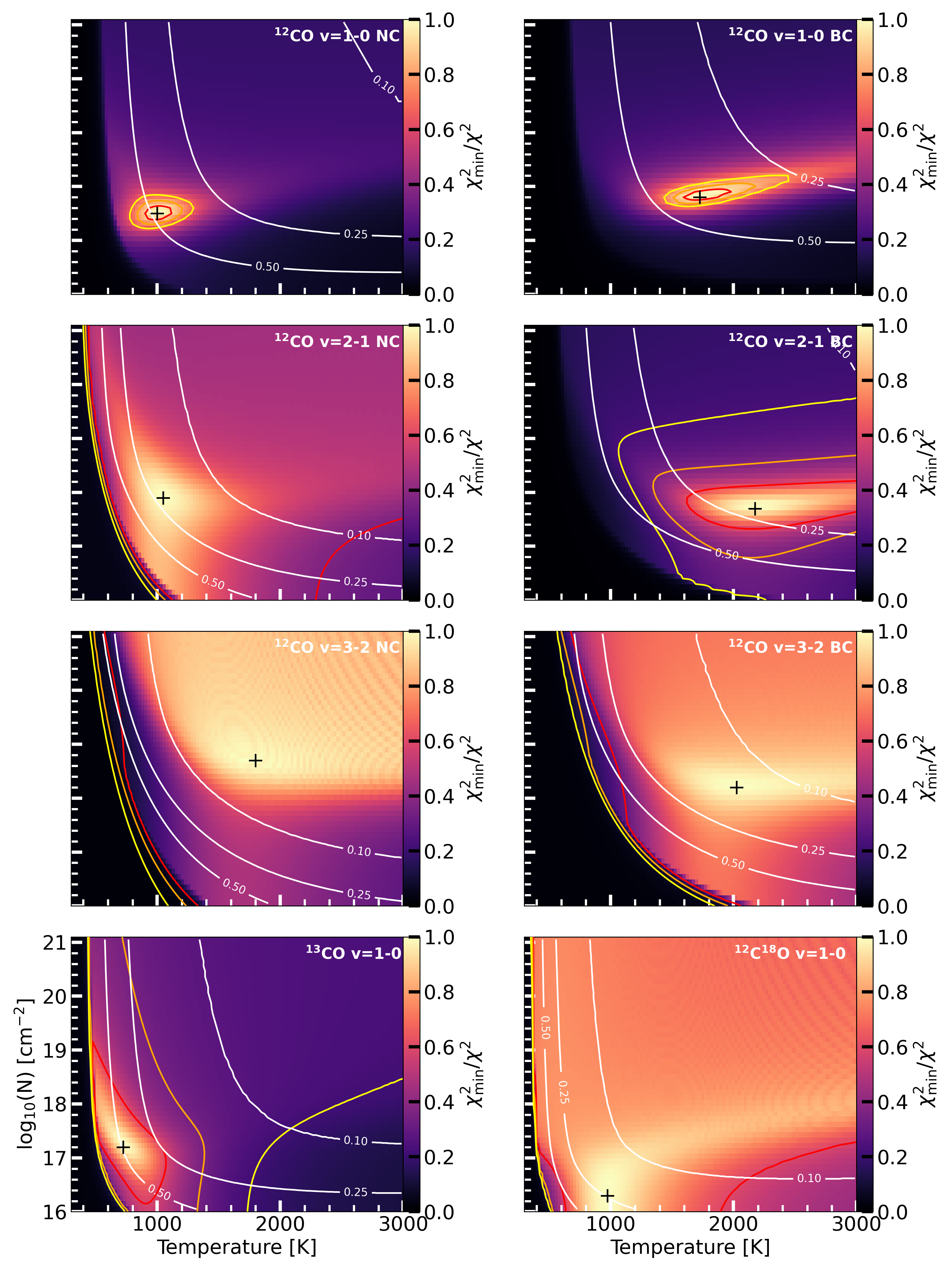}
    \caption{Similar as Figure \ref{fig:MIRI-Chi2}, but for the iSHELL \ce{CO} rotational diagrams. The top row shows those for the narrow (left) and broad (right) component of \ce{^{12}CO} v=1-0 transitions. Those for the \ce{^{12}CO} v=2-1 and v=3-2 transitions are shown on the second and third row, respectively, whereas the ones for the \ce{^{13}CO} and \ce{C^{18}O} (using VLT-CRIRES data) are shown on the bottom row.}
    \label{fig:iSHELL-RD-Chi2}
\end{figure*}

\subsection{Optical depth of \ce{^{12}CO}}
\begin{figure*}[ht!]
    \centering
    \includegraphics[width=0.5\columnwidth]{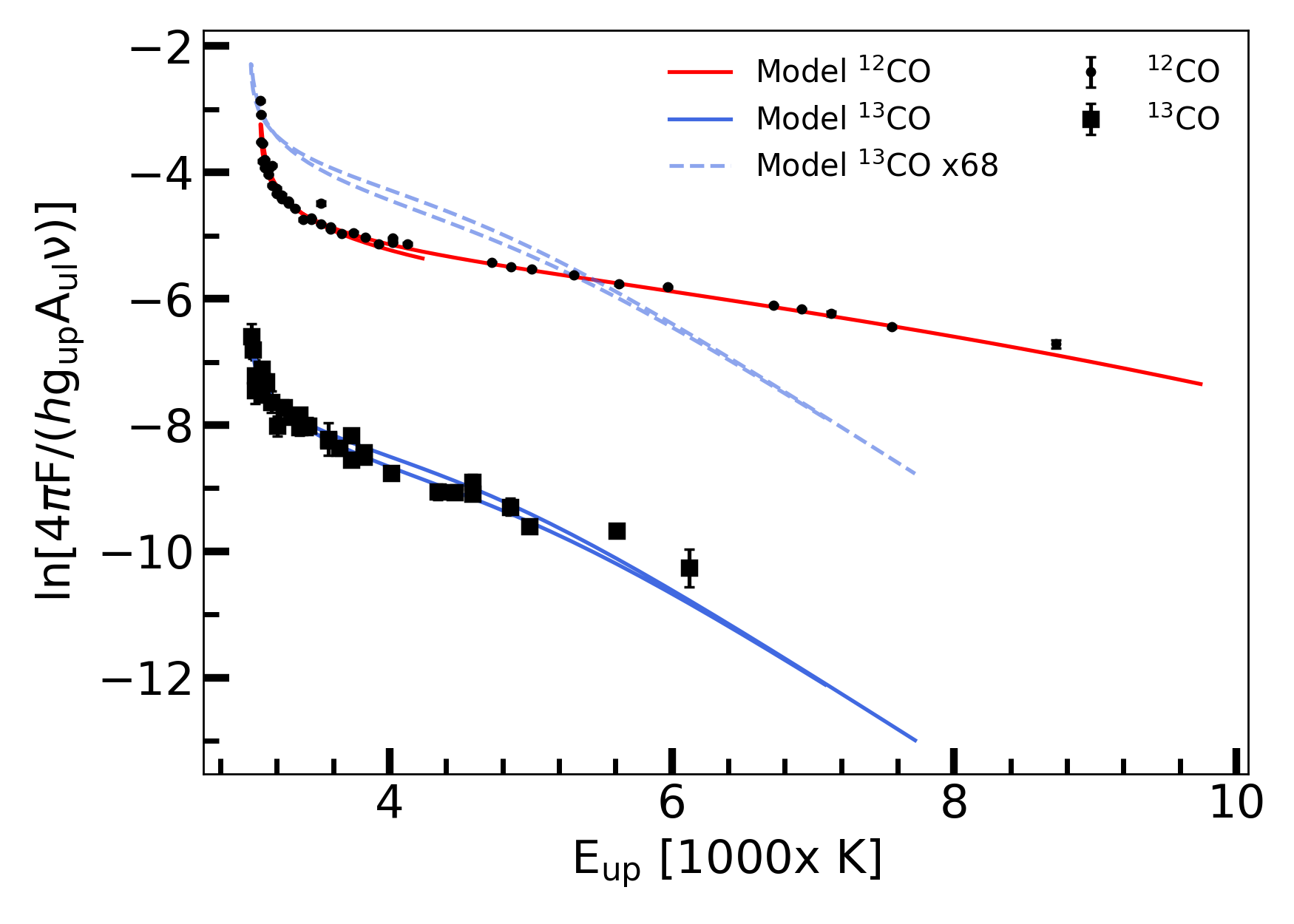}
    \caption{The rotational diagrams for \ce{^{12}CO} v=1-0 (dots) and \ce{^{13}CO} v=1-0 (squares). The blue dashed line indicates the model of \ce{^{13}CO} ($T$=750 K) multiplied by a factor of 68 to account for the isotopologue ratio. The fluxes shown for the \ce{^{12}CO} are a summation of the narrow ($T$=1000 K) and broad ($T$=1725 K) components.}
    \label{fig:RD-OD}
\end{figure*}

\subsection{LTE versus non-LTE}
\begin{figure*}[ht!]
    \centering
    \includegraphics[width=\textwidth]{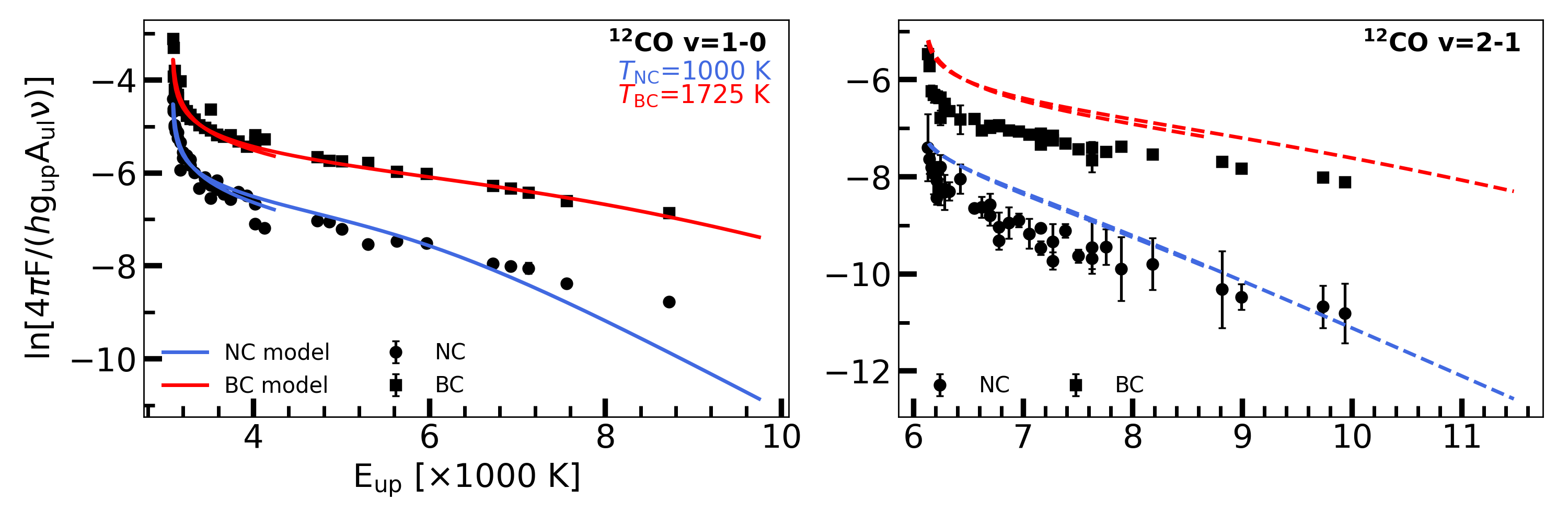}
    \caption{The left panel shows the \ce{^{12}CO} v=1-0 rotational diagrams (see also Figure \ref{fig:iSHELL-RDs}). The right panel shows the non-LTE test, where the \ce{^{12}CO} v=2-1 integrated fluxes are shown together with expected the models based on the results of the \ce{^{12}CO} v=1-0 rotational diagrams. As the models overproduce the observed \ce{^{12}CO} v=2-1 fluxes, invalidating the assumption that the levels can be characterised by a single temperature.}
    \label{fig:12CO-nonLTE}
\end{figure*}

\clearpage
\subsection{Convolution iSHELL to JWST-MIRI resolution}
\begin{figure*}[ht!]
    \centering
    \includegraphics[width=\textwidth]{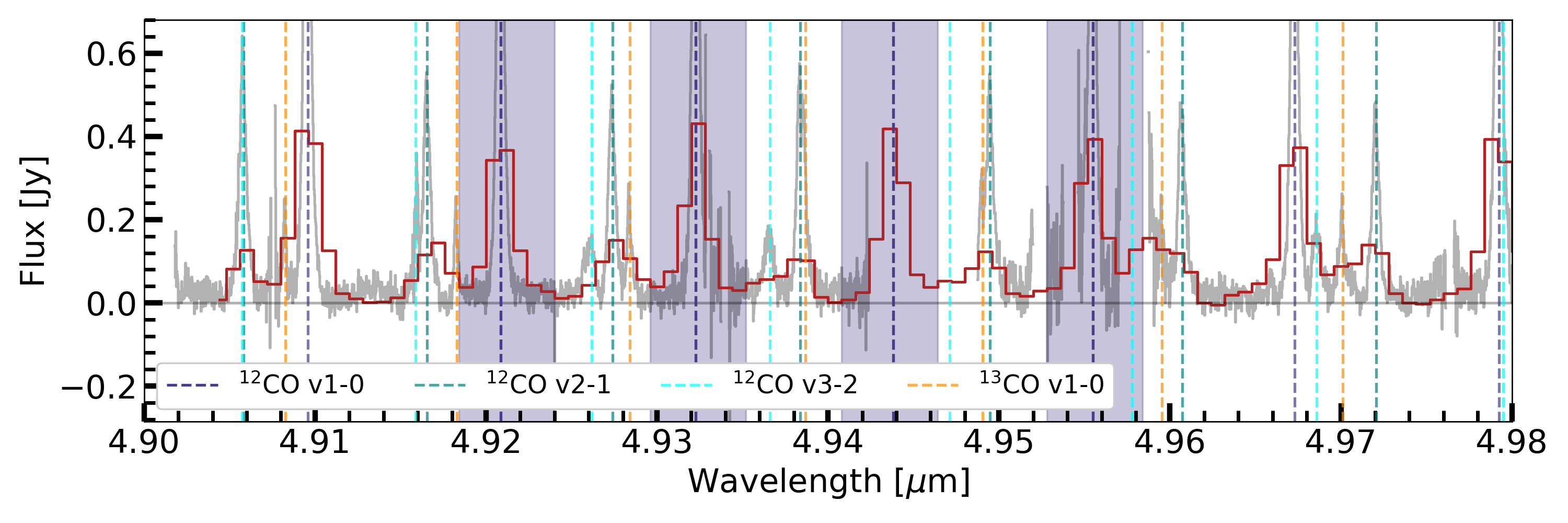}
    \caption{A zoom-in of Figure \ref{fig:iSHELL-MIRI} on the JWST-MIRI data (red) within the wavelength region 4.90-4.98 $\mu$m. The iSHELL observations are shown here in grey. The different transitions in this wavelength region of the \ce{^{12}CO} v=1-0 (dark blue), v=2-1 (teal), v=3-2 (cyan), and the \ce{^{13}CO} v=1-0 (orange) transitions are shown to identify the different lines and highlight corresponding line overlaps. The dark blue shaded areas indicate the (relatively) isolated \ce{^{12}CO} v=1-0 transitions that have been used to determine the pseudo-Voigt line profile of the JWST-MIRI data.}
    \label{fig:iSHELL-MIRI-SelLines}
\end{figure*}
\begin{figure*}[h!]
    \centering
    \includegraphics[width=12cm]{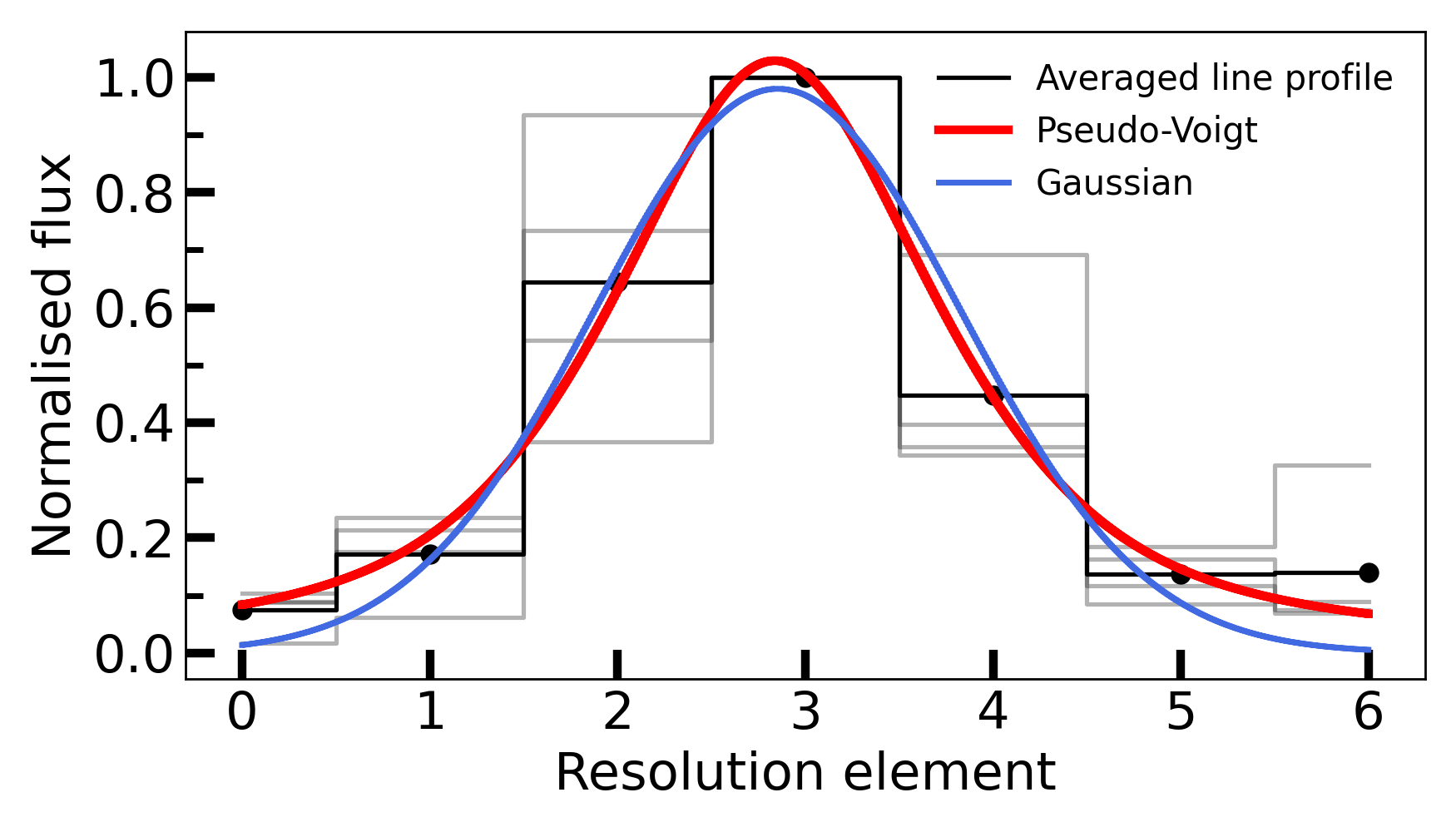}
    \caption{The averaged, normalised line profile of the 4 selected, unblended \ce{^{12}CO} v=1-0 transitions observed with JWST-MIRI (see also Figure \ref{fig:iSHELL-MIRI-SelLines}. The faint grey lines indicate the individual, normalised lines, whereas the red and blue lines display, respectively, the fitted pseudo-Voigt and Gaussian profiles.}
    \label{fig:PdV-LineProfile}
\end{figure*}

\clearpage
\subsubsection{Gaussian line profile}
\begin{figure}[ht!]
    \centering
    \includegraphics[width=12cm]{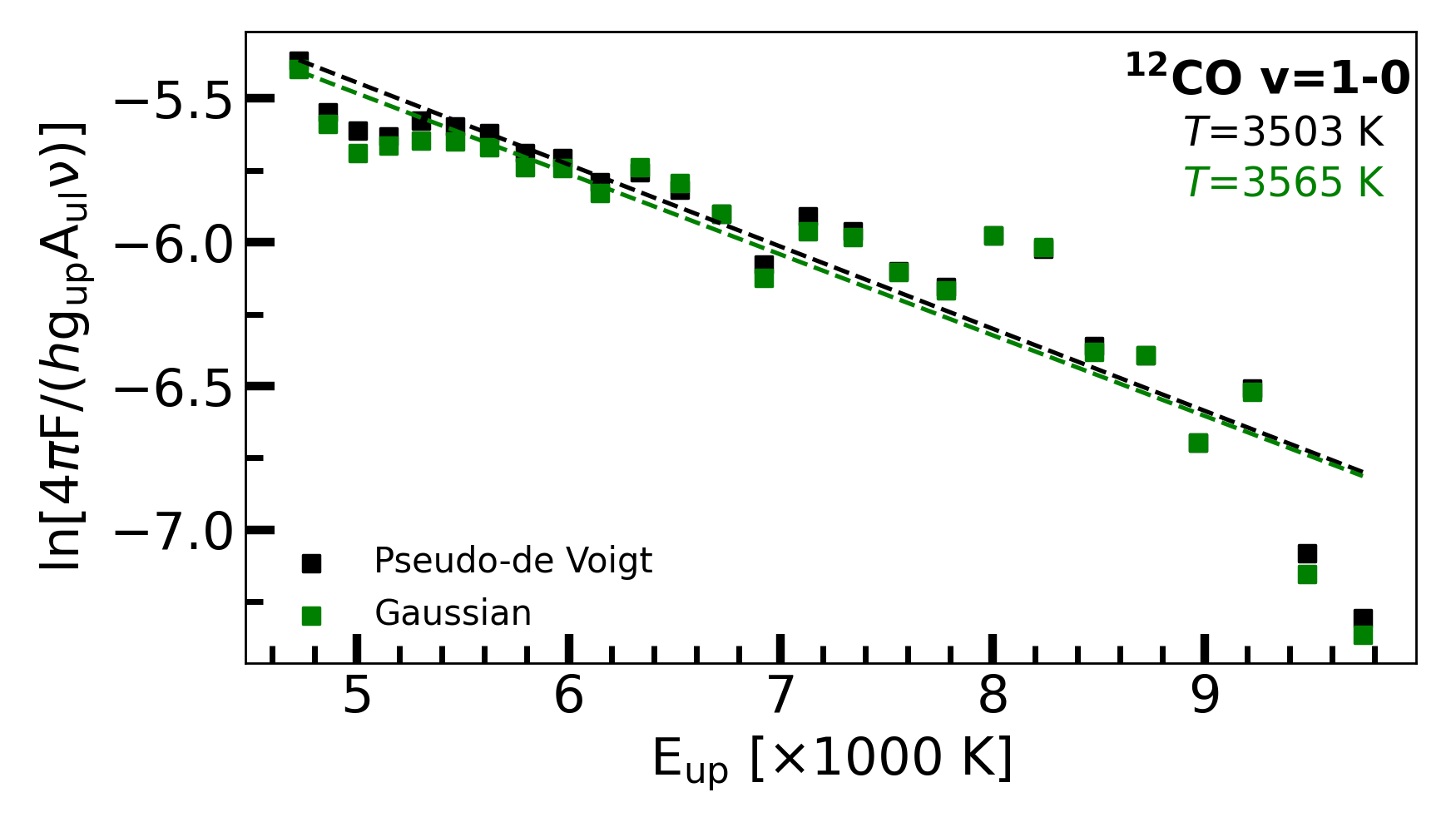}
    \caption{Simple rotational diagrams of the JWST-MIRI integrated fluxes when considering a pseudo-Voigt (black) or a Gaussian (green) line profile. Straight line fits suggest excitation temperatures of $T$=3503 K when considering the pseudo-Voigt profile and $T$=3565 when using a Gaussian.}
    \label{fig:RD-MIRI-LP}
\end{figure}

\begin{figure*}[ht!]    
    \centering
    \includegraphics[width=14cm]{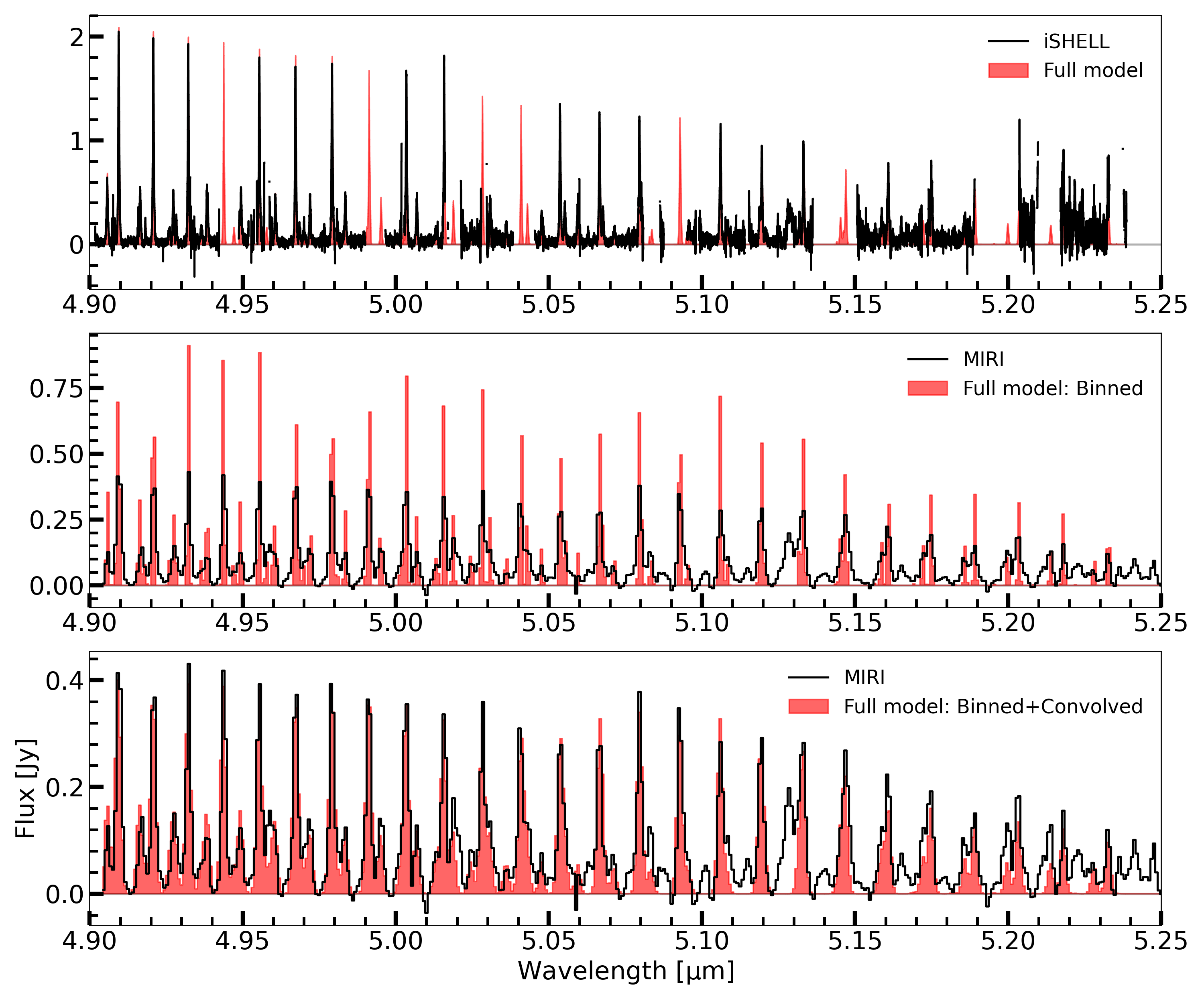}
    \caption{The same as Figure \ref{fig:iSHELL-MIRI-BC}, but for the Gaussian line profile.}
    \label{fig:iSHELL-MIRI-BC-Gauss}
\end{figure*}

\end{appendix}

\end{document}